\documentclass[twocolumn]{aastex631}

\usepackage{amsmath}
\usepackage{float}
\usepackage{mleftright}
\graphicspath{{./}{figures/}}
\usepackage{graphicx}
\usepackage[caption=false]{subfig}
\usepackage{url}
\usepackage{natbib}
\usepackage{mathtools,nccmath}
\usepackage{placeins}
\usepackage{tablefootnote}
\usepackage{tabularx}
\def\fdg{\hbox{$.\!\!^\circ$}}

\graphicspath{{./}{figures/}}

\begin{document}

\title{Testing the correlation between bending angle and polarization properties of bent radio galaxies}

% \correspondingauthor{S. Vanderwoude}
% \email{vanderwoude@astro.utoronto.ca}

\author[0009-0004-7773-1618]{S. Vanderwoude}
\affiliation{David A. Dunlap Department of Astronomy and Astrophysics, University of Toronto, Toronto, ON M5S 3H4, Canada}
\affiliation{Dunlap Institute for Astronomy and Astrophysics, University of Toronto, Toronto, ON M5S 3H4, Canada}

\author[0000-0002-5815-8965]{E. Osinga}
\affiliation{Dunlap Institute for Astronomy and Astrophysics, University of Toronto, Toronto, ON M5S 3H4, Canada}
\affiliation{Leiden Observatory, Leiden University, PO Box 9513, NL-2300 RA Leiden, The Netherlands}

\author[0000-0002-3382-9558]{B. M. Gaensler}
\affiliation{Department of Astronomy and Astrophysics, University of California, Santa Cruz, Santa Cruz, CA 95064, USA}
\affiliation{Dunlap Institute for Astronomy and Astrophysics, University of Toronto, Toronto, ON M5S 3H4, Canada}
\affiliation{David A. Dunlap Department of Astronomy and Astrophysics, University of Toronto, Toronto, ON M5S 3H4, Canada}

\author[0000-0001-7722-8458]{J. L. West}
\affiliation{Dominion Radio Astrophysical Observatory, Herzberg Astronomy and Astrophysics, National Research Council Canada, P.O. Box 248, Penticton, BC V2A 6J9, Canada}

\author[0000-0002-0587-1660]{R. J. van Weeren}
\affiliation{Leiden Observatory, Leiden University, PO Box 9513, NL-2300 RA Leiden, The Netherlands}

\begin{abstract}
%%250 words%%
The bending of radio galaxies in galaxy clusters is expected to be caused by interactions with the local environment. The physical processes responsible for jet bending, and their influence on the polarization properties of radio galaxies, remain poorly understood, leading to the question of whether jet properties in bent radio galaxies differ from those in linear radio galaxies. Using a sample of 24 polarized bent radio galaxies, observed with the Karl G. Jansky Very Large Array at 1--2 GHz, we test for correlation of bending angle with polarization parameters measuring Faraday rotation, intrinsic fractional polarization, and Faraday rotation dispersion, used here as a measure of turbulence along the line of sight. We find no statistically significant correlations. At the spatial resolution of our dataset (3--46 kpc, median 18.4 kpc), our results indicate that we are primarily probing larger-scale intracluster medium effects not related to bending angle. The absence of a statistically significant correlation suggests that bent radio galaxies are reliable probes of intracluster magnetic fields, because their intrinsic properties do not appear to introduce systematic biases into measured polarization parameters. We do detect a preference for source magnetic field vectors to align with the direction of jet bending. Finally, we estimate that the POSSUM and SKA surveys will contain $\gtrsim$300 and $\gtrsim$1000 polarized radio galaxies, respectively, providing large future samples with a range of bending angles and similar redshift distribution and number of beams per source as in our sample, enabling our results to be tested with greater statistical power.

\end{abstract}


\section{Introduction} \label{sec:intro}

Radio galaxy morphologies are widely varied, and it is generally assumed that a bent or disturbed radio galaxy morphology results from the interaction between the galaxy's jets and its surrounding environment. The typical model for a radio galaxy consists of two collimated, synchrotron-emitting jets of plasma ejected to very large distances from opposite sides of a central active galactic nucleus (AGN) of a host galaxy. With no interactions with the surrounding environment, the jets should propagate out linearly. However, the presence of density variations in the intracluster medium (ICM) in galaxy clusters and ram pressure from the motion of radio galaxies through the ICM can cause radio galaxy jets to bend and their lobes to be distorted, resulting in a variety of morphologies \citep{Begelman+79,Jones&Owen1979,Vallee+81,Hardcastle+05,Freeland+08,Morsony+13,Garon+19,ODea&Baum2023}. Under this assumption, bent radio galaxies have been used to search for and identify galaxy clusters (e.g. \citealt{Blanton+00,Wing&Blanton2011,Banfield+16,Paterno-Mahler+17}).

Radio galaxies are broadly categorized using the Fanaroff-Riley (FR) classification system \citep{Fanaroff&Riley}, where FRI galaxies are brighter at their cores and decrease in brightness with distance from the central AGN, while FRII galaxies are brighter in their lobes and fainter closer to the AGN. Bent radio galaxies require additional classifications. These include wide-angle tail (WAT) and narrow-angle tail (NAT) bent radio galaxies (often grouped together under the broader classification of head-tail radio galaxies; \citealt{Rudnick&Owen1976,Owen&Rudnick1979,Blanton+00,Dehghan+14,Missaglia+19,Pal+23}). Other classifications include ``winged'', or X- and Z-shaped, radio galaxies \citep{Yang+19,Bera+20,Bera+22}, and double-double radio galaxies \citep{Nandi+12,Mahatma+19}, while some radio galaxies have morphologies that are unable to be classified \citep{Sasmal+22,Bera+24}

Total intensity statistical studies of bent radio galaxies and their environments have been carried out both at low-to-intermediate redshifts \citep{Owen&Rudnick1978,Sakelliou&Merrifield2000,Blanton+01,Freeland+08,Wing&Blanton2013} as well as high redshifts \citep{Golden-Marx+19,Golden-Marx+21,Golden-Marx+23}. Additional works spanning a wider redshift range have further expanded this picture \citep{O'Brien+18,deVos+21,Morris+22,Vardoulaki+21,Vardoulaki+25,Mingo+19,Garon+19,vanderJagt+25}. 
Despite this extensive body of work in total intensity, no statistical polarization studies of bent radio galaxies have been conducted to date.
If radio galaxy morphology is related to the magnetic properties of the jets (e.g., magnetic field strength or geometry), we might expect to see a correlation between the bending angle of a radio galaxy and polarization parameters that probe these environmental variables, such as Faraday rotation measure (RM), polarized fraction ($p$), and depolarization, potentially due to turbulence in the ICM. In addition, understanding the origin of RM structures seen across radio galaxies jets and lobes (e.g. \citealt{Guidetti+11, Sebokolodi+20}) is important for cluster magnetism studies. Determining whether the observed  RM structure is related to the ICM foreground or whether it is intrinsic to the radio galaxy itself has implication for inferring magnetic field properties of the ICM \citep{Rudnick&Blundell03,Ensslin+03,O'Sullivan+13}.

In this paper, we investigate the question of whether the degree of bending of radio galaxies is correlated with measured polarization properties, using bent radio galaxies in cluster environments, observed with the Karl G. Jansky Very Large Array (VLA) in L-band (1--2 GHz). We organize the paper as follows. In Section \ref{sec:data and sample} we describe our data and our method of sample selection. In Section \ref{sec:data red} we outline our method of analysis, including spectral extraction and modeling. We present the results of our analysis in Section \ref{sec:results}, and we discuss the implications of these results and potential for future work with upcoming surveys in Section \ref{sec:disc}. The paper is summarized in Section \ref{sec:conc}. Throughout this work, we assume a  flat $\Lambda$ cold dark matter ($\Lambda$CDM) cosmology of $H_0 = 70$ km s$^{-1}$ Mpc$^{-1}$, $\Omega_m = 0.3$, and $\Omega_{\Lambda} = 0.7$ (see \citealt{Planck18results}).

\section{Data and sample selection} \label{sec:data and sample}

\subsection{Calibration and Imaging}

We analyze L-band (1--2 GHz) observations of radio galaxies in galaxy clusters obtained with the Karl G. Jansky Very Large Array (VLA) and first presented by \citet{Osinga+22}. The clusters were selected through the Sunyaev–Zel'dovich (SZ) effect, which is insensitive to cluster dynamical state and therefore provides a close-to-unbiased, mass-limited catalogue of clusters (e.g., \citealt{AndradeSantos+17}). From the Planck Early Sunyaev–Zel'dovich (PESZ) catalogue \citep{ESZ11}, we included all clusters at redshift z $<$ 0.35 and declination $>$ $-30^{\circ}$, observable with the VLA\footnote{Correcting a typographical error in \citet{Osinga+22}, we note that the declination cut is $-30^{\circ}$, not $40^{\circ}$}. This yielded 102 clusters. To this set, 24 clusters were added from the first and second Planck SZ source catalogues \citep{PSZ1,PSZ2} that met the same redshift and declination criteria and for which VLA L-band observations were available. In total, 126 clusters were targeted.

The resulting sample has a slightly higher mean M$_{500}$ (the total mass enclosed within a radius R$_{500}$, where the mean density is 500 times the critical density of the Universe at the cluster redshift) of $5.7 \times 10^{14}$ M$_{\odot}$ (standard deviation $2.1 \times 10^{14}$ M$_{\odot}$) compared to the parent PESZ sample in the same redshift range (mean $4.3 \times 10^{14}$ M$_{\odot}$, standard deviation $1.8 \times 10^{14}$ M$_{\odot}$). While this difference reflects the availability of VLA data, it is not expected to introduce major additional selection biases.

Each cluster was observed for approximately 40 minutes, giving a typical root-mean-square (rms) noise level of 20--30 $\mu$Jy/ beam in Stokes I near the field center. Two observations were excluded from further analysis due to severe imaging artefacts, leaving 124 usable clusters. The final images have angular resolutions ranging from 6.2 to 11.9 arcsec, restored to a uniform synthesized beam across frequency.

The data were calibrated by \citet{Osinga+22} using the Common Astronomy Software Application (CASA; \citealt{CASA}). The analysis of \citet{cal_flux_scale} was used to set the flux scale, 3C138 and 3C286 were used for polarization calibration, and 3C147 was used for on-axis leakage correction.

Each galaxy cluster was observed with 16 spectral windows over 1008--2032 MHz, each with a bandwidth of 64 MHz. In each spectral window, the first 6 MHz and the last 10 MHz were removed due to low quality. The spectral windows were then averaged to 6 frequency channels with 8 MHz channel width. Spectral window 8 was lost to radio frequency interference (RFI) in all cases, leaving 90 frequency channels and a total bandwidth of 1008 MHz for each observation. Additional RFI detection with \textit{AOflagger} \citep{AOfluxcalref} flagged individual channels in individual observations. All calibration and imaging was performed by \citet{Osinga+22}, and a more detailed description of the process can be found therein, as well as the final images of all 124 galaxy clusters.

After imaging, all channels in the Stokes \textit{IQU} data cubes for each observation were convolved to the resolution of the lowest frequency channel of that observation, using a circular synthesized beam. In each case, the galaxy cluster was observed toward the center of the primary beam. The largest angular extent of any galaxy the clusters is $\sim$4.4 arcmin. This is sufficiently smaller than the 30 arcmin FWHM of the primary beam, and therefore we do not expect the off-axis leakage to vary significantly across our sources. Stokes \textit{QU} off-axis leakage levels are expected to be $\sim 1\%$ \citep{Jagannathan+17} at the full width half maximum (FWHM) of the primary beam. Thus, no correction for off-axis leakage was applied.

Unexpected flux variations between spectral windows were observed by \citet{Osinga+22} in all observations after calibration. They corrected for this by fitting a power law to the brightest sources in each observation and deriving spectral window-specific correction factors which were then applied to each of the Stokes \textit{IQU} cubes. We apply these same correction factors to the data to correct for the flux variations in the same manner.

\subsection{Sample Selection}\label{subsec: sample}

The goal of this work is to probe for a correlation between the degree of bending of radio galaxy jets and the galaxy's magnetoionic environment. We begin with the \citet{Osinga+22} catalogue and identify all extended sources (resolved such that it is larger than at least 3 synthesized beam widths) that show some detectable polarization at any location across the source. This yields 156 radio galaxies. From this initial set of radio galaxies we then apply the following requirements:

\begin{itemize}
    \item An optical host galaxy with redshift has been identified in the Legacy Survey \citep{Legacy}, or in Pan-STARRS \citep{PanSTARRS} for fields outside of the Legacy survey coverage. This reduces the sample to 114 radio galaxies.
    \item Comparison of the host galaxy redshift ($z_g$) and cluster redshift ($z_c$) show that the galaxy is a cluster member and not background or foreground to the cluster. Following \citet{Osinga+22}, we determine a galaxy to be a cluster member if both $z_c - (z_g + \delta z_g) \leq 0.04(1+z)$ and $(z_g - \delta z_g) - z_c \leq 0.04(1+z)$ are satisfied, where $\delta z_g$ is the error associated with the host galaxy redshift. This reduces the sample to 60 radio galaxies.
    \item Radio galaxy morphology is generally discernible. Any object where it is not clear which components are physically associated with a single radio galaxy and which components are randomly associated (projected nearby but not physically local to each other) is rejected. This leaves us with a final sample of 48 radio galaxies.
\end{itemize}

The resulting sample of 48 radio galaxies covers 35 galaxy clusters. The clusters spanned by our sample cover a mass (M$_{500}$) range of $\sim$2--12 $\times 10^{14}$ M$_{\odot}$, with a mean of 5.8 $\times 10^{14}$ M$_{\odot}$ and a standard deviation of 2.3 $\times 10^{14}$ M$_{\odot}$. The mean, median, and standard deviation are the same, or nearly the same, as those of the larger \citet{Osinga+22} sample. An Anderson-Darling test shows no significant difference between the M$_{500}$ distributions of the two samples.

For each galaxy in our sample, we then calculate a bending angle, $\theta$, from the data following the same method as \citet{vanderJagt+25}. \citet{vanderJagt+25} used a subsample of the \citet{Osinga+22} data to probe the correlation between bending angle and location in projected cluster-centric distance-velocity phase space of the radio galaxies. They measured a bending angle for 109 radio galaxies with varying degrees of bending by calculating the angle between the lines connecting the host galaxy position and the two brightest peaks not coincident with the host position. Following the approach of \citet{Garon+19}, we do not provide per-source uncertainties on bending angle, since these are dominated by projection effects and structure in the radio emission and can only be characterized statistically across large samples. As \citet{Garon+19} discuss in more detail, the uncertainties scale with angular size and cannot be robustly assigned to individual sources.

Approximately 10\% of our sample of radio galaxies are brightest cluster galaxies (BCGs), with bending angles ranging from $32^{\circ}$ to $72^{\circ}$; notably, none of these sources are NATs ($>$90$^{\circ}$) or completely straight ($\sim$0$^{\circ}$) sources. We also find no dependence of bending angle on whether the galaxy is a BCG, nor any correlation between bending angle and jet power (calculated at 1.4 GHz).

\begin{figure}
\centering
    \includegraphics[width=0.47\textwidth]{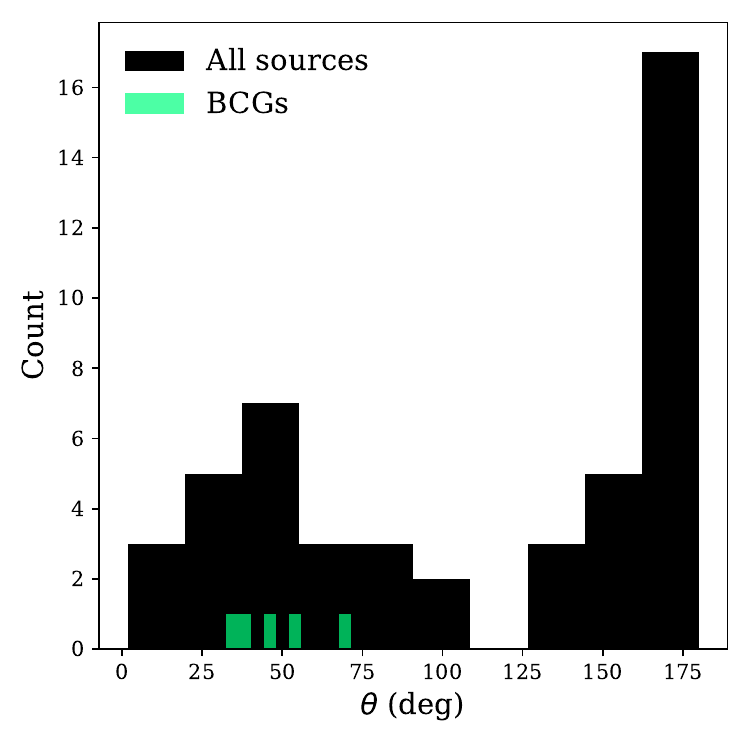}
    \caption{Distributions of bending angles for our sample of 48 radio galaxies. The distribution of bending angles for the BCG galaxies are overlaid in green.}
\label{fig: BA hist}
\end{figure}

The angular resolutions of the sample range over 6.2--11.9 arcsec. The sample has redshifts $0.023 \leq z \leq 0.318$ and bending angles $1\fdg7 \leq \theta \leq 180^{\circ}$, where 0$^{\circ}$ corresponds to a perfectly linear galaxy and increasing $\theta$ indicates increasing bending. We plot the distribution of bending angles in Figure \ref{fig: BA hist}. The peak at 180$^{\circ}$ is due to head-tail sources where the individual jet/lobe structures are not resolved, and for which we manually set the bending angle to 180$^{\circ}$. Cutouts of the total intensity multi-frequency synthesis (MFS) images of the final sample are shown in Figure \ref{fig: MFS cutouts 1}, with the 10$\sigma$ contours shown in red and the position of the host galaxy indicated by the cyan markers. Note that the images are rotated such that the bending angle is pointing toward the top of the page. The direction of North through East in RA and Dec coordinates is shown by an arrow in each cutout.

\begin{figure*}
\centering
    \includegraphics[width=0.76\textwidth]{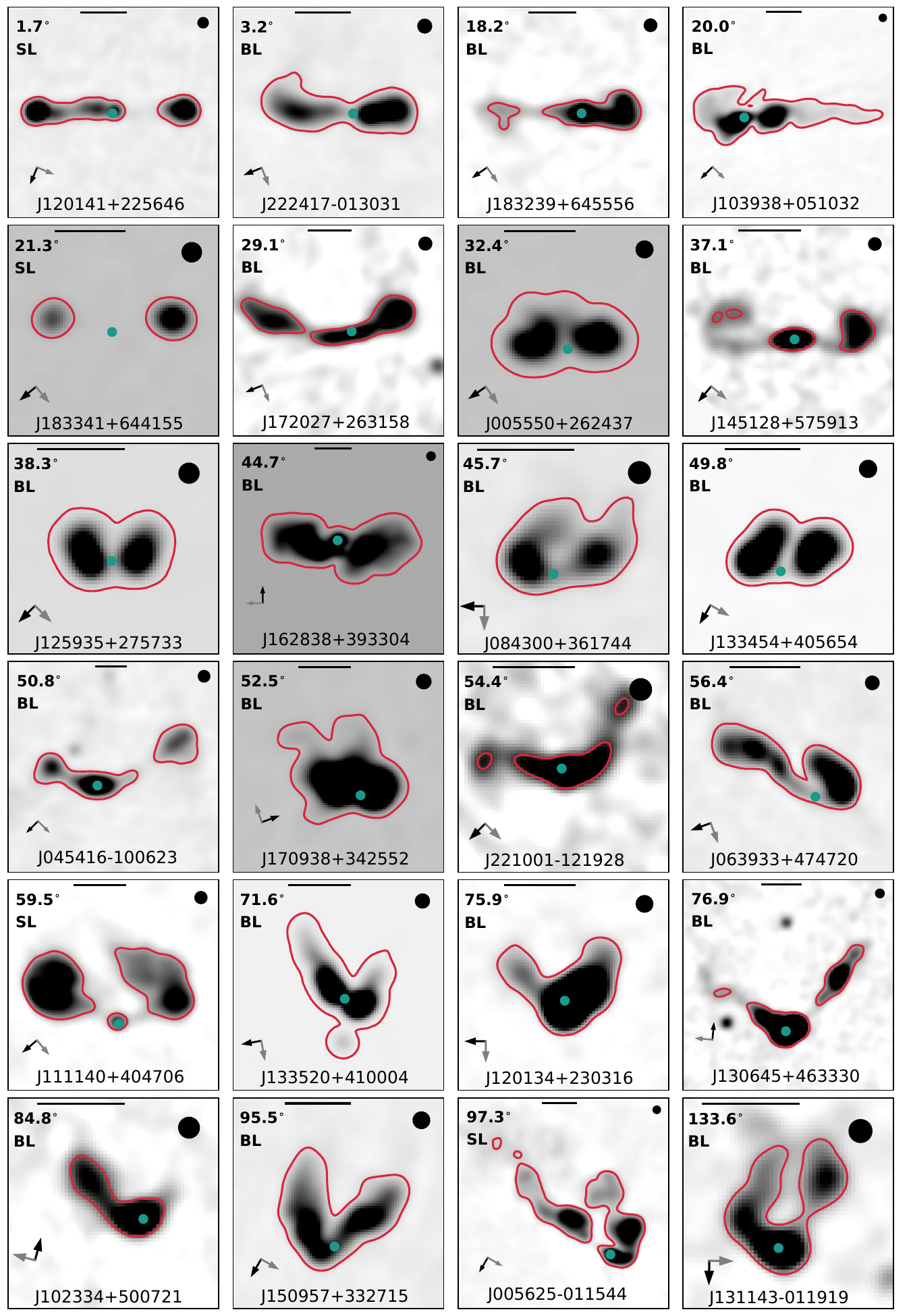}
    \caption{Cutouts of the Stokes I MFS images of the first 24 radio galaxies in our sample, ordered by increasing bending angle. The cutouts have been rotated such that the bisector is pointing approximately toward the top of the page. The grayscale maps total intensity, and the red outlines indicate the 10$\sigma$ contour lines in the image. The black arrow indicates the direction of increasing declination (north), and the gray arrow points east. The position of the host galaxy is indicated by a cyan circle. The black circles in the top right of the cutouts show the synthesized beam area and the black line at the top of the cutouts span 30 arcsec. The bending angles are noted in the top left corners, with the source integration method (``SL'', ``BL'', or ``FS'') indicated below (see text in Section \ref{sec:data red} for a description of these integration methods). Sources with bending angle 180$^{\circ}$ are in brackets to indicate that these are head-tail sources for which we are not able to resolve the individual jet/lobe structures.}
\label{fig: MFS cutouts 1}
\end{figure*}

\begin{figure*}
\ContinuedFloat
\centering
    \includegraphics[width=0.76\textwidth]{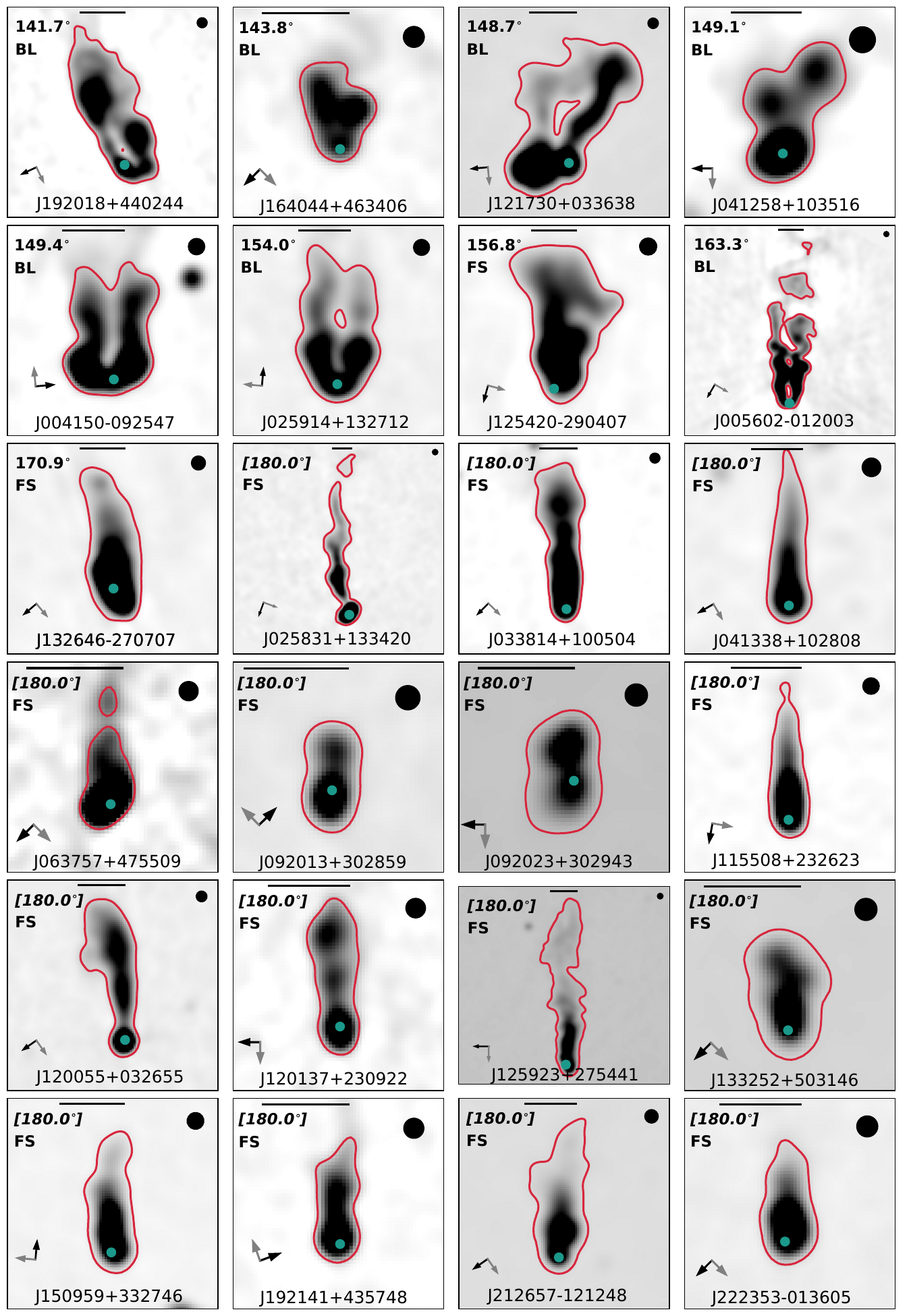}
    \caption{\textbf{continued.} The remaining 24 radio galaxies in our sample.}
\end{figure*}

We note that our sample is subject to a selection effect, described in Appendix A of \citet{Osinga+25}, who analyze the same dataset as \citet{Osinga+22}. All of the observations of the galaxy clusters that contain our sources have a common finite field of view of 30 arcmin at 1.5 GHz, leading to a correlation between the projected distance of a source from its cluster center and redshift (see Figure A.1 of \citealt{Osinga+25}). Since at lower redshift we are biased toward galaxies with smaller projected distances to their cluster center, this may potentially bias our results. We investigate this selection effect in more detail in Section \ref{subsec: z-lim pol vs theta}.

\section{Polarization measurement and analysis} \label{sec:data red}

We obtained Stokes \textit{IQU} spectra for each of our sources on which we performed our analysis. Due to the range in angular and physical resolution of our sample, we chose to integrate over the entire jet/lobe structure as an attempt to mitigate resolution effects and maintain a consistent comparison between sources. We defined the region within which we integrated the intensity as the 10$\sigma$ contour of the source in the total intensity MFS image. We chose a 10$\sigma$ contour to balance including a significant fraction of the source intensity with minimizing the impact of noise. Because we integrate all pixels within the contour, and polarized intensity is typically much lower than total intensity, pixels in Stokes \textit{QU} outside of the 10$\sigma$ contour will most likely be unpolarized and can introduce additional noise into our integrated spectrum. We used a similar method as \citet{Golden-Marx+23}, who investigated the relationship between environment and bending angle of bent radio galaxies in high-redshift clusters in total intensity using a 10$\sigma$ contour for intensity integration.

To obtain the per-channel integrated flux density, we placed the 10$\sigma$ contour from the total intensity image on each channel of the Stokes \textit{IQU} data cubes and integrated all pixels within the boundary of the contour. The per-channel uncertainty was calculated as the rms of a 30$\times$30 pixel off-source region (the same region in each channel) multiplied by the square root of the number of independent beams sampled by the integration.

There were three general cases that each required different methods of integrating the flux within the 10$\sigma$ contours:

\begin{enumerate}
    \item Radio galaxies where the individual jet/lobe structures have separate 10$\sigma$ contours (e.g., J$111140+404706$ in Figure \ref{fig: MFS cutouts 1}). In such cases, we integrated the pixels within each contour both individually and all together. This gave us spectra from the individual lobes as well as from the full source, so that, if each lobe is polarized but with different properties, we could analyze the individual lobe spectra as well as the combined spectrum (similar to the source being unresolved within the synthesized beam). In our sample, there are 4 sources that fall into this category, and they are indicated by an ``SL'' (``Separate Lobes'') in Figure \ref{fig: MFS cutouts 1}.
    \item Radio galaxies where the 10$\sigma$ contour encloses the entire source (e.g., J$084300+361744$ in Figure \ref{fig: MFS cutouts 1}). In such cases, we defined a bisector line that separated the two jet/lobe structures, passing through the position of the optical host and following a path of minimum flux between the two structures. We did this visually, guided by higher level contour lines (10$\sigma$, 20$\sigma$, and 25$\sigma$). We then integrated all pixels on each side of the bisector line as well as all pixels within the 10$\sigma$ contour, which again gave us spectra from the individual lobes as well as from the full source. In our sample, there are 27 sources that fall into this category, and they are indicated by a ``BL'' (``Bisector Line'') in Figure \ref{fig: MFS cutouts 1}.
    \item Radio galaxies where the individual jet/lobe structures could not be resolved (see Section \ref{subsec: sample}), or where to draw a bisector line between the two jet/lobe structures is not obvious. This includes sources with a bending angle of 180$^{\circ}$ (e.g., J$041338+102808$ in Figure \ref{fig: MFS cutouts 1}), or sources with a large bending angle and no clear indication from the 10$\sigma$ contour where the two jets separate (e.g., J$125420-290407$ in Figure \ref{fig: MFS cutouts 1}). In these cases we only integrated all pixels inside the 10$\sigma$ contour. These sources were compared only to the full-source integrated spectra from the other sources in the sample. In our sample, there are 17 sources that fall into this category, and they are indicated by an ``FS'' (``Full Source'') in Figure \ref{fig: MFS cutouts 1}.
\end{enumerate}

\subsection{Physical resolution effects}

The radio galaxies in our sample span a range of redshifts and angular sizes. We first wanted to determine whether we are able to compare our sample at their original resolutions, without experiencing effects due to variation in physical and angular resolutions. To do this, we selected a subsample of radio galaxies with clearly resolved jet/lobe structures as well as one radio galaxy where the jets/lobes were not individually resolved (a source with bending angle 180$^{\circ}$). We degraded the resolution from that of the original observation (7--10 arcsec) to 30 arcsec, at which point the jet/lobe structures in each of the observations became fully unresolved and began to merge with each other. We performed 1D RM synthesis on the spectra of each source at each resolution using the \texttt{RM-Tools}\footnote{\url{https://github.com/CIRADA-Tools/RM-Tools}} software package (\textit{v1.4.6}; \citealt{RM-tools2020}) to obtain the peak RM and the polarized fraction, $p$.

We found that, up to a resolution where the jet/lobes structures do not merge (typically $\sim$15--20 arcsec), the measured RMs typically remained consistent within uncertainty, and the measured polarized fraction varied by $\sim$1\% in the more resolved galaxies and by $\lesssim$10\% in the 180$^{\circ}$ galaxy. This is in agreement with resolution testing done by \citet{Sebokolodi+20} on Cygnus A, in which they found that beyond a certain resolution, fractional polarization along a given line-of-sight varied little with decreasing resolution (see Figure 7 of \citealt{Sebokolodi+20}). These results suggested that physical resolution does not have a strong effect on our results, and so we performed all further analysis at the original angular resolutions of each observation.

\subsection{RM synthesis}

We performed 1D RM synthesis on the integrated spectra of all of our sources. We found that 24 out of 48 (50\%) of the radio galaxies were polarized above a signal-to-noise in polarization, S/N$_{\rm{pol}}$, threshold of 6 in at least one lobe or over the full source. We adopted a threshold of S/N$_{\rm{pol}}$ $ = 6$ to ensure robust polarization detections, as lower S/N values can be affected by statistical bias and uncertainty in polarization measurements \citep{Marquart+12}. Of the 15 total sources with $\theta = 180^{\circ}$ (sources where we are unable to resolve two distinct jets/lobes), 8 (53\%) were polarized. Of the 33 total sources with two distinct jets/lobes, 7 (21\%) had only one polarized lobe, 5 (15\%) were polarized in both lobes, and 4 (12\%) were polarized only when integrating all pixels inside the 10$\sigma$ contour. We also found that 18 of the 24 (75\%) unpolarized sources had a projected distance to the cluster center R $<$ R$_{500}$, where R$_{500}$ is the radius enclosing an overdensity of 500 at the cluster redshift. Of the polarized sources, 14 out of 24 (58\%) had R $<$ R$_{500}$. This agrees with the findings of \citet{Osinga+22}, who found increasing depolarization with decreasing distance to the cluster center. The measured fractional polarization of the unpolarized (polarized) sources ranges over 0.02--6\% (0.02--14\%), with a mean of 1\% (3\%) and standard deviation of 1\% (3\%). We summarize these subsamples and their statistics in Table \ref{tab:subset_stats}.

\begin{table}
\centering
\caption{Summary of statistics of sample subsets.}
\begin{tabular}{lll}
\hline
Subset & Count (\%) & Statistics \\
\hline
Total & 48 (100) & 50\% polarized \\[3pt]
$\theta$ = 180$^{\circ}$ & 15 (31) & 53\% polarized \\[3pt]
$\theta$ $\neq$ 180$^{\circ}$ & 33 (69) & 21\% one lobe polarized \\
 &  & 15\% both lobes polarized \\
 &  & 12\% full source polarized \\[3pt]
Unpolarized & 24 (50) & 75\% at R $<$ R$_{500}$ \\
 &  & $p$: 0.02--6\%, $\mu = 1\%$, $\sigma = 1\%$ \\[3pt]
Polarized & 24 (50) & 58\% at R $<$ R$_{500}$ \\
 &  & $p$: 0.02--14\%, $\mu = 3\%$, $\sigma = 3\%$ \\
\hline
\end{tabular}
\label{tab:subset_stats}
\end{table}

\begin{figure}
    \centering
    \subfloat{
    \includegraphics[width=0.47\textwidth]{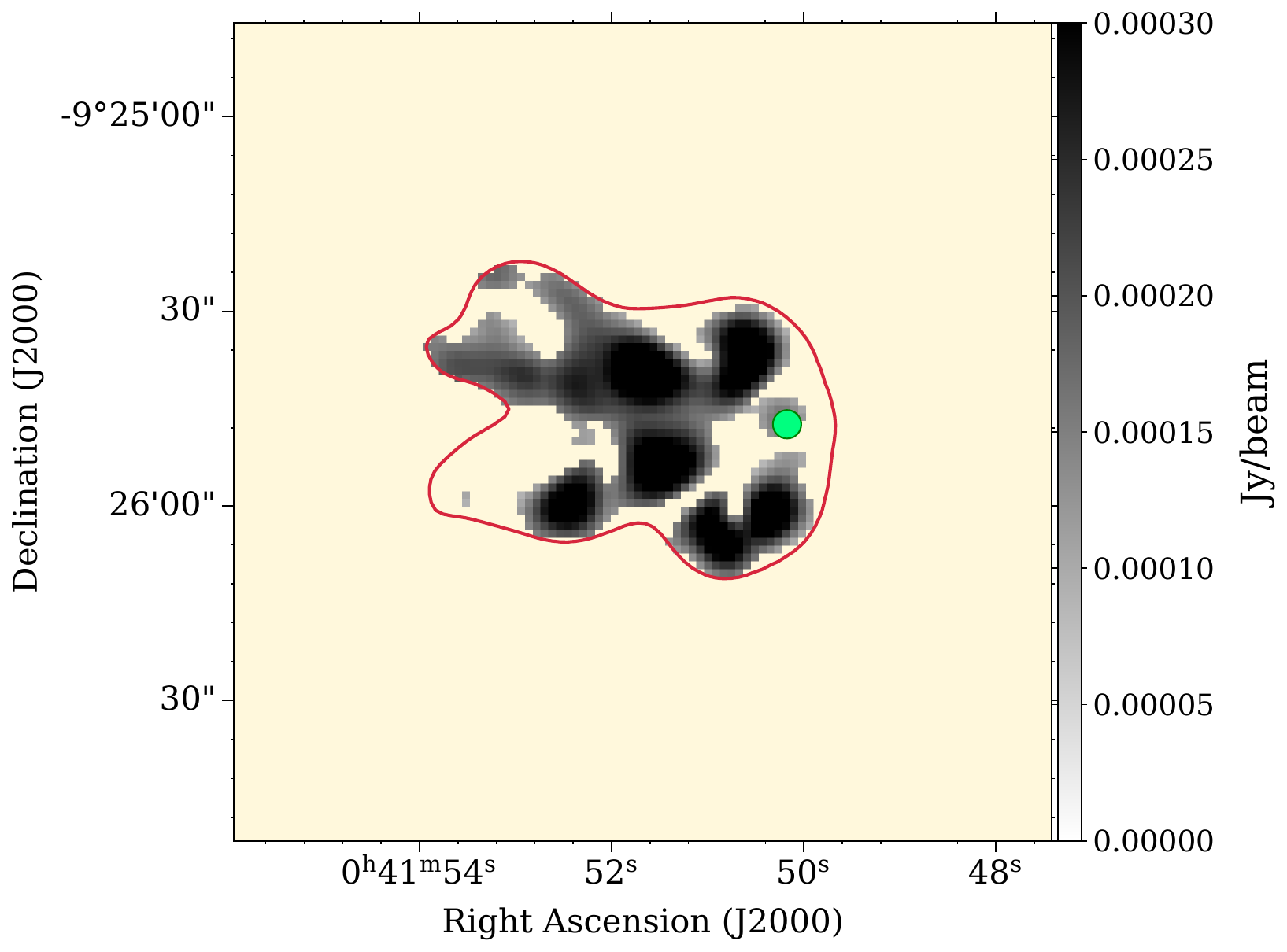}
        \label{fig: ex pi 2d}}\\
    \subfloat{
    \includegraphics[width=0.448\textwidth]{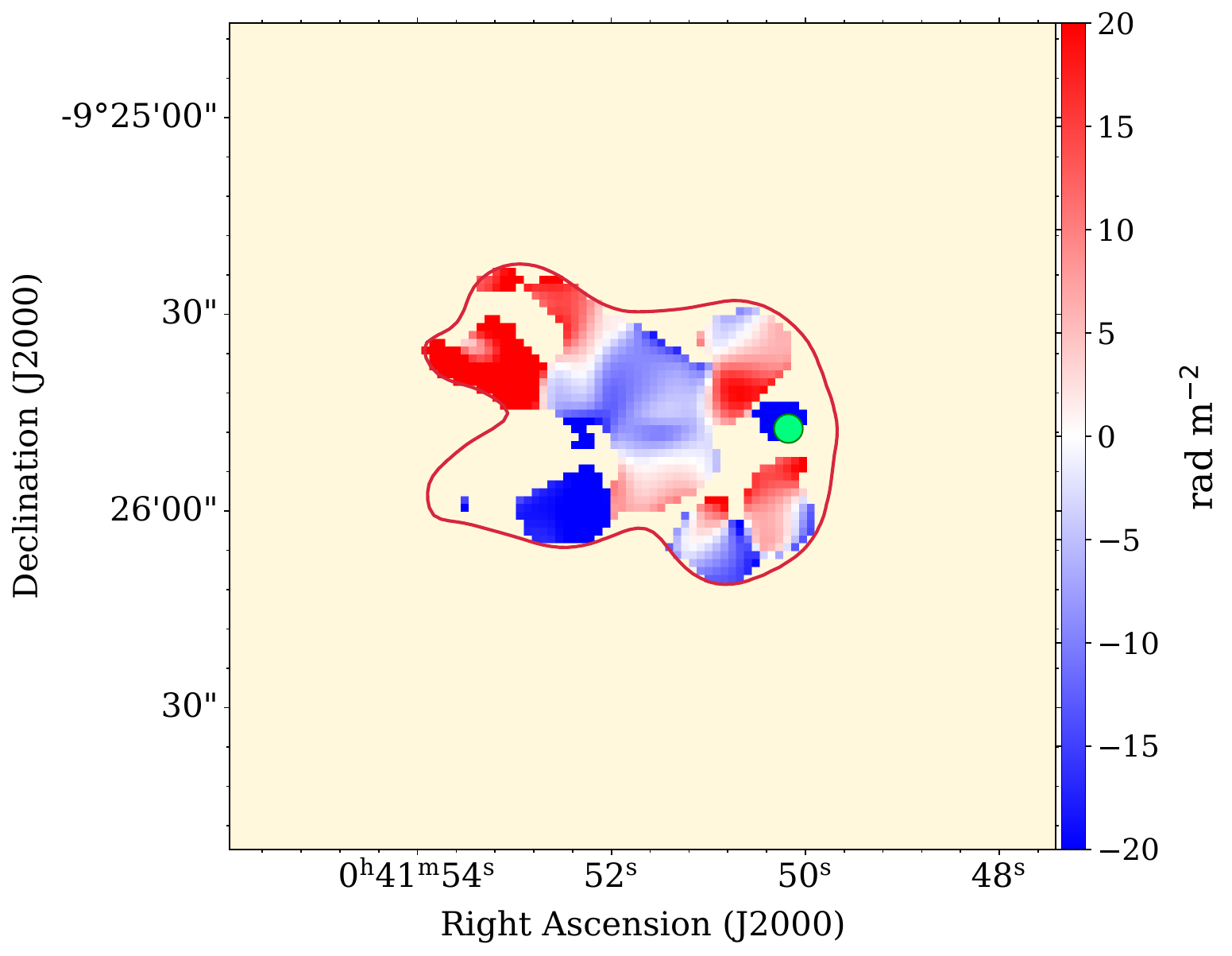}
        \label{fig: ex rm 2d}}
    \caption{Example 2D maps of polarized intensity (\textit{top}) and RM (\textit{bottom}) for source J$004150-092547$, derived from 3D RM synthesis. The 10$\sigma$ contours from the Stokes I MFS image is shown in red, and the position of the host galaxy is indicated by the green circle. Masked pixels are colored light yellow for clarity.}
    \label{fig: ex 2ds}
\end{figure}

We also performed 3D RM synthesis on cutouts of each of our sources using the \texttt{RM-Tools} package. This returned continuous 2D maps of several polarization properties, including polarized intensity, RM, S/N$_{\rm{pol}}$, and derotated polarization angle ($\psi_0$). In Figure \ref{fig: ex 2ds} we show examples of the 2D polarized intensity (top panel) and RM maps (bottom panel) for source J$004150-092547$.

\subsection{Modeling the polarized spectra}

After extracting the integrated Stokes \textit{IQU} spectra for each of our sources, we fit the integrated fractional Stokes \textit{qu} spectra (\textit{Q/I}, \textit{U/I}) with a model that describes depolarization due to an external, non-emitting, turbulent Faraday screen (isotropic turbulence). With this model, the complex fractional Stokes \textit{qu} function, $p(\lambda)$, is defined as

\begin{equation}\label{eq: m2}
    p(\lambda) = p_0 \exp{(2i \,[\psi_0 + \rm{RM} \lambda^2])} \, \exp{(-2 \sigma^2_{\rm{RM}} \lambda^4)} \;,
\end{equation}

\noindent where $p_0$ is the intrinsic fractional polarization of the emission, $\psi_0$ is the derotated polarization angle of the emission, RM is the total Faraday rotation experienced by the polarized emission along the line-of-sight, and $\sigma_{\rm{RM}}$ is the dispersion in RM. This model is adapted from \citet{Burn1966} and \cite{Sokoloff1998}, and has been used in several previous studies (e.g., \citealt{Anderson+15,O'Sullivan+12,O'Sullivan+17,Ma+19,Osinga+25}).

We fit our data with this model because \citet{Osinga+22} showed that this model was an accurate fit to $\sim$75\% of their data (of which our data are a subset), and 88\% of our sample are among those well fit by the model. We note that \citet{Osinga+22} fit individual components identified by a source finder, while we fit the lobe- and source-integrated spectra (which may include multiple components), making a direct comparison between the fits difficult. We fit all source spectra with a single model to limit the number of free parameters and to simplify the analysis and comparison between sources. This approach minimized the risk of overfitting the data while also maintaining consistency across the data set.

To perform the model fitting, we used the Stokes $QU$ fitting module in \texttt{RM-Tools}. The module uses multinest fitting and the \texttt{dynesty} sampler to explore a defined parameter space. We defined the parameter space for the model fit as follows: $p_0$ over 0.001--1, $\Psi_0$ over 0$^{\circ}$--180$^{\circ}$, RM over $-1000$ to $+1000$ rad m$^{-2}$, and $\sigma_{\rm{RM}}$ over 0--100 rad m$^{-2}$.

\subsubsection{Correcting model parameters for redshift}\label{subsubsec: z-dep}

Equation \ref{eq: m2} has four free parameters: $p_0$, $\psi_0$, RM, and $\sigma_{\rm{RM}}$. RM and $\sigma_{\rm{RM}}$ have a dependence on redshift that needs to be corrected. The total measured RM from an extragalactic source can be written in terms of the individual contributions:

\begin{equation}
    \rm{RM} = \rm{RM_{host}} + \rm{RM_{ICM}} + \rm{RM_{IGM}} + \rm{RM_{MW}} \, ,
\end{equation}

\noindent where $\rm{RM_{host}}$ is the contribution to the total RM from the host galaxy itself, $\rm{RM_{ICM}}$ is the contribution from the intracluster medium (if the galaxy is located within a galaxy cluster), $\rm{RM_{IGM}}$ is the contribution from the intergalactic medium, and  $\rm{RM_{MW}}$ is the contribution from the path length through the Milky Way. Any other intervening magnetized plasma will also contribute to the total RM of a source (see \citealt{Akahori&Ryu2010} for a more detailed discussion). Upper limits on $\lvert \rm{RM_{IGM}} \rvert$ are estimated to be $\sim$1--4 rad m$^{-2}$ \citep{Akahori&Ryu2010,O'Sullivan+20,Amaral+21}, and so it is not expected to contribute meaningfully to the total source RM due to its relatively small magnitude.

We are interested in the RM from the source and its host environment, and not the contribution from our Galaxy, $\rm{RM_{MW}}$. To correct for $\rm{RM_{MW}}$, we subtracted the Galactic contribution along the source line-of-sight using the value from the \citet{Hutschenreuter+22} Galactic Faraday sky map, defining the result as RM$_{\rm{sub}}$. We then corrected RM$_{\rm{sub}}$ for redshift dependence using the equation $\rm{RM}_{\rm{corr}} = \rm{RM}_{\rm{sub}} (1 + z_c)^2$. We note that this method of redshift correction assumes that the  total contribution to the RM from the intergalactic medium (IGM) along the line of sight is negligible with respect to the RM contribution from the ICM.

We corrected the model-fit $\sigma_{\rm{RM}}$ for redshift dependence in a similar manner, using the equation $\sigma_{\rm{RM,corr}} = \sigma_{\rm{RM}} (1 + z_c)^2$. None of our sources were observed along lines of sight near the Galactic plane (all sources are at Galactic latitude $b < -28^{\circ}$ or $b > +17^{\circ}$), which likely minimizes the Galactic contribution to the total value of $\sigma_{\rm{RM}}$, and a plot of $\sigma_{\rm{RM}}$ versus Galactic latitude shows no obvious correlation. We currently have no way of separating the Galactic contribution to the total $\sigma_{\rm{RM}}$, so this is a potential confounding factor in our analysis \citep{Leahy1987}. The other model parameters, $\psi_0$ and $p_0$, have no dependence on redshift and required no correction.

\section{Results} \label{sec:results}

In this section we present the results of Kendall’s rank correlation coefficient ($\tau$) tests between the polarization parameters of the polarized sources and bending angle. We also present the results of tests for coherence in the derotated magnetic field vector angles of the sources, and alignment with the direction of bending.

\subsection{Parameters as a function of bending angle}\label{subsec: pol vs theta}

\begin{figure*}
\centering
    \includegraphics[width=1.0\textwidth]{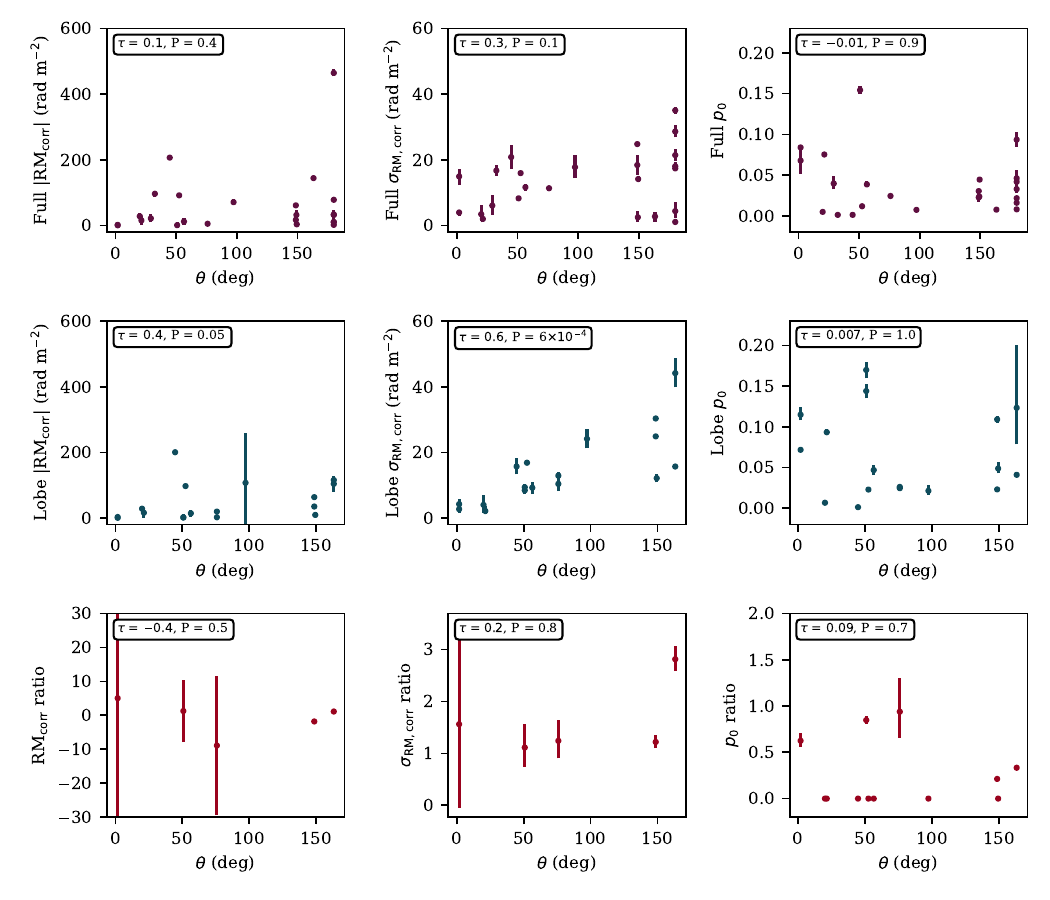}
    \caption{Model parameters versus bending angle for the full-source integrated spectra (top row), the individual lobe-integrated spectra (middle row), and the lobe ratios (bottom row). In each panel, the Kendall's tau correlation coefficient, $\tau$, and corresponding P-value for the data is given in the top left corner. We use a modified P-value of $3.1\times10^{-5}$ as our threshold for correlation.}
\label{fig: corr subplots BA}
\end{figure*}

In Figure \ref{fig: corr subplots BA} we present plots of our primary model parameters, $|\rm{RM_{corr}}|$, $\sigma_{\rm{RM,corr}}$, and $p_0$, as a function of bending angle for the full-source integrated spectra (top row), the individual lobe-integrated spectra (middle row), and, where possible, the lobe ratios (bottom row). We note that when calculating the polarized fraction ratio, if one of the two lobes has S/N$_{\rm{pol}} < 6$, we consider it depolarized and the ratio is set to 0. In this way, sources with only one detected polarized lobe are still assigned a ratio and included in the polarized fraction ratio plot, whereas for $|\rm{RM_{corr}}|$ and $\sigma_{\rm{RM,corr}}$ both lobes must be polarized to form a ratio. This results in more data points in the polarized fraction ratio plot than in the $|\rm{RM_{corr}}|$ and $\sigma_{\rm{RM,corr}}$ ratio plots.

Each subplot includes the Kendall's $\tau$ correlation statistic, $\tau$, and the corresponding P-value for the test. We set an initial stringent P-value threshold of 0.001, and, because we are conducting a large number (32) of correlation tests in this work, we also apply a P-value correction to account for the increased chance of a false positives, or the ``look-elsewhere'' effect \citep{Algeri+16}. We therefore set the threshold $\rm{P} = \frac{0.001}{32} = 3.1\times10^{-5}$ as the maximum likelihood threshold for a potential correlation.

Figure \ref{fig: corr subplots BA} does not indicate a statistically significant a correlation or anti-correlation between any of our parameters and $\theta$. 
The statistic for the lobe $\sigma_{\rm{RM}}$ versus $\theta$ ($\tau$ = 0.6, P = $6 \times 10^{-4}$) suggests a relationship between the two parameters, but it does not meet the threshold of our modified P-value for a correlation. We analyze this potential correlation further in Section \ref{subsec: z-lim pol vs theta}. We summarize the Kendall's tau statistics from Figure \ref{fig: corr subplots BA} in Table \ref{tab: corr stats}.

\begin{table}
\centering
\caption{Summary of correlation statistics}
\label{tab: corr stats}
\begin{tabular}{p{4.5cm}p{3.4cm}} \hline
Parameters  & Kendall's $\tau$\footnote{For all tests except lobe $\sigma_{\rm{RM}}$ vs $\theta$, we find P $>$ 0.01. We use P = $3.1\times10^{-5}$ as the maximum likelihood threshold for a potential correlation.} \\\hline
Full $\lvert$RM$_{\rm{corr}}\rvert$ vs $\theta$ & 0.1 \\
Lobe $\lvert$RM$_{\rm{corr}}\rvert$ vs $\theta$ & 0.4 \\
Full $\sigma_{\rm{RM}}$ vs $\theta$ & 0.3 \\
Lobe $\sigma_{\rm{RM}}$ vs $\theta$ & 0.6 (P = $6 \times 10^{-4}$) \\
Full $p_0$ vs $\theta$ & $-0.01$ \\
Lobe $p_0$ vs $\theta$ & $-0.007$ \\
Lobe RM ratio vs $\theta$ & $-0.4$ \\
Lobe $\sigma_{\rm{RM}}$ ratio vs $\theta$ & 0.2 \\
Lobe $p_0$ ratio vs $\theta$ & $-0.09$ \\\hline
\end{tabular}
\end{table}

As noted in Section \ref{subsec: sample}, our sample is a subsample of the data used by both \citet{Osinga+22} and \citet{Osinga+25}. \citet{Osinga+22} report increased depolarization (traced by $\sigma_{\rm{RM}}$) as a function of projected distance to the cluster center, and \citet{Osinga+25} identify increased scatter in RM with decreasing distance to the cluster core. The authors attribute this trend primarily to radial variations in electron density and magnetic field strength in the ICM, and their results are consistent with the expectation that denser, more magnetized plasma near the cluster center produces higher RMs \citep{Govoni&Feretti2004,Bonafede+10}. Similarly, \citet{vanderJagt+25}, who also analyze a subset of the \citet{Osinga+22} data, identify a significant correlation between jet bending angle and R/R$_{500}$, suggesting enhanced ram pressure closer to the cluster core. In our lobe spectra sample, we find tentative correlations between both $\lvert$RM$\rvert$ and R/R$_{500}$ and between bending angle and R/R$_{500}$ (P = 0.002 and P = 0.004, respectively), though neither meets our modified threshold for statistical significance. In the full spectra sample, bending angle and R/R$_{500}$ again show a weak trend (P = 0.003), but the $\lvert$RM$\rvert$–R/R$_{500}$ relation is statistically insignificant (P = 0.07).

\subsection{Redshift-limited sample parameters versus bending angle}\label{subsec: z-lim pol vs theta}

As described in Section \ref{subsec: sample}, our data are subject to a selection effect due to the finite field of view of our observations. We investigate the effects of this selection bias here.

\begin{figure}
\centering
    \includegraphics[width=0.47\textwidth]{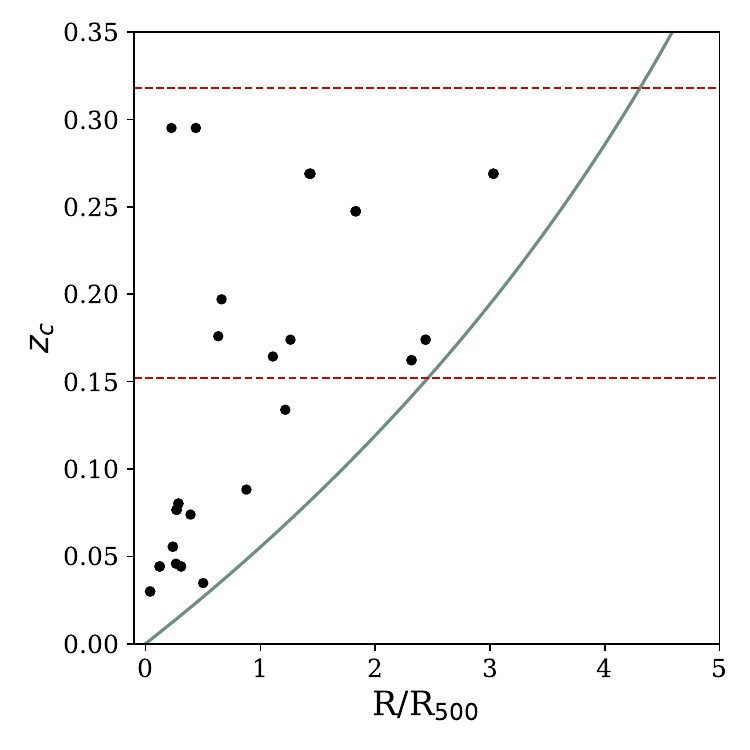}
    \caption{Plot of redshift as a function of the ratio of projected distance to the cluster center and R$_{500}$. The dashed lines show the redshift bounds within which we select our redshift-limited sample, with which to test the strength of the field-of-view selection effect described in the text on our results. The solid line shows the FWHM of the primary beam of our observations (30 arcsec) for a cluster with a representative R$_{500}$ of 1 Mpc.}
\label{fig: z vs r500}
\end{figure}

To test whether the redshift-projected distance selection bias noted above has a significant effect on our results, we take a subset of our sources within a range of redshifts where the distance to the cluster center is not biased by the field of view (i.e., there is no correlation between the two parameters). This is done by selecting sources in the redshift range 0.152–0.301, which yields a reduced sample of 12 radio galaxies. We illustrate this choice in Figure \ref{fig: z vs r500}, where we plot redshift versus R/R$_{500}$ for our polarized sources, and we include a curve (in black) showing the FWHM of the primary beam of our observations for a cluster with a representative R$_{500}$ of 1 Mpc.

With this reduced sample, we again calculate the Kendall's $\tau$ correlation statistics for our three model parameters, $\lvert \rm{RM_{corr}} \rvert$, $\sigma_{\rm{RM,corr}}$, and $p_0$, and bending angle for the full-source integrated spectra and the lobe integrated spectra. We do not include the lobe ratio results because the redshift-limited sample contains only 3 sources with two polarized lobes, and is not large enough to calculate any meaningful statistics with.

\begin{figure*}
\centering
    \includegraphics[width=1.0\textwidth]{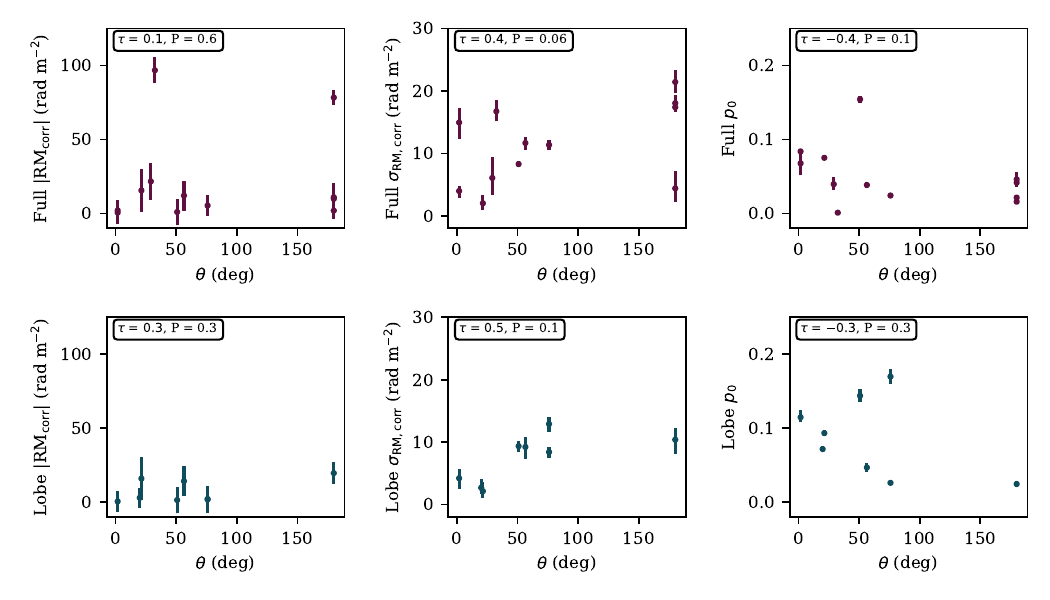}
    \caption{Model parameters versus bending angle for the redshift-limited sample. The full-source integrated spectra are shown in the top row and the individual lobe-integrated spectra are shown in the bottom row. In each panel, the Kendall's tau correlation coefficient, $\tau$, and corresponding P-value for the data are given in the top left corner.}
\label{fig: corr subplots BA trim}
\end{figure*}

We present the results in Figure \ref{fig: corr subplots BA trim}. No correlation is found between the polarization parameters and bending angle in the redshift-limited sample. In the case of the lobe $\sigma_{\rm{RM,corr}}$ versus bending angle, the possible correlation that we see in Figure \ref{fig: corr subplots BA} is now absent ($\tau$ = 0.5, P = 0.1). This suggests that the apparent trend in $\sigma_{\rm RM, corr}$ versus bending angle was influenced by selection effects, although the smaller size of the redshift-limited sample may also play a role. For $|{\rm RM}|$ and $p_0$, the bias toward sources with a smaller projected distance to the cluster center at smaller redshifts does not appear to have a significant effect on the correlation with bending angle

\subsection{Parameters as a function of integrated physical area}\label{subsec: pol vs area}

For each source, the total integrated physical area is different, depending on the redshift and physical size of each galaxy. Since each spectrum that we model is from a different integrated physical area, it is important to test whether there is a correlation between physical area and polarization properties, to determine if area is a confounding factor in our tests of correlation with bending angle.

\begin{figure*}[]
\centering
    \includegraphics[width=1.0\textwidth]{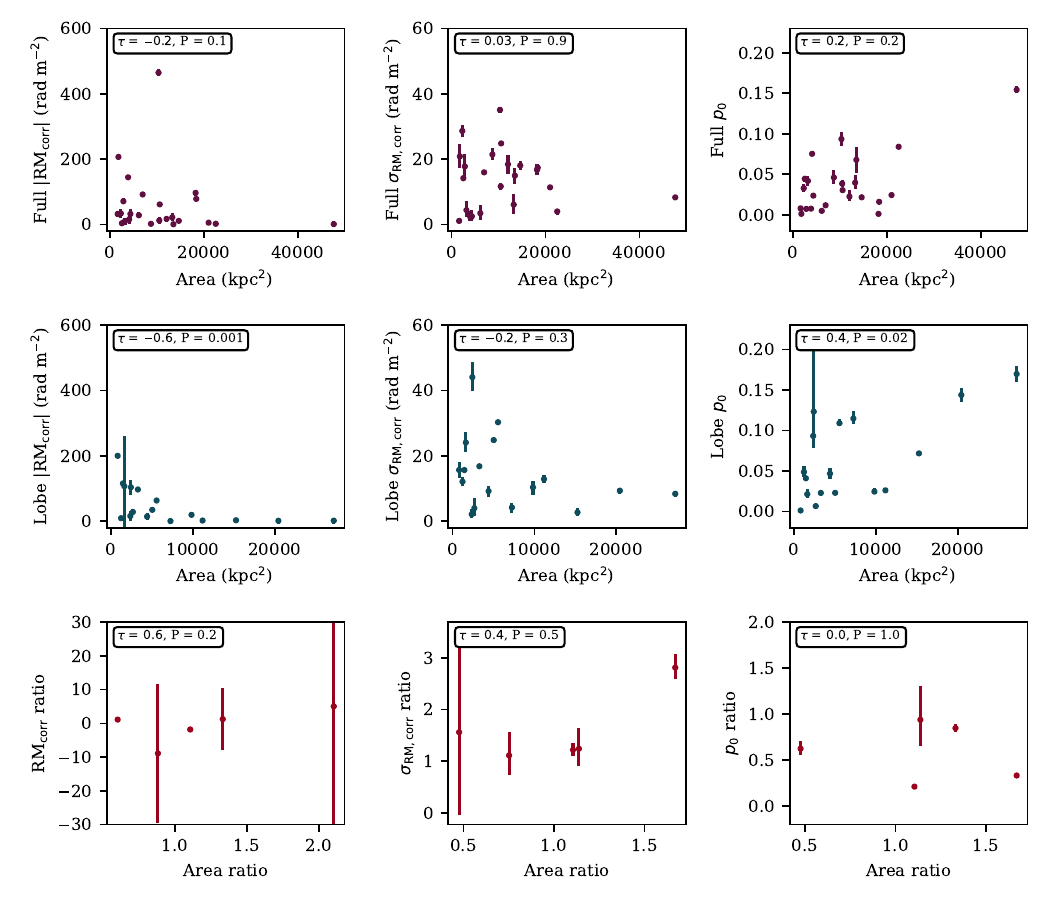}
    \caption{Model parameters versus total integrated physical area for the full-source integrated spectra (top row), the individual lobe-integrated spectra (middle row), and the lobe ratios (bottom row). In each panel, the Kendall's tau correlation coefficient, $\tau$, and corresponding P-value for the data is given in the top left corner.}
\label{fig: corr subplots area}
\end{figure*}

We plot the results in Figure \ref{fig: corr subplots area}, for the full-source integrated spectra (top row), the individual lobe-integrated spectra (middle row), and the lobe ratios where possible (bottom row). The Kendall's tau correlation statistic is calculated for each pair of variables and is included at the top of each subplot. We again use a modified P-value of $3.1\times10^{-5}$ as our threshold for correlation.

The physical area of the synthesized beams in our sample range over $\sim$24--2400 kpc$^2$. The typical integrated physical area of the source spectra, both lobe and full source, is $\lesssim$20 000 kpc$^2$, and spans $\sim$800--47 000 kpc$^2$. No pair of variables meets our threshold for correlation. A Spearman correlation test also shows no correlation between the polarization parameters and integrated physical area.

\subsection{Magnetic field vector coherence}\label{subsec: bvec coherence}

After performing 3D RM synthesis, we have a map of the intrinsic magnetic field vector angles for all of our sources. From visual inspection, we find that there appears to be coherence in the vectors in many of the sources, although the coherence can vary between regions (e.g. between a jet and a lobe). To test the coherence of the magnetic field vectors in our sources, we perform a two-sample Anderson-Darling (AD) test on each source, comparing the distribution of the vectors with a randomly drawn distribution of the same sample size.

To obtain reliable results from the AD test, we only consider sources with at least 50 independent magnetic field vectors per source. Additionally, we include only one vector per synthesized beam area to avoid coherence due solely to resolution. To determine which sources meet this criterion, we first mask all pixels in the magnetic field vector map that lie outside of the 10$\sigma$ contour drawn from the total intensity MFS image. We also mask all pixels inside of the 10$\sigma$ contour where S/N$_{\rm{pol}} < 6$. We then randomly select an unmasked pixel, note the magnetic field vector value at that position, then mask that pixel and all other pixels within one synthesized beam radius. We continue this process until no pixels are left unmasked. The number of independent magnetic field vectors per source from this selection process range over 1--225, with most sources having less than 20 independent vectors. The distribution of the magnetic field vector angle uncertainties lies primarily between 5$^{\circ}$ and 13$^{\circ}$, with $\mu = 9\fdg4$ and $\sigma = 2\fdg5$.

\begin{figure*}
    \centering
    \subfloat[J005625$-011544$.\\AD $=$ 0.8, P $=$ 0.2.]{
    \includegraphics[width=0.31\textwidth]{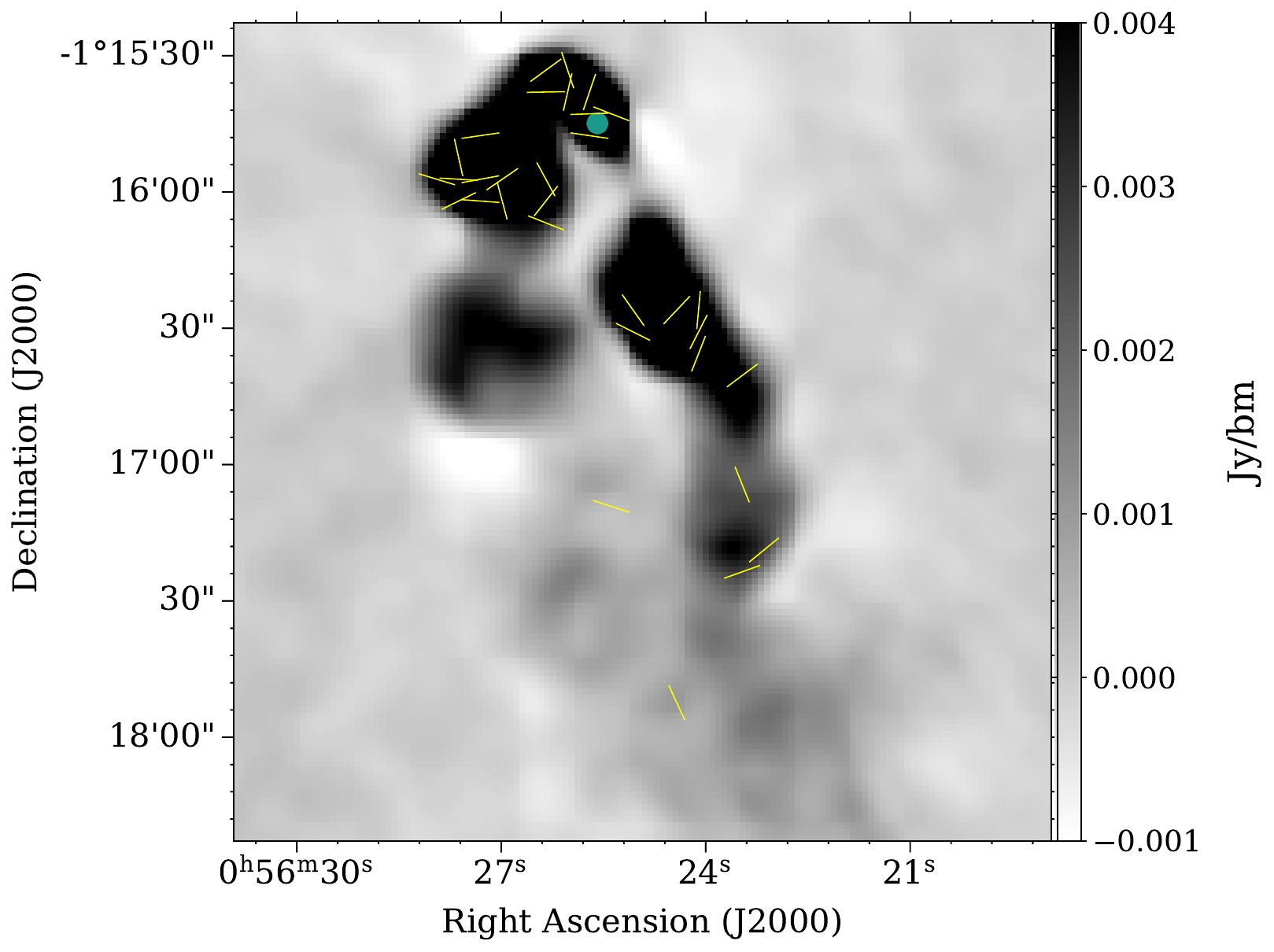}
        \label{fig: bvec src45}}
    \subfloat[J192018$+440244$.\\AD $=$ $-0.2$, P $=$ 0.5.]{
    \includegraphics[width=0.31\textwidth]{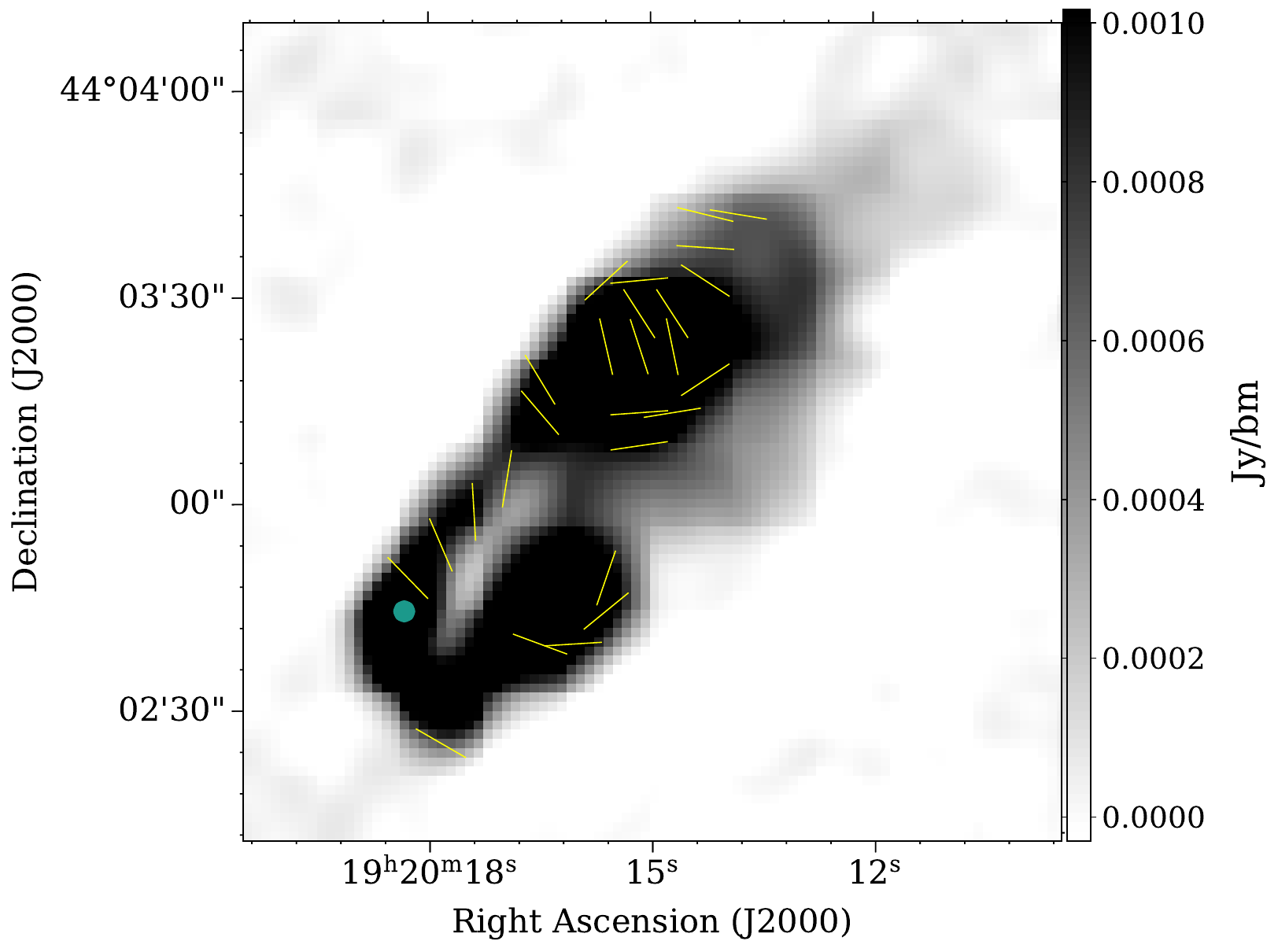}
        \label{fig: bvec src23}}
    \subfloat[J121731$+033656$.\\AD $=$ $-0.1$, P $=$ 0.4.]{
    \includegraphics[width=0.31\textwidth]{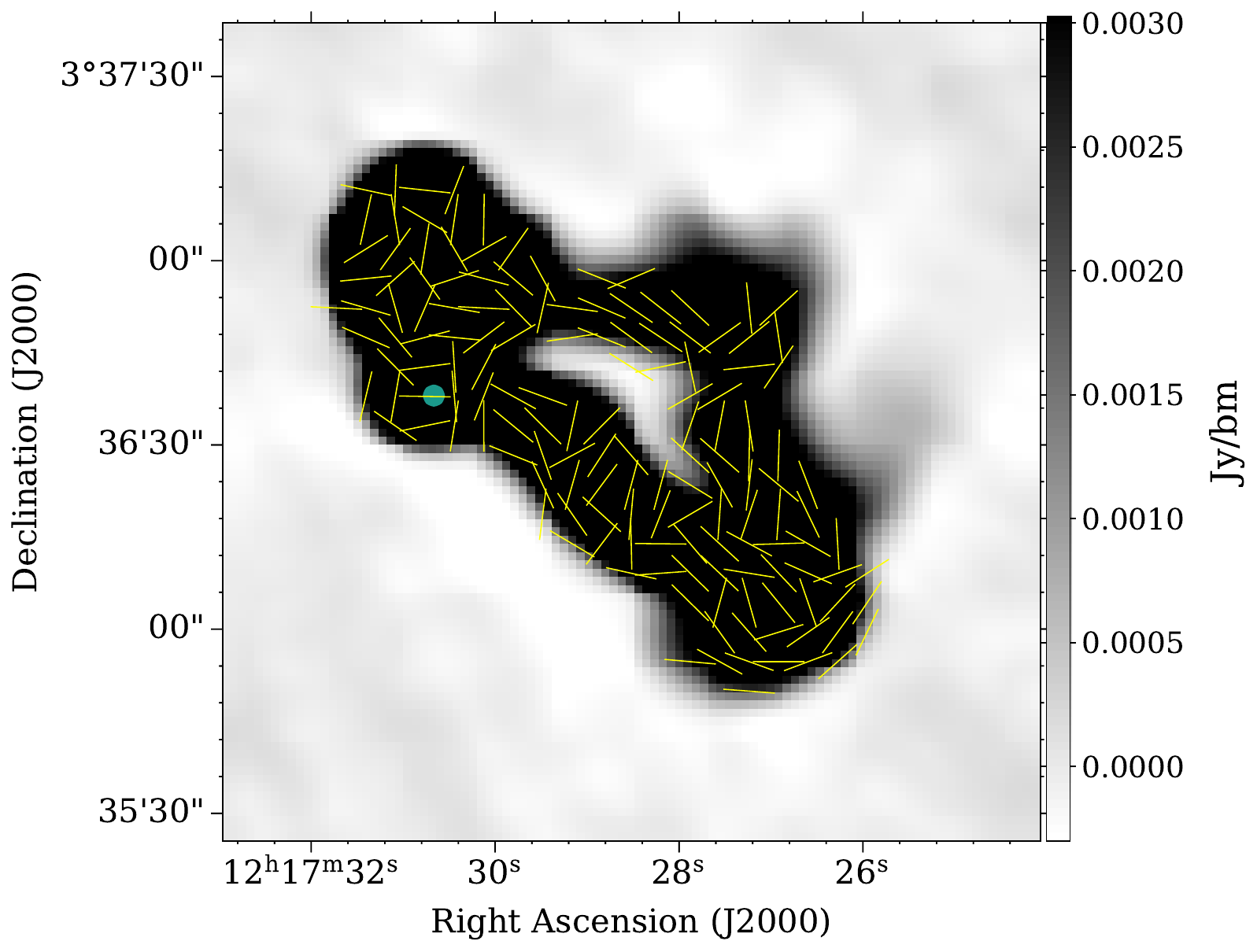}
        \label{fig: bvec src78}}\\
    \centering
    \subfloat[J004150$-092547$.\\AD $=$ 4.6, P $=$ $5\times10^{-3}$.]{
    \includegraphics[width=0.32\textwidth]{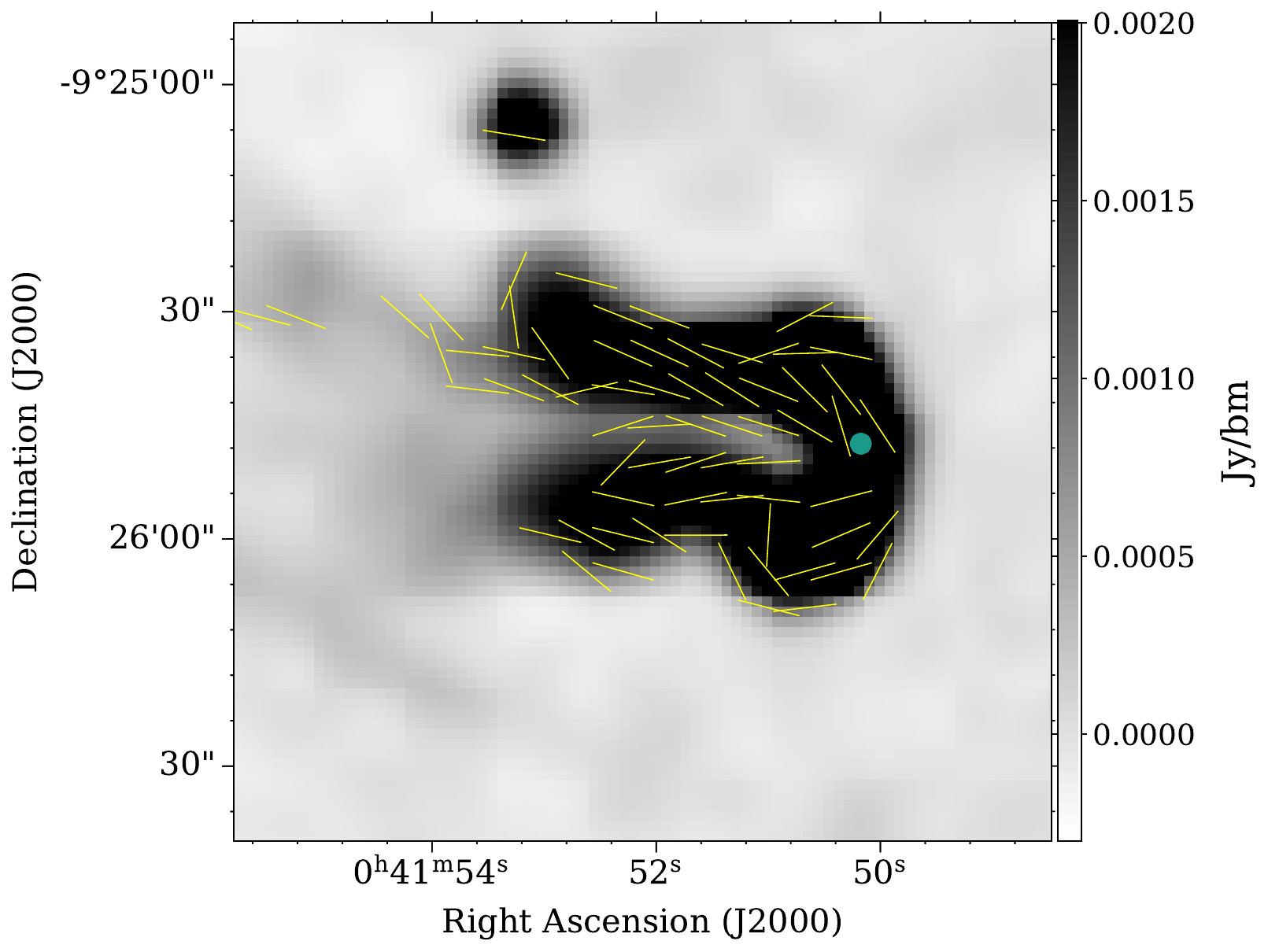}
        \label{fig: bvec src39}}
    \subfloat[J005602$-012003$.\\AD $=$ 0.1, P $=$ 0.3.]{
    \includegraphics[width=0.31\textwidth]{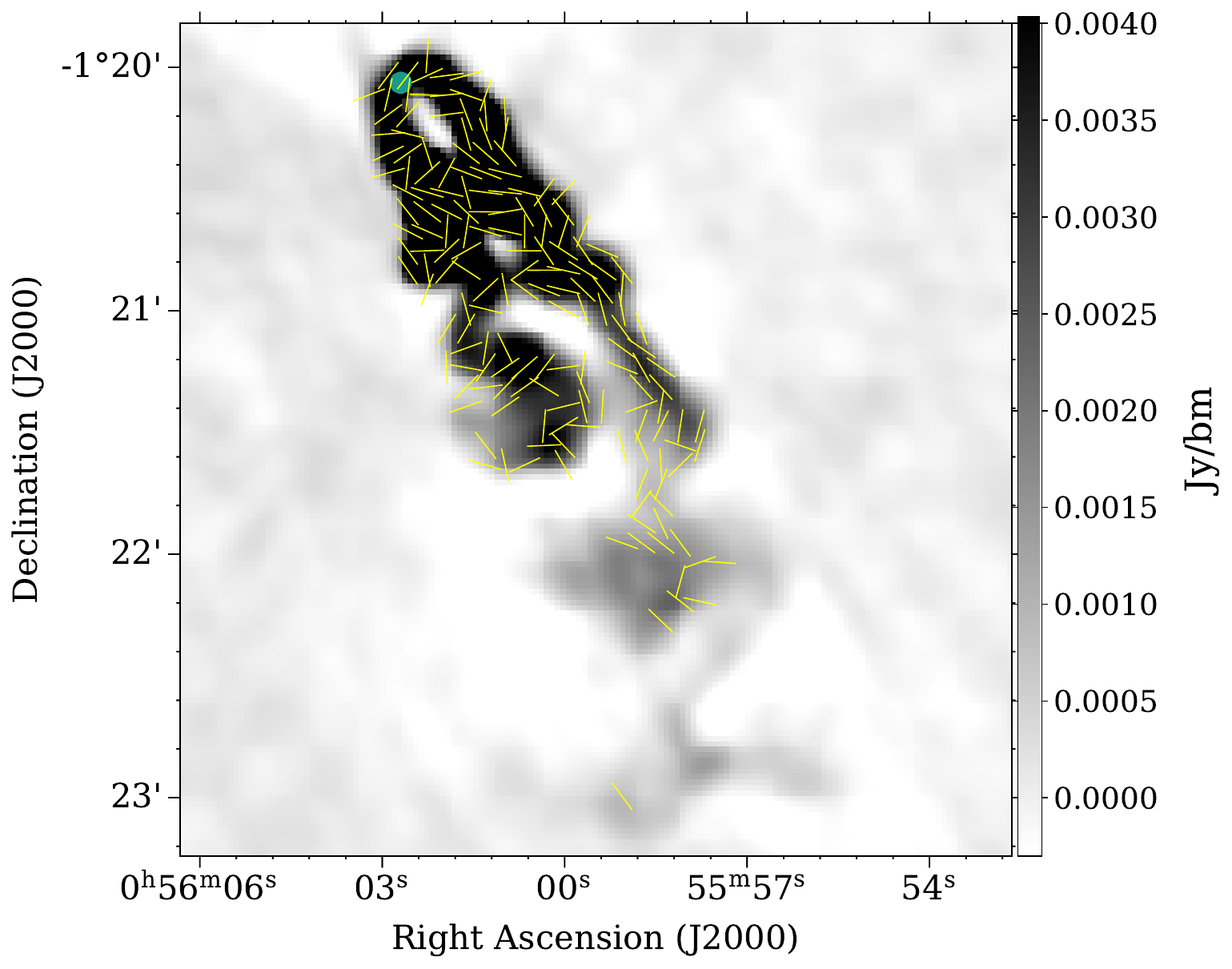}
        \label{fig: bvec src44}}
    \subfloat[J025831$+133420$.\\AD $=$ 2.4, P $=$ 0.04.]{
    \includegraphics[width=0.31\textwidth]{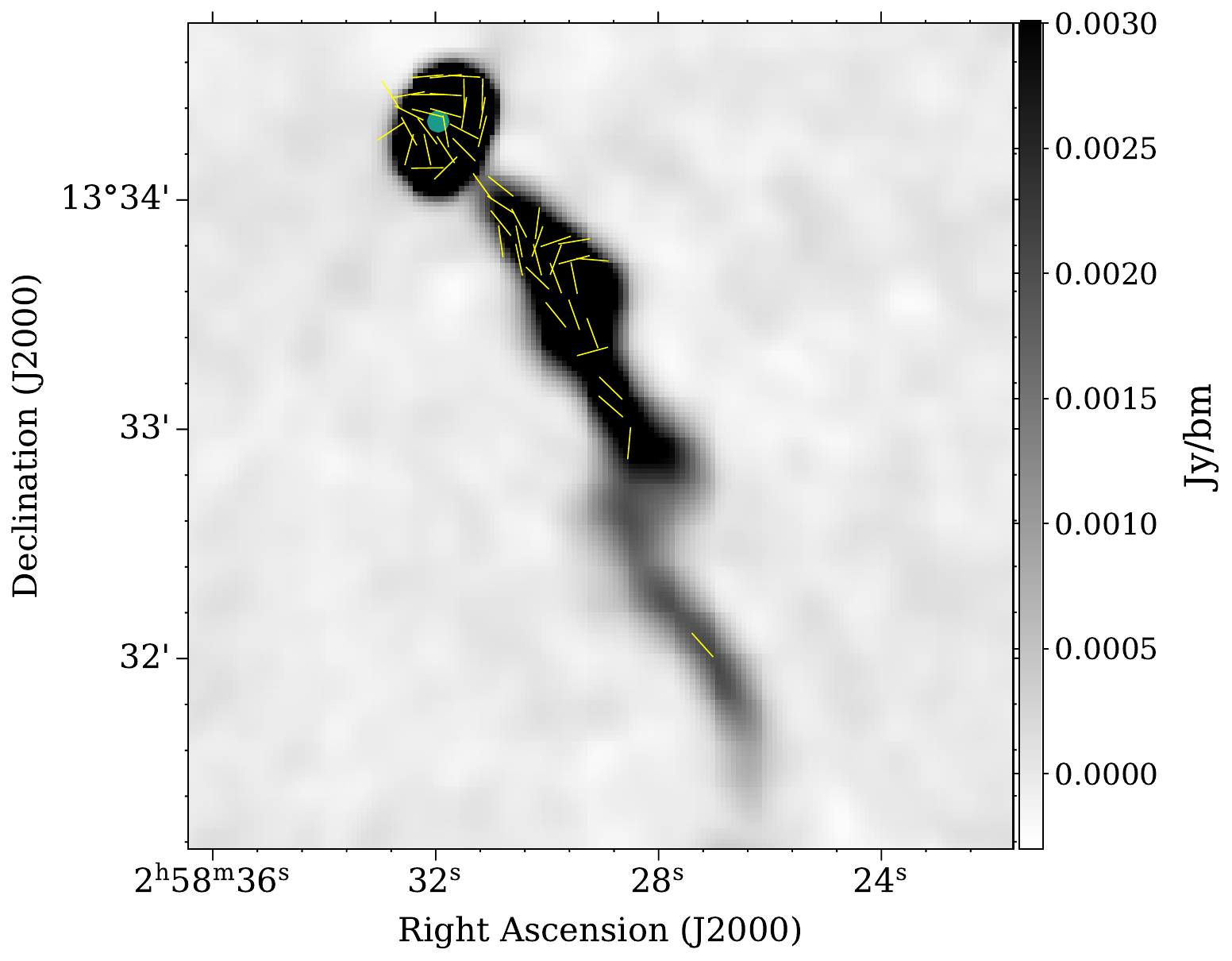}
        \label{fig: bvec src53}}\\
    \centering
    \subfloat[J033814$+100504$.\\AD $=$ 1.3, P $=$ 0.1.]{
    \includegraphics[width=0.32\textwidth]{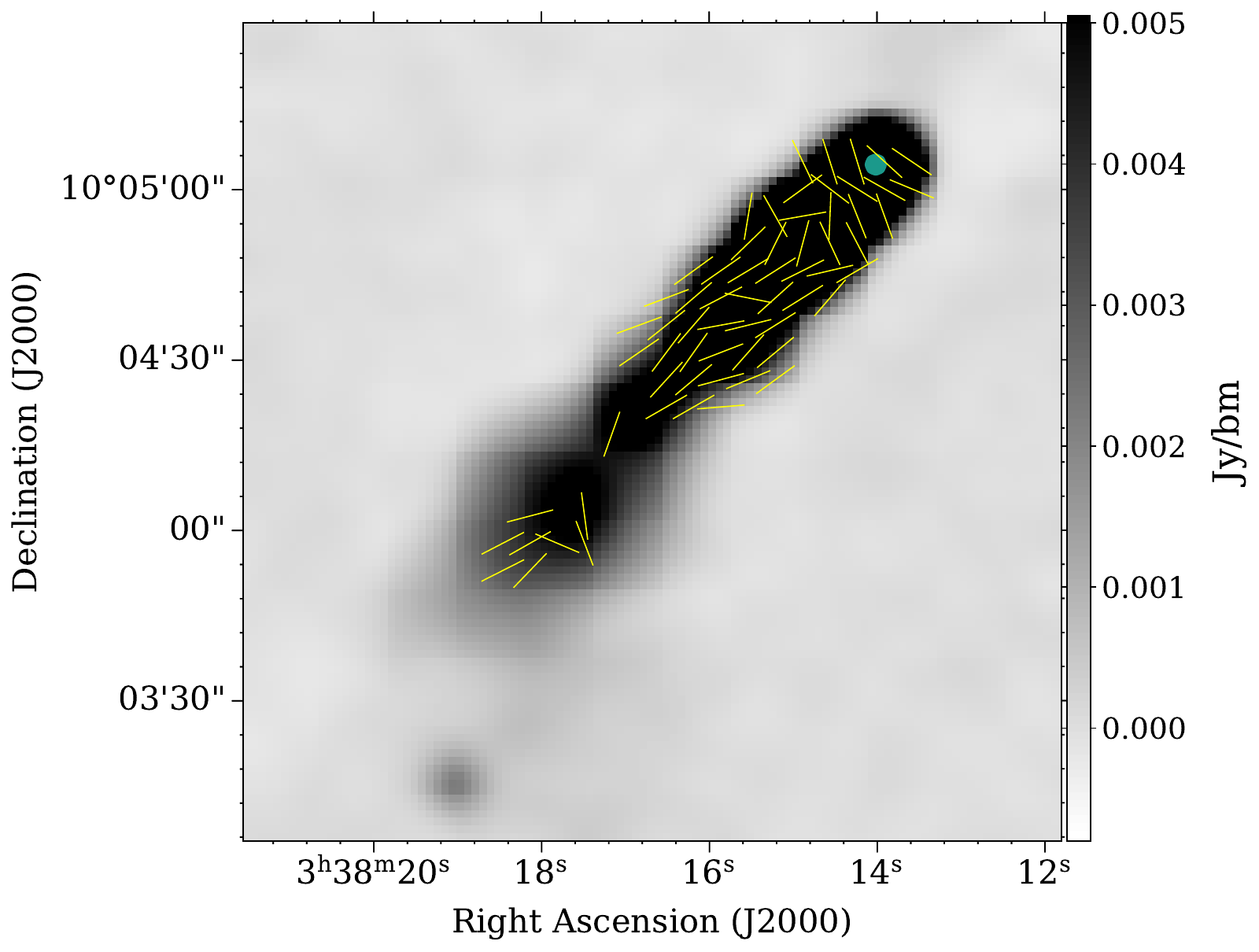}
        \label{fig: bvec src59}}
    \subfloat[J120055$+032655$.\\AD $=$ 0.9, P $=$ 0.1.]{
    \includegraphics[width=0.32\textwidth]{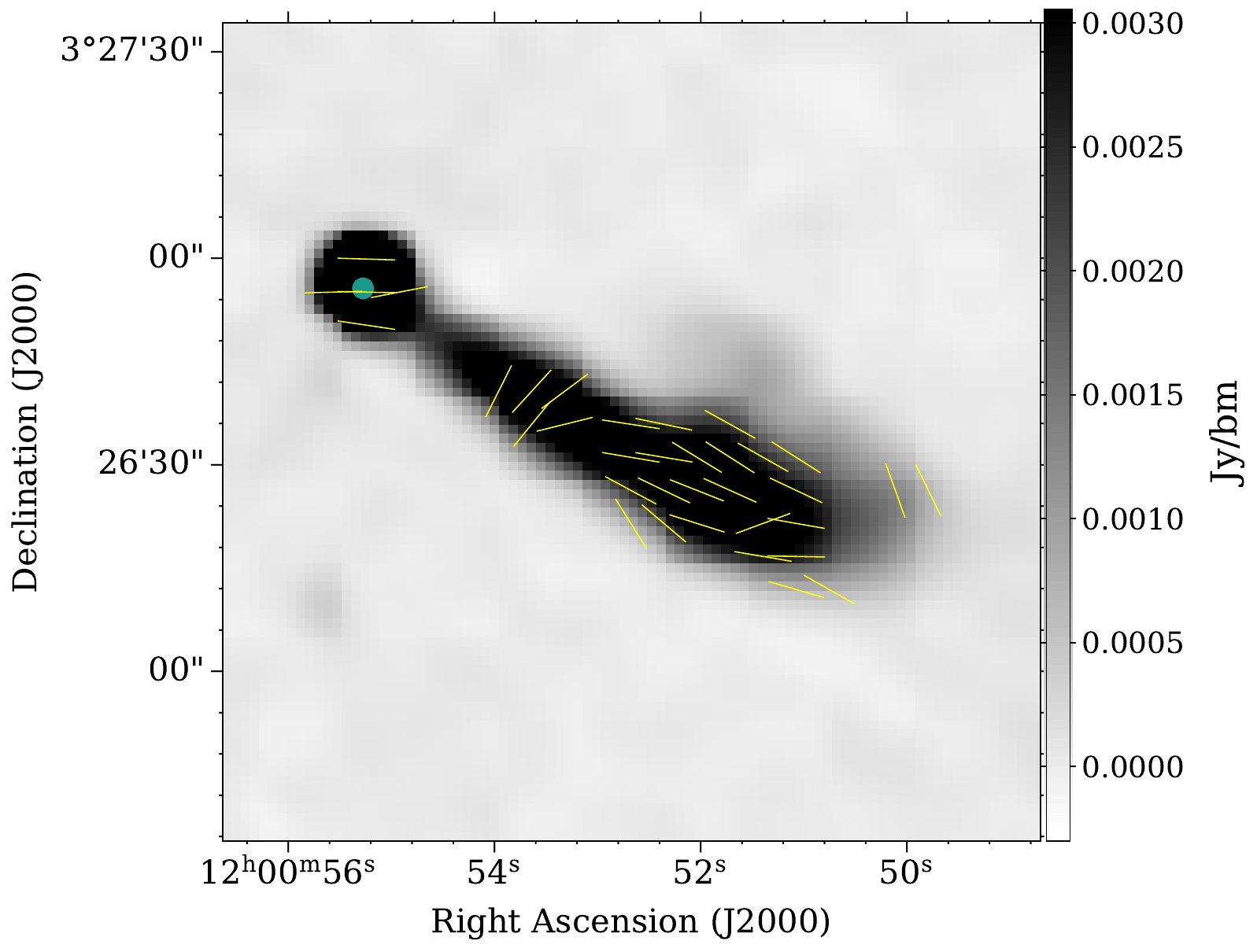}
        \label{fig: bvec src76}}
    \caption{Magnetic field vector maps of the 8 sources with greater than 50 independent vectors. We note that this figure includes the oversampled magnetic field vectors instead of the re-sampled independent field vectors to provide a clearer visualization. The position of the host galaxy in each image is indicted by a cyan circle.}
    \label{fig: bvec maps}
\end{figure*}

After this selection process, we are left with 8 sources with at least 50 independent magnetic field vectors each. These sources are shown in Figure \ref{fig: bvec maps}. We note that this figure includes the oversampled magnetic field vectors instead of the re-sampled independent field vectors to provide a clearer visualization. The two-sample AD test statistics are included in the captions below each plot. The results indicate that none of the 8 sources have independent magnetic field vector distributions that are inconsistent with a random field (i.e., the P-value of the AD test is less than $3.1\times10^{-5}$).

Sources J$121731+033656$ and J$005602-012003$ (Figures \ref{fig: bvec src78} and \ref{fig: bvec src44}, respectively) have more complicated morphologies as well as having S/N$_{\rm{pol}} > 6$ across most of the source. For these sources in particular, the magnetic field vectors appear to follow the morphology of the source, which leads to a variety of different vector angles. While regions of coherence can be identified visually in most of the 8 sources, a more broad distribution of vector angles due to changes in vector orientation across the sources is most likely why the AD test cannot distinguish between the data and random distributions. The two-sample AD test only tests for alignment of the full sample of magnetic field vectors across the entire source, however there may be local alignment that is not captured by this test. For this reason, in the next section we investigate the local alignment of the magnetic field vectors with the direction of bending of the source.

\subsection{Magnetic field vector alignment with the direction of bending}\label{subsec: bvec alignment}

The tendency of the magnetic field vectors in our sources to show some coherence leads to the question of whether the vectors are typically aligned in a common direction. As noted above, in many of the sources in Figure \ref{fig: bvec maps}, the magnetic field vectors appear to trace the morphology of the radio galaxy. In particular, here we test whether the direction of the magnetic field vectors aligns with the direction of bending in the sources. We perform this analysis on the 24 polarized sources in our sample. We only consider magnetic field vectors derived from pixels with S/N$_{\rm{pol}} > 6$, and we select only one pixel per synthesized beam area, using the same selection method as described in Section \ref{subsec: bvec coherence}.

We define the direction of bending to be the line of peak brightness in total intensity along the source. This is akin to the use of ridgelines, where radio brightness is expected to trace the direction of fluid flow in a radio galaxy (e.g., \citealt{Perucho+12,Vega-Garcia+19,Barkus+22}). To define the direction of bending of our sources, we initially tried using the Python package \textsc{FilFinder} \citep{FilFinder}, however we found that the complicated morphology of many of our sources made it difficult to automate the ridgeline identification process. In addition, not all of the ridgelines were identified by the package in the more complicated sources.

Instead, to identify the ridgelines, we take the derivative of the total intensity image of each source and manually define the ridgelines using the path of the minimum in the gradient image as a guide. We provide plots of the ridgelines for each of our 24 polarized radio galaxies in Figure \ref{fig: example ridgelines} in Appendix \ref{app: ridgelines}. \citet{Wezgowiec+24} perform a similar analysis using a filter to highlight intensity gradients in the image of their source. Some sources require diverging ridgelines instead of a single line running along the entire source (e.g., source J$005602-012003$ of Figure \ref{fig: example ridgelines} in Appendix \ref{app: ridgelines}). In all cases, the gradient images show a clear path of the minimum derivative along the source, making identification of a ridgeline straightforward.

\begin{figure}
\centering
    \includegraphics[width=0.47\textwidth]{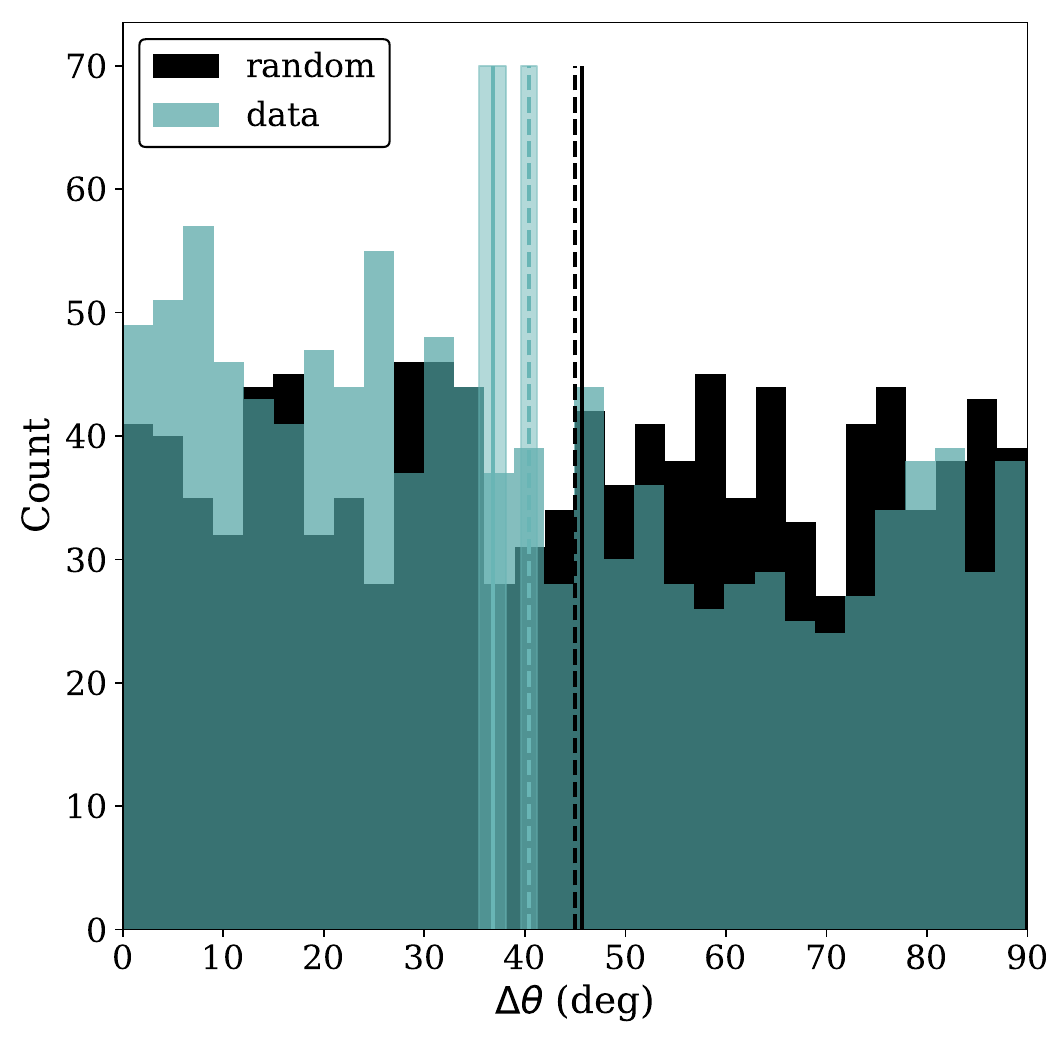}
    \caption{Distribution of the difference in angle, $\Delta\theta$, between the intrinsic magnetic field vectors and the slope of the nearest point on the source ridgeline for each of the 24 radio galaxies in our sample (blue), compared to a randomly generated distribution of angle differences (black). The random distribution is approximately uniform, with a mean(dashed black line) and median (solid black line) of $45\fdg0$ and $45\fdg7$, respectively. The observed distribution is skewed toward smaller angular differences, with a mean (dashed blue line) and median (solid blue line) of $40\fdg4$ and $36\fdg8$, respectively. Shaded bands around the blue lines indicate the 1$\sigma$ uncertainty levels, calculated from 1000 realizations of the random distribution.}
\label{fig: cos-diff hist}
\end{figure}

We calculate the difference between each independent magnetic field vector (using the random selection processes described above to exclude correlated magnetic field vectors within the same synthesized beam area) and the slope of the nearest point on the ridgeline. The results are shown in Figure \ref{fig: cos-diff hist}.

The distribution of the data in Figure \ref{fig: cos-diff hist} is skewed to smaller angle differences, indicating that the direction of the magnetic field vectors has a preference for alignment with the direction of bending along the sources. The mean and median of the data is $40\fdg4$ and $36\fdg8$, respectively, while the mean and median of the randomly drawn distribution is $45\fdg0$ and $45\fdg7$, respectively. We draw 1000 random samples and calculate the standard deviation of the means and medians of each distribution ($\sim$0.77 and $\sim$1.5, respectively) and find a 6$\sigma$ significance level, supporting the conclusion that magnetic field vectors preferentially align with the jet bending direction. We discuss the implications of this result in Section \ref{subsec: disc bvecs}. Because of the generally small number of independent magnetic field vectors in our sources, our ability to assess alignment in the sources individually is limited. A larger sample with an overall higher S/N$_{\rm{pol}}$ is needed to further investigate alignment with direction of bending. However, as shown by \citet{Osinga+22}, sources in galaxy clusters see increased depolarization depending on their location within the cluster. As a result, building such a sample will require considerably deeper data to recover polarized emission across more sources and over more extended regions within each source (e.g., \citealt{Loi+25}).

Although outside the aim of this work, an interesting avenue for future investigation would be to compare direction of bending direction with the location of the brightest cluster galaxy in each cluster, as has been done in previous studies of bent radio galaxy trajectories (e.g., \citealt{Sakelliou&Merrifield2000,Golden-Marx+21}). The availability of magnetic field vector information in our sample could provide valuable additional insight into such analyses.

\subsection{RM and derotated polarization angle gradient alignment}\label{subsec: res rm psi0}

To test whether the RM observed across our bent radio galaxies originates local to the source or is dominated by foreground contributions, we analyze the alignment between the spatial gradients of the RM and of the intrinsic polarization angle, $\psi_0$, maps for each of our sources. A correlation between the gradients of these two quantities may indicate a shared physical origin.

We follow the approach outlined by \citet{Ensslin+03}, which builds upon work by \citet{Rudnick&Blundell03}. \citet{Ensslin+03} introduced the gradient alignment statistic (GAS), which compares the spatial gradients of RM and of $\psi_0$ to test for alignment while correcting for noise-related biases. The alignment statistic, $A$, is defined as:

\begin{equation}
    A[\textbf{p,q}] = \frac{\int d^2x \, \langle \textbf{p}(\textbf{x}),\textbf{q}(\textbf{x}) \rangle}{\int d^2x \, \left|\textbf{p}(\textbf{x})\right| \left|\textbf{q}(\textbf{x})\right|} \; ,
\end{equation}

\noindent where $\textbf{p}(\textbf{x}) = \nabla \mathrm{RM}$ and $\textbf{q}(\textbf{x}) = \nabla \psi_0$\footnote{The gradient of $\psi_0$ is calculated using subtraction modulo 180$^{\circ}$ to account for the cyclic nature of the parameter.}, and $\langle \textbf{p}(\textbf{x}),\textbf{q}(\textbf{x}) \rangle$ is defined as:

\begin{equation}
    \langle \textbf{p}(\textbf{x}),\textbf{q}(\textbf{x}) \rangle = \frac{(p_x^2 - p_y^2)(q_x^2 - q_y^2) + 4 p_x p_y q_x q_y}{\sqrt{p_x^2 + p_y^2} \sqrt{q_x^2 + q_y^2}} \; .
\end{equation}

\noindent $A$ measures the correlation in gradient directions between RM and $\psi_0$, and will have a value of 0 for uncorrelated maps and a value of 1 for full correlated maps. The gradient vector product statistic, $V$, is defined as:

\begin{equation}
    V[\textbf{p,q}] = \frac{\int d^2x \, \textbf{p}(\textbf{x}),\textbf{q}(\textbf{x})}{\int d^2x \, \left|\textbf{p}(\textbf{x})\right| \left|\textbf{q}(\textbf{x})\right|} \; ,
\end{equation}

\noindent and accounts for spurious alignment due to correlated noise. The sum $A + V$ provides a noise-corrected measure of alignment between the gradients, with increasing values indicating stronger structural correspondence between RM and $\psi_0$. 

\begin{figure}
\centering
    \includegraphics[width=0.47\textwidth]{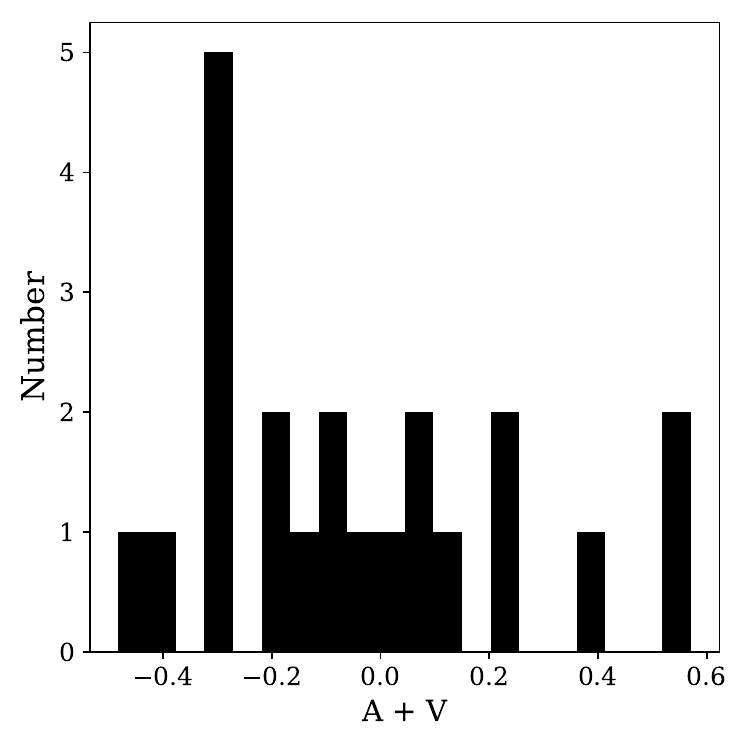}
    \caption{Distribution of the $A+V$ alignment statistic for 22 of the 24 radio galaxies in our sample.}
\label{fig: av hist}
\end{figure}

We compute $A$, $V$, and $A + V$ for 22 of the 24 polarized radio galaxies in our sample (two of our 24 sources are excluded due to an insufficient number of polarized pixels with which to calculate a gradient). As in Sections \ref{subsec: bvec coherence} and \ref{subsec: bvec alignment}, we only use pixels within the 10$\sigma$ total intensity contour with S/N$_{\rm{pol}}$ $>$ 6 to calculate the statistics. A histogram of the resulting $A + V$ values is shown in Figure \ref{fig: av hist}. The distribution has a mean value of $-0.03$ and a standard deviation of 0.29.

Of the 22 sources, 19 have $A + V < 0.2$, consistent with no structural alignment between RM and $\psi_0$ and indicating that the origin of the observed RMs across these sources is foreground to the source. While there is no strict threshold distinguishing co-aligned from unaligned sources, a value of $\sim$0.3 is consistent with the interpretation of \citet{Ensslin+03} that such sources exhibit little to no magnetic field alignment. The remaining 3 sources have $A + V > 0.3$, which we interpret as indicating co-alignment and which suggests that the origin of the observed RMs is local or intrinsic to these sources. We discuss these results in more detail in Section \ref{subsec: disc rm psi0}.

\section{Discussion} \label{sec:disc}

In this section, we first interpret our key observational results, then consider the implications for RM grid experiments, and finally discuss expectations for future wide-field polarization surveys.

\subsection{Lack of Correlation Between Polarization Parameters and Bending Angle}\label{subsec: RM_sigRM_vs_theta}

We find no statistically significant correlation between bending angle and either $\sigma_{\rm{RM}}$ or $\lvert$RM$\rvert$ in our sample of polarized bent radio galaxies. This result is somewhat surprising in light of previous work by \citet{Osinga+22,Osinga+25} and \citet{vanderJagt+25}, who find correlations between RM and bending angle and R/R$_{500}$ (see Section \ref{subsec: pol vs theta}). These studies imply that both polarization properties and radio galaxy morphology are sensitive to the ICM environment, either locally or integrated along the line of sight, and might be expected to correlate with each other. However, in our sample, neither $\sigma_{\rm{RM}}$ nor $\lvert$RM$\rvert$ show a statistically significant correlation with bending angle (see Figure \ref{fig: corr subplots BA}).

Two broad interpretations can account for this result: (1) a correlation may exist but remains undetectable given the resolution of our observations or the size of our sample, or (2) there is no intrinsic correlation between bending angle and these polarization parameters. The first explanation is supported by our correlation statistics. In Section \ref{subsec: pol vs theta}, we showed that although \citet{Osinga+25} (a sample of $\sim$800 polarized components) and \citet{vanderJagt+25} (a sample of 109 bent radio galaxies) report significant correlations between RM, bending angle, and R/R$_{500}$, our data reveal only weak, statistically insignificant relationships among these parameters. The weakened correlations between RM and position in the cluster in our sample may therefore explain the lack of correlation between RM and bending angle. To test this explicitly, we drew 1000 random subsamples of 24 sources from the \citet{Osinga+25} and \citet{vanderJagt+25} samples individually. For the \citet{Osinga+25} sample, we selected only components located inside the cluster. The full sample shows a strong correlation between $|$RM$|$ and R/R$_{500}$ ($\tau=-0.4$, $p=8\times10^{-24}$), but the 1000 random subsamples typically yielded weaker results (median $p=0.01$, mean $p=0.1$, with standard deviation $p=0.1$). We note, however, that \citet{Osinga+25} measures RMs of individual components rather than integrated spectra, in contrast to our approach, making direct comparison with our results more difficult. Similarly, the \citet{vanderJagt+25} sample shows a correlation between bending angle and R/R$_{500}$ in the full dataset ($\tau=-0.3$, $p=6\times10^{-5}$), while the 1000 random subsamples of 24 sources also yielded weaker correlation results (median $p=0.1$, mean $p=0.2$, with a standard deviation $p=0.2$). These results demonstrate that when restricted to small samples, the correlations reported in larger studies can weaken substantially, consistent with the lack of correlation we observe.

Alternatively, the lack of correlation may be understood in terms of the distinct spatial scales and physical quantities that these observables probe. The ICM is expected to be turbulent across a broad range of scales in both merging and relaxed clusters \citep{Subramanian+06,Dolag+08,Vazza+12}. RM and $\sigma_{\rm{RM}}$ reflect integrated properties of the magnetized plasma along the entire line of sight, such as total electron column density and fluctuations in the line-of-sight magnetic field. Bending angle, by contrast, is governed by ram pressure \citep{Jones&Owen1979,Vallee+81,Garon+19}, $P_{\rm{ram}} \propto \rho_{\rm{ICM}}v^2$, which depends on the local ICM density ($\rho_{\rm{ICM}}$) and the galaxy’s velocity relative to the ICM ($v$). As such, it reflects local conditions at the galaxy’s position within the cluster. If two bent radio galaxies have the same bending angle and are located at the same projected distance from the cluster center, but one lies on the near side of the cluster and the other on the far side, their measured RMs may differ significantly due to the depth and structure of the intervening ICM. Similarly, turbulence in the ICM introduces local fluctuations in density, potentially leading to spatial and temporal variations in the ram pressure experienced by the jet \citep[e.g.,][]{Tonnesen&Bryan2008}. These differences in how RM, $\sigma_{\rm{RM}}$, and bending angle probe the cluster environment could obscure any direct relationship between them.

Projection effects further complicate the interpretation of any potential correlation. We observe bending angle only as projected in the sky plane. A galaxy moving with a significant line-of-sight velocity through a dense region of the ICM will experience strong ram pressure and enhanced RM at its leading edges as magnetic field lines become compressed \citep{Pfrommer&Dursi2010}, but its projected bending may appear small due to its projection onto the 2D plane. Conversely, the same galaxy moving largely in the plane of the sky will exhibit its full bending. Thus, even if two galaxies experience comparable physical ram pressure, their observed bending angles may differ substantially depending on orientation. Similarly, RM and $\sigma_{\rm{RM}}$ are affected by the galaxy’s depth in the cluster and the structure of the ICM along the line of sight, introducing additional scatter between galaxies at similar projected radii.

In addition, differences in sample selection may contribute to the absence of correlation. \citet{vanderJagt+25} select their sample in total intensity, while our sample is selected based on detected polarization. This difference in selection could introduce a bias in RM properties. Polarized sources are more easily detected when Faraday depolarization is low, which may favor sources located on the near side of the cluster or along less dense, less turbulent sightlines. As a result, our polarization-selected sample may underrepresent the full range of RM values observed near the cluster center. If so, this would further limit the dynamic range in RM and could weaken any potential correlation with bending angle. However, low polarization can also result from intrinsic source properties, which may limit the extent to which Faraday effects alone account for the observed detection bias.

Jet bending is a time-dependent process, reflecting the dynamical evolution of radio galaxies as they move through the ICM. Simulations by \citet{O'Neill+19} show that a NAT morphology evolves through an earlier WAT morphology, supporting the idea that bending angle increases with time. While \citet{Morsony+13} demonstrate that the bending angle (or radius of curvature) is primarily determined by environmental and jet properties—such as jet luminosity, jet velocity, AGN velocity, and ICM density—they also find that the radius of curvature can vary by up to 25 percent for fixed input parameters. Moreover, they show that the length of the radio tail trailing the galaxy grows over time, consistent with an extended emission history. These results suggest that although bending angle is largely determined by environmental and jet properties, morphological features may evolve over time, which could weaken any observed correlation with turbulent quantities like $\sigma_{\rm{RM}}$ in a statistical sample.

Finally, we note that $\sigma_{\rm{RM}}$ may be more strongly correlated with ICM turbulence that varies systematically with position in the cluster, rather than with bending angle. We perform a Kendall's $\tau$ correlation test between $\sigma_{\rm{RM}}$ and R/R$_{500}$ in the lobe spectra and find $\tau = -0.7$ and P $= 5\times10^{-4}$. This does not meet our threshold for a potential correlation, but it does indicate a possible dependence between these two parameters. This suggests that $\sigma_{\rm{RM}}$ is correlated with the position of a radio galaxy within the cluster, and that we may be tracing turbulence from the intervening ICM, which is largely dependent on the distance from the cluster center and not bending angle. This agrees with cluster magnetism studies that use resolved radio galaxies as probes of the ICM (e.g., \citealt{Bonafede+10}).

In summary, we find no statistically significant correlation between $\sigma_{\rm{RM}}$ and bending angle in our sample of bent radio galaxies. This result may reflect a genuine lack of connection between radio galaxy morphology and small-scale turbulence in the ICM. While our data are consistent with a scenario in which $\sigma_{\rm{RM}}$ traces turbulence along the line-of-sight and where bending angle reflects local dynamical conditions or time-evolution, a more robust examination of any potential correlation will require larger samples and higher-resolution observations.

\subsection{Magnetic field vector coherence and alignment with direction of bending}\label{subsec: disc bvecs}

In Figure \ref{fig: bvec maps} we showed that magnetic field vector coherence is present in many of our more polarized sources, and Figure \ref{fig: cos-diff hist} hints at a preference for magnetic field vectors to align with the direction of bending, or fluid flow. This implies that the bending or distortion of radio galaxies from a linear morphology does not disorder the large-scale magnetic field within the source jets and lobes. \citet{O'Neill+19} conducted the first high-resolution 3D magnetohydrodynamic simulation focusing on the evolution of a bipolar-jet radio galaxy interacting with a constant crosswind. The authors found that the jets generally remain coherent during and after the bending process, although if instabilities arise they can disrupt the jets. \citet{O'Neill+19} also found that the stress from the jet-bending process can convert the initial toroidal magnetic field into a predominantly poloidal configuration. Helical magnetic fields are also thought to be present in the jets of radio galaxies, which can manifest (among many other ways) as an orthogonal polarization pattern along the jet of a galaxy \citep{Gabuzda+18}. If helical magnetic fields are present in some of the sources and multiple periods of the helical path are averaged within the synthesized beam, this could potentially obscure any relationship between magnetic field vector orientation and direction of bending.

To our knowledge, there are few examples of magnetic field vector maps of well resolved bent radio galaxies. In a map of the asymmetric radio galaxy 4C 70.19 (believed to be a member of a galaxy group and not a galaxy cluster) from \citet{Wezgowiec+24}, the magnetic field vectors show a high level of coherence with regions where the vectors are generally aligned with the direction of fluid flow as well as region where the vectors appear nearly perpendicular to the direction of flow. Together these previous studies suggest that alignment or anti-alignment with the direction of bending may be dependent on which part of the source we are probing (e.g., jet or lobe, inner lobe or the edges). Many of our sources are depolarized across much of the source. In each of our 8 sources with well-sampled magnetic field vectors, we may be probing the magnetic field vectors in different regions of the source (e.g., inner jet versus outer jet versus lobe).

The results of our magnetic field vector analysis are highly dependent on resolution and angular size of the source. Lower resolution combined with small angular size will average more magnetic field vectors together within the synthesized beam. We are also limited by S/N, which limits the number of pixels with a reliably measured magnetic field vector. The majority of our sources have few magnetic field vectors with which to test coherence and alignment (see Section \ref{subsec: bvec coherence}). A larger sample of sources with higher S/N$_{\rm{pol}}$ and better-resolved spatial structure would be ideal for future analysis.

\subsection{Implications of RM–$\psi_0$ gradient alignment analysis}\label{subsec: disc rm psi0}

The gradient alignment analysis in Section \ref{subsec: res rm psi0} indicates that the Faraday rotation in most of our bent radio galaxies is dominated by foreground material. Of the 22 sources examined, 19 have $A+V \lesssim 0.3$, pointing to no significant physical correlation between RM and $\psi_0$ and therefore to a foreground RM origin. Only 3 sources reach $A+V > 0.3$, suggestive of a local RM origin. The distribution of $A+V$ values is reasonably narrow and centered on zero ($\mu = -0.03$ and $\sigma = 0.29$), reinforcing the view that large-scale RM–$\psi_0$ alignment is generally absent in our sample. These findings are similar to those of \citet{Ensslin+03}, who reported $A+V < 0.2$ for cluster-member radio galaxies PKS 1246$-$410, Cygnus A, Hydra A, and 3C 465.

\begin{figure}
\centering
    \includegraphics[width=0.47\textwidth]{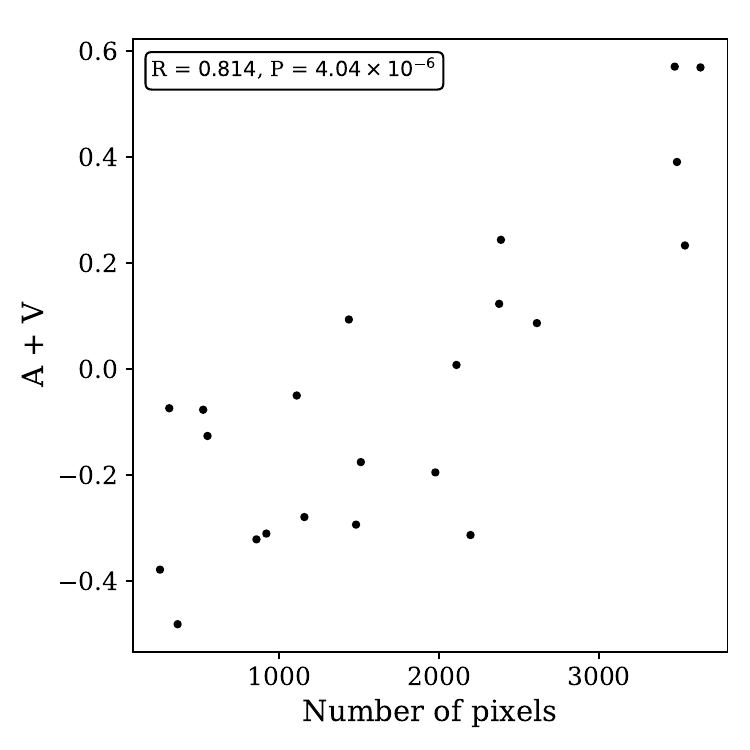}
    \caption{$A+V$ alignment statistic versus the number of pixels used in the calculation of the statistic. A statistically significant correlation is detected between the two quantities.}
\label{fig: av vs pix}
\end{figure}

Figure \ref{fig: av vs pix} plots $A+V$ against the number of polarized pixels present in each of the source gradient maps. Pixel count serves as a proxy for how finely a source is sampled in polarized intensity: a larger tally of pixels above $\mathrm{S/N_{pol}}>6$ means more independent measurements of RM and $\psi_0$ across the lobes. However, pixel count also depends on the resolution and angular size, so this is not a perfect proxy. A Pearson correlation test identifies a statistically significant relationship between $A+V$ and pixel count (R = $0.814$, P = $4.04 \times 10^{-6}$). This indicates that the apparent presence (or absence) of RM–$\psi_0$ gradient co-alignment may be sensitive to both resolution and polarized S/N. The dependence of the $A+V$ statistic on number of pixels comes from the $V$ term, which shows the same dependence on pixel count as the sum, while the $A$ term shows no dependence.

Our gradient alignment results are consistent with the lack of a statistically significant correlation between RM and bending angle seen in Section \ref{subsec: RM_sigRM_vs_theta}, which indicates that we are primarily probing foreground ICM properties rather than source properties. This supports the interpretation that bent radio galaxies are not a biased population in terms of Faraday depth and can be reliably used to probe the magnetized ICM. However, we note that this conclusion may be affected by the dependence of the $A + V$ statistic on pixel count, and we therefore caution against overinterpreting its results. A larger, more uniformly sampled set of bent radio galaxies will be essential to confirm these findings and to better isolate observational biases from genuine differences in Faraday rotation origin.

\vspace{5mm}

Overall, our results lead to the conclusion that we are likely primarily probing large-scale ICM effects that are not dependent on bending angle. This is consistent with the findings of \citet{Osinga+22}, who show that radio galaxies embedded in a cluster show similar depolarization profiles to radio galaxies that are background to the cluster. This implies that local ICM effects are not dominant in statistical samples.

\section{Bent radio galaxies as a population of RM grid sources}

Maps of RMs on the sky, known as RM grids \citep{Gaensler+04}, are an invaluable method of probing magnetic field geometry in many different environments, including the large-scale Galactic magnetic field \citep{Mao+10,VanEck+11,Hutschenreuter+22}, molecular clouds \citep{Tahani+18}, the jets and lobes of radio galaxies \citep{Feain+09,O'Sullivan+18}, galaxy clusters \citep{Bonafede+10,Anderson+21}, and the cosmic web \citep{Vernstrom+19,Amaral+21,Carretti+22}. Current surveys such as the Polarisation Sky Survey of the Universe's Magnetism (POSSUM; \citealt{Gaensler+25}), as well as the upcoming Square Kilometre Array, will greatly increase the number of polarized source detections that can be used in RM grid experiments \citep{Johnston-Hollitt+15}.

Since bent radio galaxies typically reside in overdense environments like galaxy clusters and groups, they will be a subset of the population of sources used to probe the magnetic field strength and geometry in clusters. While \citet{Osinga+25} find a correlation between RM scatter and projected distance to the cluster center, and \citet{vanderJagt+25} find a correlation between bending angle and projected distance to cluster center, our work shows no correlation between RM and bending angle at the physical resolutions of our data (synthesized beam extends of 3--46 kpc, with mean and median values 18.8 and 18.4 kpc, respectively). This lack of statistically significant correlation between bending angle and RM is evidence that we are probing foreground ICM effects, and that bent radio galaxies are likely not biasing the RM-distance correlation. Therefore, bent radio galaxies at the typical resolution of this study are likely a useful population of sources for probing ICM magnetic field properties with RM grid studies of galaxy clusters.

\subsection{Estimated bent radio galaxy sample from POSSUM}\label{subsubsec: POSSUM sample}

As noted above, a larger statistical sample size would provide a more robust determination of the presence or absence of correlation between bending angle and polarization properties of a source. POSSUM is currently surveying 20 000 deg$^2$ of the sky at Declination (J2000) $<$ 0$^{\circ}$ with the Australian Square Kilometre Array Pathfinder (ASKAP; \citealt{Hotan+21}). Here we estimate the number of bent radio galaxies that will be present in the full POSSUM survey catalogue.

The all-sky Planck catalogue of Sunyaev-Zeldovich sources (PSZ2; \citealt{PSZ2}) contains 818 galaxy cluster detections at Declination $<$ 0$^{\circ}$. Of these 818 clusters, 499 are previously optically-identified clusters with known redshifts, and of this subset, 348 (70\%) are at a redshift of $z \leq 0.3$, where 0.3 is approximately the highest redshift in our sample. We find that the 24 polarized sources in our sample are distributed across 19 clusters from the 124 observed by \citet{Osinga+22} (see Section \ref{sec:data and sample}). However, the POSSUM synthesized beam width will be 20 arcsec, which is $\sim$2--3$\times$ larger than the beam widths in our sample. Of the 24 polarized sources, 16 would be resolved enough in the POSSUM survey to still have at least 3 synthesized beam widths across the source. These 16 sources are found in 15 galaxy clusters, corresponding to 12\% of the total number of clusters originally observed ($\frac{15}{124}$), and 1.1 polarized galaxies per cluster.

From these calculations, we estimate that the POSSUM survey catalogue will contain 348 PSZ2 clusters $\times$ 0.12 $\times$ 1.1 polarized bent radio galaxies per cluster, which is 46 bent galaxies with a range of bending angles. We include the caveat that the mass distribution of clusters in the full PSZ2 survey will not match that of our sample, and is likely to extend to lower-mass systems. While the occurrence of WATs and NATs as a function of cluster mass is not well constrained, bent radio galaxies are known to be preferentially associated with over-dense environments. They have, however, also been detected in lower-mass groups in both observations (e.g., \citealt{Freeland+08}) and simulations (e.g., \citealt{Morsony+13}), indicating that bending can occur across a range of environments. Nonetheless, if bent sources are more common in higher-density systems, it is plausible that lower-mass clusters will host fewer tailed radio galaxies, making our estimate an upper limit. However, additional polarized bent radio galaxies can likely be added to the sample once redshifts are measured for the new PSZ cluster detections. If the redshift distribution is similar in the newly discovered galaxy clusters (which is not necessarily expected), we could expect up to an additional 30 bent radio galaxies. As well, all galaxy clusters in the POSSUM survey area should be inspected. Any galaxies with a larger angular extent (at least 3 synthesized beam widths across the source) in higher redshift ($z > 0.3$) clusters that also meet our sample selection criteria should be included in the sample.

If we instead consider the Spectrum-Roentgen Gamma mission with the extended ROentgen Survey with an Imaging Telescope Array (SRG/eROSITA; \citealt{Sunyaev+21,Predehl+21}) catalogue \citep{ERASS1}, we estimate a significantly larger sample. SRG/eROSITA surveyed $\sim$20 000 deg$^2$ of the sky at X-ray frequencies. Only regions of the survey that overlapped with the DESI Legacy Survey Data Releases 9 and 10 (for optical follow-up of cluster detections) and that did not contain high Galactic-latitude supernova remnants were used, leaving a final survey area for cluster detection of 13 116 deg$^2$. The catalogue contains $10\,248$ galaxy clusters with redshift at declination $< 0^{\circ}$, the majority of which are new detections \citep{ERASS1}. \citet{ERASS1} construct a more robust, securely confirmed catalogue of 5259 galaxy clusters, which we here use to estimate sample size in the POSSUM survey. Of the 5259 securely confirmed cluster detections, 2337 are at declination $< 0^{\circ}$ and are at a redshift of $z \leq 0.3$. Using the same polarized source density per cluster and percentage of cluster with polarized sources that meet our selection requirements as above, we estimate a sample of polarized bent radio galaxies in POSSUM of at least 2337 $\times$ 0.12 $\times$ 1.1 $=$ 308 bent radio galaxies. This number may be even larger, since the SRG/eROSITA survey does not cover the entire sky at declination $< 0^{\circ}$, and because we are only considering the smaller catalogue of securely confirmed cluster detections. Similar to the caveat described for the PSZ2 clusters, the mass distribution of the eROSITA clusters is not expected to match that of our sample, so our estimates here should likewise be considered an upper limit.

\subsection{Estimated bent radio galaxy sample from the SKA}

The forthcoming all-sky survey with the Square Kilometre Array, particularly SKA1-Mid Band 2\footnote{\url{https://www.skao.int/en/science-users/118/ska-telescope-specifications}} (SKA1-Mid2), will deliver major improvements in both angular resolution and sensitivity over existing wide-area radio surveys. SKA1-Mid2 is expected to observe $\sim$$30\,000$ deg$^2$ of the sky over 950--1760 MHz with a resolution of approximately 2 arcsec and a sensitivity of 4 $\mu$Jy/beam \citep{Heald+20}, representing an order-of-magnitude improvement over the POSSUM survey in both respects.

These enhancements are expected to significantly increase the number of bent radio galaxies that meet the selection criteria for studies such as this one. In our current dataset, we find an average of 1.1 polarized radio galaxies per cluster that are sufficiently resolved (see Section \ref{subsubsec: POSSUM sample}) and meet our S/N${\rm{pol}}$ threshold. The fivefold improvement in sensitivity between POSSUM ($\sim$20 $\mu$Jy/beam) and SKA1-Mid2 (4 $\mu$Jy/beam) implies a larger number of detectable sources. Using the cumulative number count distribution for polarized sources, which scales approximately as $P_{det}^{-0.6}$ \citep{Rudnick&Owen2014}, where $P_{det}$ is the detection limit in polarization, we estimate that SKA1-Mid2 will detect approximately three times as many polarized radio galaxies per cluster as POSSUM, based on sensitivity alone. Updating our estimates from Section \ref{subsubsec: POSSUM sample}, we estimate that SKA1-Mid2 will be able to detect $3.3 \times 308 = 1016$ polarized radio galaxies at $z \leq 0.3$ and declination $< 0$ with the SRG/eROSITA catalogue.

If we remove the redshift constraint, SRG/eROSITA has 4516 securely detected clusters out to $z \sim 0.8$, nearly double the number present at $z \leq 0.3$. Given its sensitivity, we expect SKA1-Mid2 to detect polarized radio galaxies in all of these clusters (e.g. \citealt{Johnston-Hollitt+15b}). Additionally, the order-of-magnitude improvement in angular resolution from POSSUM (21 arcsec) to SKA1-Mid2 (2 arcsec) will further increase the number of sources that satisfy our $\geq$3-beamwidth selection criterion.

Higher angular resolution will also enable improved mapping of the intrinsic magnetic field vectors within individual sources. In Section \ref{subsec: bvec alignment}, we report a statistically significant preference for alignment between magnetic field vectors and the direction of jet bending. However, this preference for alignment may be underestimated due to beam-averaging of sharp angle changes (e.g., a change in direction of jet-bending) or helical magnetic field structures. The increased resolution and sample size provided by SKA1-Mid2 will allow this potential alignment to be tested in a more robust and statistically powerful way.

We compare our results to other bent radio galaxy samples in the literature. We estimate that POSSUM will contain 308 polarized bent radio galaxies (0.015 deg$^{-2}$) and the SKA will contain 1016 polarized bent radio galaxies (0.034 deg$^{-2}$), requiring that sources are extended by at least three beamwidths, are polarized, and lie at $z\leq0.3$. In contrast, \citet{Wing&Blanton2011} identified 272 bent and 449 straight radio galaxies through an automated search in total intensity in the VLA Faint Images of the Radio Sky at Twenty Centimeters (FIRST; \citealt{Becker+95}) survey over $\sim$9000 deg$^{2}$. They also identified an additional 166 bent radio galaxies through a visual search over a $\sim$3000 deg$^{2}$ region of the same survey, giving a total average sky density of 0.068 deg$^{-2}$. Similarly, \citet{Paterno-Mahler+17} reported 646 bent radio galaxies in total intensity in the high-redshift Clusters Occupied by Bent Radio AGN (COBRA) Survey survey (a sky density of 0.072 deg$^{-2}$), while \citet{Vardoulaki+25} found 19 bent radio galaxies in total intensity in the 2 deg$^{2}$ extragalactic COSMOS field (a sky density of 9.5 deg$^{-2}$).

Assuming that only $\sim$20\% of extended sources are polarized (we find that 156 of the 704 total extended sources in the \citealt{Osinga+22} catalogue show some level of polarization) the \citet{Wing&Blanton2011} and \citet{Paterno-Mahler+17} densities both reduce to 0.014 deg$^{-2}$, broadly consistent with our POSSUM estimate. The COSMOS polarized bent radio galaxy density remains substantially higher, likely due to the sub-arcsecond resolution of some of the \citet{Vardoulaki+25} observations as well as the absence of a redshift limit. Variations in redshift range (the majority of the \citealt{Paterno-Mahler+17} and \citealt{Vardoulaki+25} sample is at $z>0.3$), no requirement of cluster membership, and angular resolution and sensitivity make direct comparison between these samples and ours difficult, but they also indicate that the true sample size of polarized bent radio galaxies in the POSSUM and SKA surveys will be larger than we estimate if our redshift restrictions are relaxed.

\section{Conclusion and future work}\label{sec:conc}

In this work, we present the first polarization study of a large number (24) of polarized radio galaxies with varying degrees of bending. We test for correlations between bending angle and the polarization properties $\lvert \rm{RM_{corr}} \rvert$, $\sigma_{\rm{RM}}$, and fractional polarization. We find no statistically significant correlation between these properties and bending angle or integrated physical area. We do find a potential correlation between lobe $\sigma_{\rm RM}$ and bending angle (P = $6\times10^{-4}$), which does not meet our significance threshold. Future larger samples of bent radio galaxies from POSSUM and SKA will be able to perform more robust statistical tests to determine whether this potential correlation is a result of selection effects or a true physical relationship between the parameters. Overall, our results lead to the conclusion that we are primarily probing large-scale ICM effects, which are not dependent on bending angle, and not the intrinsic polarization properties of the sources.

We find coherence in the polarization angles in 8 of our more highly polarized sources, suggesting that the large-scale magnetic fields are generally organized as a radio galaxy bends. We calculate the difference between the direction of magnetic field vectors and the direction of bending with our sources, and we find that the magnetic field vectors show a preference for alignment with the direction of bending, with a 6$\sigma$ significance level.

We discuss the use of bent radio galaxies in RM grid studies of galaxy cluster magnetic fields. We conclude that the use of bent radio galaxies at the typical physical resolutions of our sample in the construction of cluster RM grids should not bias values of magnetic field strength estimated from the relation between RM and depolarization and distance to the cluster center. Finally, using galaxy cluster detections from PSZ and SRG/eROSITA, we estimate that the $20\,000$ deg$^2$ POSSUM survey area and the anticipated $\sim$$30\,000$ deg$^2$ SKA-Mid2 survey will contain $\gtrsim$300 and $\gtrsim$1000 polarized radio galaxies, respectively, with a range of bending angles. These estimates assume a similar redshift distribution and number of beamwidths across each source as our sample, and they represent a 1--2 order of magnitude increase over the current dataset.

Further analysis to determine if the origins of the measured RMs are foreground to the galaxies is needed. A larger sample of bent radio galaxies is needed to perform a more robust statistical test of correlation between polarization properties and bending angle, as well as to further investigate the coherence and alignment of magnetic field vectors in these radio galaxies.

\section*{Acknowledgments}

We thank Sarah Bradbury for her guidance in the early stages of using and troubleshooting the \textsc{FilFinder} package. The Dunlap Institute is funded through an endowment established by the David Dunlap family and the University of Toronto. S.V. acknowledges the support of the Natural Sciences and Engineering Research Council of Canada (NSERC) through grant RGPIN-2022-03163, and of the Canada Research Chairs program. The National Radio Astronomy Observatory and Green Bank Observatory are facilities of the U.S. National Science Foundation operated under cooperative agreement by Associated Universities, Inc. We thank the anonymous reviewer for their careful review of the manuscript and their thoughtful feedback.

\appendix

\vspace{-5mm}
\section{Ridgelines}\label{app: ridgelines}

We here show our ridgeline tracing method. In Figure \ref{fig: example ridgelines} we plot the derivative maps of the total intensity maps of each of the 24 polarized radio galaxies in our sample. As described in Section \ref{subsec: bvec alignment}, the ridgeline of each source is traced by minimum path along each jet/lobe region.

\begin{figure*}
\centering
    \includegraphics[width=0.76\textwidth]{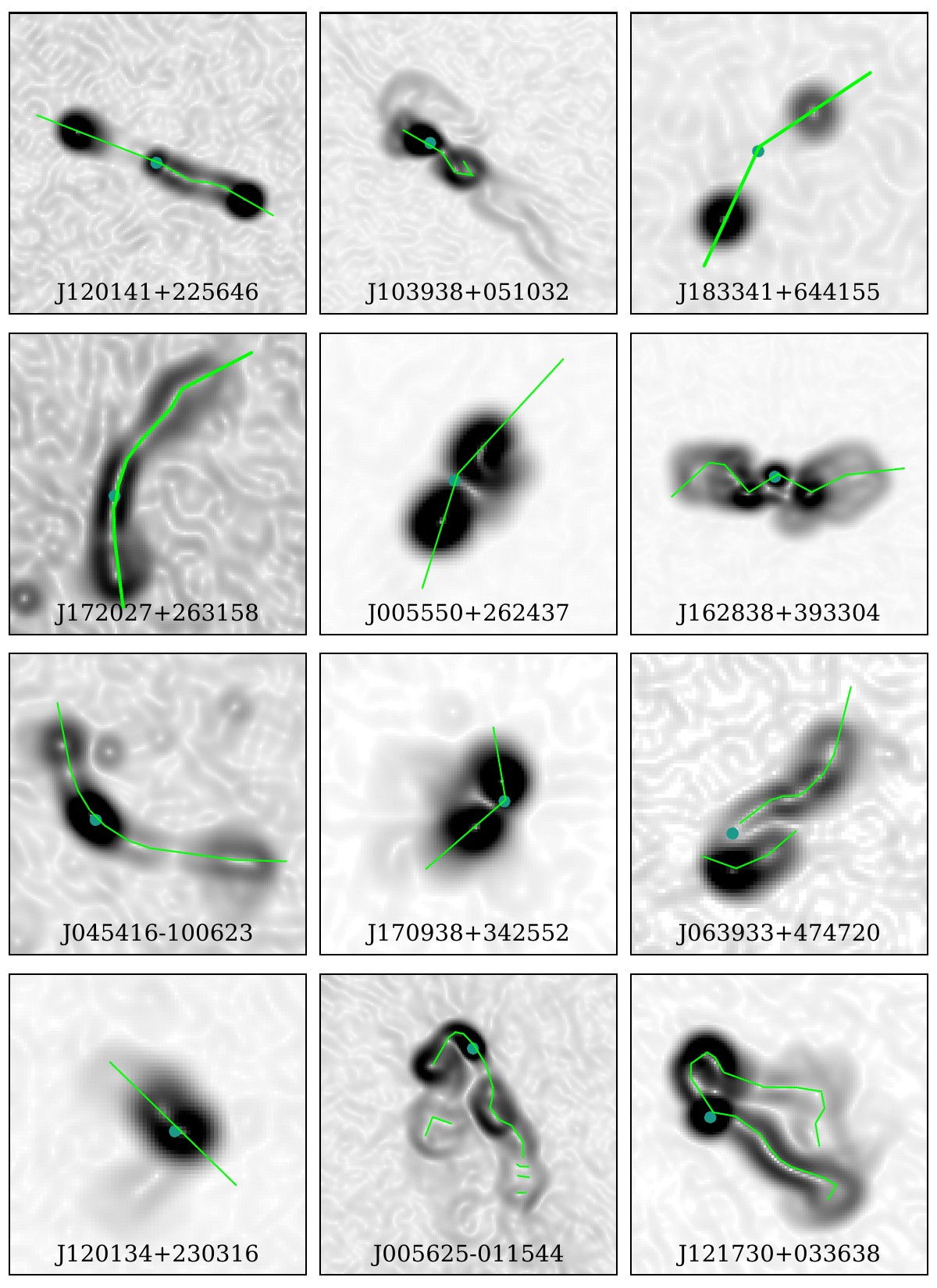}
    \caption{Source ridgelines for the first (ordered by increasing bending angle) 12 polarized radio galaxies in our sample. For each source, the greyscale image is of the derivative of the total intensity MFS image from Figure \ref{fig: MFS cutouts 1}, and the green lines indicate the ridgeline that we have manually traced, following the minimum line along each jet/lobe. The position of the host galaxy in each image is indicted by a cyan circle, and the source identifier is noted at the bottom of each subplot.}
\label{fig: example ridgelines}
\end{figure*}

\begin{figure*}
\ContinuedFloat
\centering
    \includegraphics[width=0.76\textwidth]{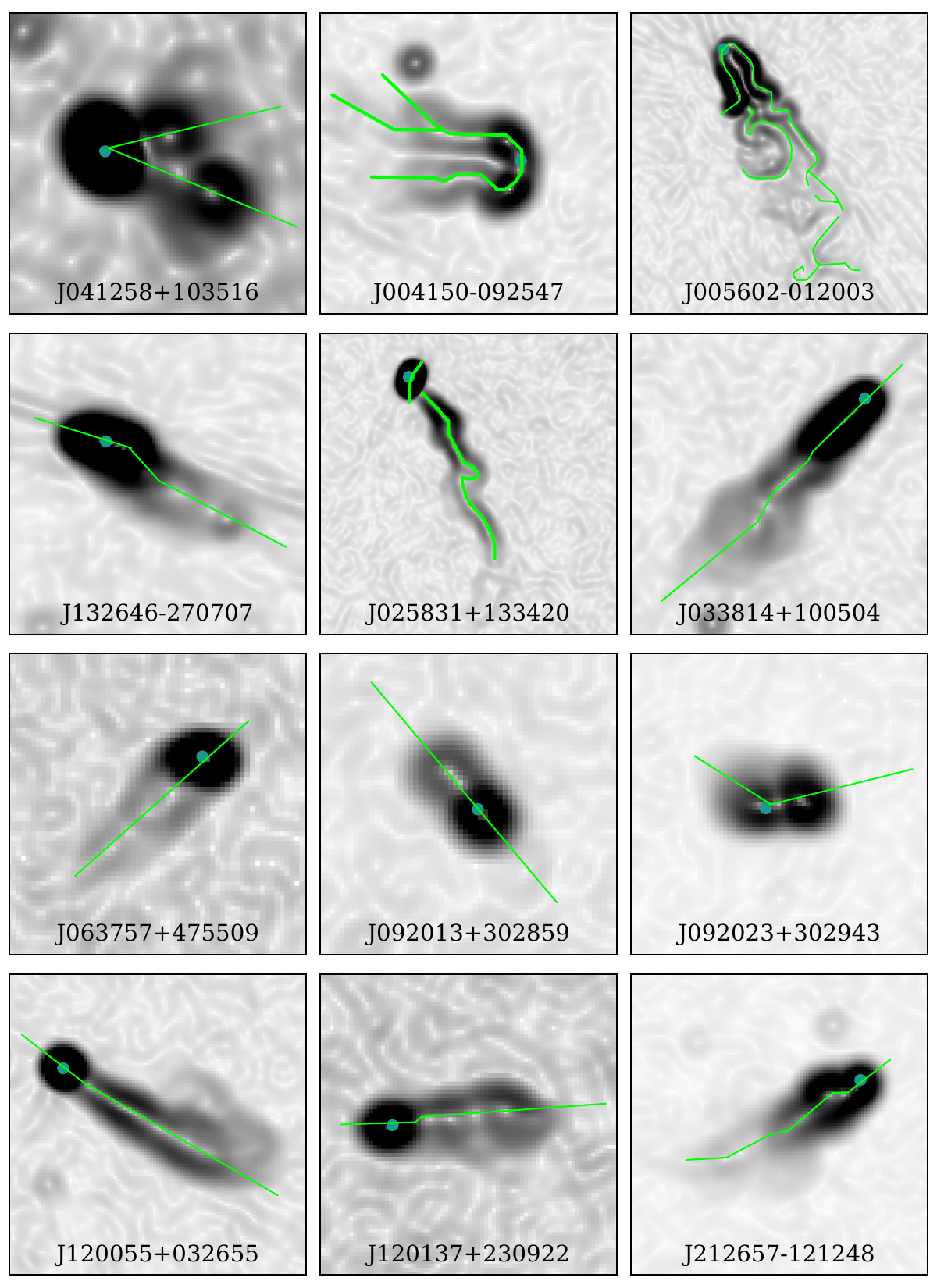}
    \caption{\textbf{continued.} Ridgelines for the remaining 12 polarized radio galaxies in our sample.}
\end{figure*}

\bibliography{bent_jets}{}

@ARTICLE{Burn1966,
       author = {{Burn}, B.~J.},
        title = "{On the depolarization of discrete radio sources by Faraday dispersion}",
      journal = {\mnras},
         year = "1966",
        month = "Jan",
       volume = {133},
        pages = {67},
          doi = {10.1093/mnras/133.1.67},
       adsurl = {https://ui.adsabs.harvard.edu/abs/1966MNRAS.133...67B}
}

@ARTICLE{Gaensler+04,
       author = {{Gaensler}, B.~M. and {Beck}, R. and {Feretti}, L.},
        title = "{The origin and evolution of cosmic magnetism}",
      journal = {New Astronomy Reviews},
         year = "2004",
        month = "Dec",
       volume = {48},
        pages = {1003-1012},
          doi = {10.1016/j.newar.2004.09.003},
archivePrefix = {arXiv},
       eprint = {astro-ph/0409100},
 primaryClass = {astro-ph},
       adsurl = {https://ui.adsabs.harvard.edu/\#abs/2004NewAR..48.1003G}
}

@ARTICLE{Sokoloff1998,
       author = {{Sokoloff}, D.~D. and {Bykov}, A.~A. and {Shukurov}, A. and
         {Berkhuijsen}, E.~M. and {Beck}, R. and {Poezd}, A.~D.},
        title = "{Depolarization and Faraday effects in galaxies}",
      journal = {\mnras},
         year = "1998",
        month = "Aug",
       volume = {299},
        pages = {189-206},
          doi = {10.1046/j.1365-8711.1998.01782.x},
       adsurl = {https://ui.adsabs.harvard.edu/abs/1998MNRAS.299..189S}
}

@MISC{RM-tools2020,
       author = {{Purcell}, C.~R. and {Van Eck}, C.~L. and {West}, J. and {Sun}, X.~H. and {Gaensler}, B.~M.},
        title = "{RM-Tools: Rotation measure (RM) synthesis and Stokes QU-fitting}",
 howpublished = {Astrophysics Source Code Library, record ascl:2005.003},
         year = 2020,
        month = may,
          eid = {ascl:2005.003},
        pages = {ascl:2005.003},
archivePrefix = {ascl},
       eprint = {2005.003},
       adsurl = {https://ui.adsabs.harvard.edu/abs/2020ascl.soft05003P}
}

@ARTICLE{Hotan+21,
       author = {{Hotan}, A.~W. and {Bunton}, J.~D. and {Chippendale}, A.~P. and {Whiting}, M. and {Tuthill}, J. and {Moss}, V.~A. and {McConnell}, D. and {Amy}, S.~W. and {Huynh}, M.~T. and {Allison}, J.~R. and {Anderson}, C.~S. and {Bannister}, K.~W. and {Bastholm}, E. and {Beresford}, R. and {Bock}, D.~C. -J. and {Bolton}, R. and {Chapman}, J.~M. and {Chow}, K. and {Collier}, J.~D. and {Cooray}, F.~R. and {Cornwell}, T.~J. and {Diamond}, P.~J. and {Edwards}, P.~G. and {Feain}, I.~J. and {Franzen}, T.~M.~O. and {George}, D. and {Gupta}, N. and {Hampson}, G.~A. and {Harvey-Smith}, L. and {Hayman}, D.~B. and {Heywood}, I. and {Jacka}, C. and {Jackson}, C.~A. and {Jackson}, S. and {Jeganathan}, K. and {Johnston}, S. and {Kesteven}, M. and {Kleiner}, D. and {Koribalski}, B.~S. and {Lee-Waddell}, K. and {Lenc}, E. and {Lensson}, E.~S. and {Mackay}, S. and {Mahony}, E.~K. and {McClure-Griffiths}, N.~M. and {McConigley}, R. and {Mirtschin}, P. and {Ng}, A.~K. and {Norris}, R.~P. and {Pearce}, S.~E. and {Phillips}, C. and {Pilawa}, M.~A. and {Raja}, W. and {Reynolds}, J.~E. and {Roberts}, P. and {Roxby}, D.~N. and {Sadler}, E.~M. and {Shields}, M. and {Schinckel}, A.~E.~T. and {Serra}, P. and {Shaw}, R.~D. and {Sweetnam}, T. and {Troup}, E.~R. and {Tzioumis}, A. and {Voronkov}, M.~A. and {Westmeier}, T.},
        title = "{Australian square kilometre array pathfinder: I. system description}",
      journal = {\pasa},
         year = 2021,
        month = mar,
       volume = {38},
          eid = {e009},
        pages = {e009},
          doi = {10.1017/pasa.2021.1},
archivePrefix = {arXiv},
       eprint = {2102.01870},
 primaryClass = {astro-ph.IM},
       adsurl = {https://ui.adsabs.harvard.edu/abs/2021PASA...38....9H}
}

@ARTICLE{Anderson+15,
       author = {{Anderson}, C.~S. and {Gaensler}, B.~M. and {Feain}, I.~J. and {Franzen}, T.~M.~O.},
        title = "{Broadband Radio Polarimetry and Faraday Rotation of 563 Extragalactic Radio Sources}",
      journal = {\apj},
         year = 2015,
        month = dec,
       volume = {815},
       number = {1},
          eid = {49},
        pages = {49},
          doi = {10.1088/0004-637X/815/1/49},
archivePrefix = {arXiv},
       eprint = {1511.04080},
 primaryClass = {astro-ph.GA},
       adsurl = {https://ui.adsabs.harvard.edu/abs/2015ApJ...815...49A}
}

@ARTICLE{Hutschenreuter+22,
       author = {{Hutschenreuter}, S. and {Anderson}, C.~S. and {Betti}, S. and {Bower}, G.~C. and {Brown}, J. -A. and {Br{\"u}ggen}, M. and {Carretti}, E. and {Clarke}, T. and {Clegg}, A. and {Costa}, A. and {Croft}, S. and {Van Eck}, C. and {Gaensler}, B.~M. and {de Gasperin}, F. and {Haverkorn}, M. and {Heald}, G. and {Hull}, C.~L.~H. and {Inoue}, M. and {Johnston-Hollitt}, M. and {Kaczmarek}, J. and {Law}, C. and {Ma}, Y.~K. and {MacMahon}, D. and {Mao}, S.~A. and {Riseley}, C. and {Roy}, S. and {Shanahan}, R. and {Shimwell}, T. and {Stil}, J. and {Sobey}, C. and {O'Sullivan}, S.~P. and {Tasse}, C. and {Vacca}, V. and {Vernstrom}, T. and {Williams}, P.~K.~G. and {Wright}, M. and {En{\ss}lin}, T.~A.},
        title = "{The Galactic Faraday rotation sky 2020}",
      journal = {\aap},
         year = 2022,
        month = jan,
       volume = {657},
          eid = {A43},
        pages = {A43},
          doi = {10.1051/0004-6361/202140486},
archivePrefix = {arXiv},
       eprint = {2102.01709},
 primaryClass = {astro-ph.GA},
       adsurl = {https://ui.adsabs.harvard.edu/abs/2022A&A...657A..43H}
}

@ARTICLE{Anderson+21,
       author = {{Anderson}, C.~S. and {Heald}, G.~H. and {Eilek}, J.~A. and {Lenc}, E. and {Gaensler}, B.~M. and {Rudnick}, Lawrence and {Van Eck}, C.~L. and {O'Sullivan}, S.~P. and {Stil}, J.~M. and {Chippendale}, A. and {Riseley}, C.~J. and {Carretti}, E. and {West}, J. and {Farnes}, J. and {Harvey-Smith}, L. and {McClure-Griffiths}, N.~M. and {Bock}, Douglas C.~J. and {Bunton}, J.~D. and {Koribalski}, B. and {Tremblay}, C.~D. and {Voronkov}, M.~A. and {Warhurst}, K.},
        title = "{Early Science from POSSUM: Shocks, turbulence, and a massive new reservoir of ionised gas in the Fornax cluster}",
      journal = {\pasa},
         year = 2021,
        month = apr,
       volume = {38},
          eid = {e020},
        pages = {e020},
          doi = {10.1017/pasa.2021.4},
archivePrefix = {arXiv},
       eprint = {2102.01702},
 primaryClass = {astro-ph.GA},
       adsurl = {https://ui.adsabs.harvard.edu/abs/2021PASA...38...20A}
}

@ARTICLE{Amaral+21,
       author = {{Amaral}, A.~D. and {Vernstrom}, T. and {Gaensler}, B.~M.},
        title = "{Constraints on large-scale magnetic fields in the intergalactic medium using cross-correlation methods}",
      journal = {\mnras},
         year = 2021,
        month = may,
       volume = {503},
       number = {2},
        pages = {2913-2926},
          doi = {10.1093/mnras/stab564},
archivePrefix = {arXiv},
       eprint = {2102.11312},
 primaryClass = {astro-ph.CO},
       adsurl = {https://ui.adsabs.harvard.edu/abs/2021MNRAS.503.2913A}
}

@ARTICLE{Vernstrom+19,
       author = {{Vernstrom}, T. and {Gaensler}, B.~M. and {Rudnick}, L. and {Andernach}, H.},
        title = "{Differences in Faraday Rotation between Adjacent Extragalactic Radio Sources as a Probe of Cosmic Magnetic Fields}",
      journal = {\apj},
         year = 2019,
        month = jun,
       volume = {878},
       number = {2},
          eid = {92},
        pages = {92},
          doi = {10.3847/1538-4357/ab1f83},
archivePrefix = {arXiv},
       eprint = {1905.02410},
 primaryClass = {astro-ph.CO},
       adsurl = {https://ui.adsabs.harvard.edu/abs/2019ApJ...878...92V}
}

@ARTICLE{Heald+20,
       author = {{Heald}, George and {Mao}, Sui and {Vacca}, Valentina and {Akahori}, Takuya and {Damas-Segovia}, Ancor and {Gaensler}, Bryan and {Hoeft}, Matthias and {Agudo}, Ivan and {Basu}, Aritra and {Beck}, Rainer and {Birkinshaw}, Mark and {Bonafede}, Annalisa and {Bourke}, Tyler and {Bracco}, Andrea and {Carretti}, Ettore and {Feretti}, Luigina and {Girart}, J. and {Govoni}, Federica and {Green}, James and {Han}, JinLin and {Haverkorn}, Marijke and {Horellou}, Cathy and {Johnston-Hollitt}, Melanie and {Kothes}, Roland and {Landecker}, Tom and {Nikiel-Wroczy{\'n}ski}, B{\l}a{\.z}ej and {O'Sullivan}, Shane and {Padovani}, Marco and {Poidevin}, Fr{\'e}d{\'e}rick and {Pratley}, Luke and {Regis}, Marco and {Riseley}, Christopher and {Robishaw}, Tim and {Rudnick}, Lawrence and {Sobey}, Charlotte and {Stil}, Jeroen and {Sun}, Xiaohui and {Sur}, Sharanya and {Taylor}, A. and {Thomson}, Alec and {Van Eck}, Cameron and {Vazza}, Franco and {West}, Jennifer and {SKA Magnetism Science Working Group}},
        title = "{Magnetism Science with the Square Kilometre Array}",
      journal = {Galaxies},
         year = 2020,
        month = jul,
       volume = {8},
       number = {3},
        pages = {53},
          doi = {10.3390/galaxies8030053},
archivePrefix = {arXiv},
       eprint = {2006.03172},
 primaryClass = {astro-ph.GA},
       adsurl = {https://ui.adsabs.harvard.edu/abs/2020Galax...8...53H}
}

@INPROCEEDINGS{Johnston-Hollitt+15,
       author = {{Johnston-Hollitt}, M. and {Govoni}, F. and {Beck}, R. and {Dehghan}, S. and {Pratley}, L. and {Akahori}, T. and {Heald}, G. and {Agudo}, I. and {Bonafede}, A. and {Carretti}, E. and {Clarke}, T. and {Colafrancesco}, S. and {Ensslin}, T.~A. and {Feretti}, L. and {Gaensler}, B. and {Haverkorn}, M. and {Mao}, S.~A. and {Oppermann}, N. and {Rudnick}, L. and {Scaife}, A. and {Schnitzeler}, D. and {Stil}, J. and {Taylor}, A.~R. and {Vacca}, V.},
        title = "{Using SKA Rotation Measures to Reveal the Mysteries of the Magnetised Universe}",
    booktitle = {Advancing Astrophysics with the Square Kilometre Array (AASKA14)},
         year = 2015,
        month = apr,
          eid = {92},
        pages = {92},
          doi = {10.22323/1.215.0092},
archivePrefix = {arXiv},
       eprint = {1506.00808},
 primaryClass = {astro-ph.CO},
       adsurl = {https://ui.adsabs.harvard.edu/abs/2015aska.confE..92J}
}

@ARTICLE{Mao+10,
       author = {{Mao}, S.~A. and {Gaensler}, B.~M. and {Haverkorn}, M. and {Zweibel}, E.~G. and {Madsen}, G.~J. and {McClure-Griffiths}, N.~M. and {Shukurov}, A. and {Kronberg}, P.~P.},
        title = "{A Survey of Extragalactic Faraday Rotation at High Galactic Latitude: The Vertical Magnetic Field of the Milky Way Toward the Galactic Poles}",
      journal = {\apj},
         year = 2010,
        month = may,
       volume = {714},
       number = {2},
        pages = {1170-1186},
          doi = {10.1088/0004-637X/714/2/1170},
archivePrefix = {arXiv},
       eprint = {1003.4519},
 primaryClass = {astro-ph.GA},
       adsurl = {https://ui.adsabs.harvard.edu/abs/2010ApJ...714.1170M}
}

@ARTICLE{VanEck+11,
       author = {{Van Eck}, C.~L. and {Brown}, J.~C. and {Stil}, J.~M. and {Rae}, K. and {Mao}, S.~A. and {Gaensler}, B.~M. and {Shukurov}, A. and {Taylor}, A.~R. and {Haverkorn}, M. and {Kronberg}, P.~P. and {McClure-Griffiths}, N.~M.},
        title = "{Modeling the Magnetic Field in the Galactic Disk Using New Rotation Measure Observations from the Very Large Array}",
      journal = {\apj},
         year = 2011,
        month = feb,
       volume = {728},
       number = {2},
          eid = {97},
        pages = {97},
          doi = {10.1088/0004-637X/728/2/97},
archivePrefix = {arXiv},
       eprint = {1012.2938},
 primaryClass = {astro-ph.GA},
       adsurl = {https://ui.adsabs.harvard.edu/abs/2011ApJ...728...97V}
}

@ARTICLE{Carretti+22,
       author = {{Carretti}, Ettore and {Vacca}, V. and {O'Sullivan}, S.~P. and {Heald}, G.~H. and {Horellou}, C. and {R{\"o}ttgering}, H.~J.~A. and {Scaife}, A.~M.~M. and {Shimwell}, T.~W. and {Shulevski}, A. and {Stuardi}, C. and {Vernstrom}, T.},
        title = "{Magnetic field strength in cosmic web filaments}",
      journal = {\mnras},
         year = 2022,
        month = may,
       volume = {512},
       number = {1},
        pages = {945-959},
          doi = {10.1093/mnras/stac384},
archivePrefix = {arXiv},
       eprint = {2202.04607},
 primaryClass = {astro-ph.CO},
       adsurl = {https://ui.adsabs.harvard.edu/abs/2022MNRAS.512..945C}
}

@ARTICLE{Feain+09,
       author = {{Feain}, I.~J. and {Ekers}, R.~D. and {Murphy}, T. and {Gaensler}, B.~M. and {Macquart}, J. -P. and {Norris}, R.~P. and {Cornwell}, T.~J. and {Johnston-Hollitt}, M. and {Ott}, J. and {Middelberg}, E.},
        title = "{Faraday Rotation Structure on Kiloparsec Scales in the Radio Lobes of Centaurus A}",
      journal = {\apj},
         year = 2009,
        month = dec,
       volume = {707},
       number = {1},
        pages = {114-125},
          doi = {10.1088/0004-637X/707/1/114},
archivePrefix = {arXiv},
       eprint = {0910.3458},
 primaryClass = {astro-ph.CO},
       adsurl = {https://ui.adsabs.harvard.edu/abs/2009ApJ...707..114F}
}

@ARTICLE{Bonafede+10,
       author = {{Bonafede}, A. and {Feretti}, L. and {Murgia}, M. and {Govoni}, F. and {Giovannini}, G. and {Dallacasa}, D. and {Dolag}, K. and {Taylor}, G.~B.},
        title = "{The Coma cluster magnetic field from Faraday rotation measures}",
      journal = {\aap},
         year = 2010,
        month = apr,
       volume = {513},
          eid = {A30},
        pages = {A30},
          doi = {10.1051/0004-6361/200913696},
archivePrefix = {arXiv},
       eprint = {1002.0594},
 primaryClass = {astro-ph.CO},
       adsurl = {https://ui.adsabs.harvard.edu/abs/2010A&A...513A..30B}
}

@ARTICLE{Rudnick&Owen2014,
       author = {{Rudnick}, L. and {Owen}, F.~N.},
        title = "{The Distribution of Polarized Radio Sources >15 {\ensuremath{\mu}}Jy in GOODS-N}",
      journal = {\apj},
         year = 2014,
        month = apr,
       volume = {785},
       number = {1},
          eid = {45},
        pages = {45},
          doi = {10.1088/0004-637X/785/1/45},
archivePrefix = {arXiv},
       eprint = {1402.3637},
 primaryClass = {astro-ph.GA},
       adsurl = {https://ui.adsabs.harvard.edu/abs/2014ApJ...785...45R}
}

@ARTICLE{Akahori&Ryu2010,
       author = {{Akahori}, Takuya and {Ryu}, Dongsu},
        title = "{Faraday Rotation Measure Due to the Intergalactic Magnetic Field}",
      journal = {\apj},
         year = 2010,
        month = nov,
       volume = {723},
       number = {1},
        pages = {476-481},
          doi = {10.1088/0004-637X/723/1/476},
archivePrefix = {arXiv},
       eprint = {1009.0570},
 primaryClass = {astro-ph.CO},
       adsurl = {https://ui.adsabs.harvard.edu/abs/2010ApJ...723..476A}
}

@ARTICLE{Tahani+18,
       author = {{Tahani}, M. and {Plume}, R. and {Brown}, J.~C. and {Kainulainen}, J.},
        title = "{Helical magnetic fields in molecular clouds?. A new method to determine the line-of-sight magnetic field structure in molecular clouds}",
      journal = {\aap},
         year = 2018,
        month = jun,
       volume = {614},
          eid = {A100},
        pages = {A100},
          doi = {10.1051/0004-6361/201732219},
archivePrefix = {arXiv},
       eprint = {1802.07831},
 primaryClass = {astro-ph.GA},
       adsurl = {https://ui.adsabs.harvard.edu/abs/2018A&A...614A.100T}
}

@ARTICLE{Sebokolodi+20,
       author = {{Sebokolodi}, M. Lerato L. and {Perley}, Richard and {Eilek}, Jean and {Carilli}, Chris and {Smirnov}, Oleg and {Laing}, Robert and {Greisen}, Eric W. and {Wise}, Michael},
        title = "{A Wideband Polarization Study of Cygnus A with the Jansky Very Large Array. I. The Observations and Data}",
      journal = {\apj},
         year = 2020,
        month = nov,
       volume = {903},
       number = {1},
          eid = {36},
        pages = {36},
          doi = {10.3847/1538-4357/abb80e},
archivePrefix = {arXiv},
       eprint = {2009.06554},
 primaryClass = {astro-ph.GA},
       adsurl = {https://ui.adsabs.harvard.edu/abs/2020ApJ...903...36S}
}

@ARTICLE{Osinga+22,
       author = {{Osinga}, E. and {van Weeren}, R.~J. and {Andrade-Santos}, F. and {Rudnick}, L. and {Bonafede}, A. and {Clarke}, T. and {Duncan}, K. and {Giacintucci}, S. and {Mroczkowski}, T. and {R{\"o}ttgering}, H.~J.~A.},
        title = "{The detection of cluster magnetic fields via radio source depolarisation}",
      journal = {\aap},
         year = 2022,
        month = sep,
       volume = {665},
          eid = {A71},
        pages = {A71},
          doi = {10.1051/0004-6361/202243526},
archivePrefix = {arXiv},
       eprint = {2207.09717},
 primaryClass = {astro-ph.CO},
       adsurl = {https://ui.adsabs.harvard.edu/abs/2022A&A...665A..71O}
}

@INPROCEEDINGS{CASA,
       author = {{McMullin}, J.~P. and {Waters}, B. and {Schiebel}, D. and {Young}, W. and {Golap}, K.},
        title = "{CASA Architecture and Applications}",
    booktitle = {Astronomical Data Analysis Software and Systems XVI},
         year = 2007,
       editor = {{Shaw}, R.~A. and {Hill}, F. and {Bell}, D.~J.},
       series = {Astronomical Society of the Pacific Conference Series},
       volume = {376},
        month = oct,
        pages = {127},
       adsurl = {https://ui.adsabs.harvard.edu/abs/2007ASPC..376..127M}
}

@ARTICLE{cal_flux_scale,
       author = {{Perley}, R.~A. and {Butler}, B.~J.},
        title = "{An Accurate Flux Density Scale from 50 MHz to 50 GHz}",
      journal = {\apjs},
         year = 2017,
        month = may,
       volume = {230},
       number = {1},
          eid = {7},
        pages = {7},
          doi = {10.3847/1538-4365/aa6df9},
archivePrefix = {arXiv},
       eprint = {1609.05940},
 primaryClass = {astro-ph.IM},
       adsurl = {https://ui.adsabs.harvard.edu/abs/2017ApJS..230....7P}
}

@ARTICLE{Jagannathan+17,
       author = {{Jagannathan}, P. and {Bhatnagar}, S. and {Rau}, U. and {Taylor}, A.~R.},
        title = "{Direction-dependent Corrections in Polarimetric Radio Imaging. I. Characterizing the Effects of the Primary Beam on Full-Stokes Imaging}",
      journal = {\aj},
         year = 2017,
        month = aug,
       volume = {154},
       number = {2},
          eid = {56},
        pages = {56},
          doi = {10.3847/1538-3881/aa77f8},
archivePrefix = {arXiv},
       eprint = {1706.01501},
 primaryClass = {astro-ph.IM},
       adsurl = {https://ui.adsabs.harvard.edu/abs/2017AJ....154...56J}
}

@ARTICLE{Legacy,
       author = {{Dey}, Arjun and {Schlegel}, David J. and {Lang}, Dustin and {Blum}, Robert and {Burleigh}, Kaylan and {Fan}, Xiaohui and {Findlay}, Joseph R. and {Finkbeiner}, Doug and {Herrera}, David and {Juneau}, St{\'e}phanie and {Landriau}, Martin and {Levi}, Michael and {McGreer}, Ian and {Meisner}, Aaron and {Myers}, Adam D. and {Moustakas}, John and {Nugent}, Peter and {Patej}, Anna and {Schlafly}, Edward F. and {Walker}, Alistair R. and {Valdes}, Francisco and {Weaver}, Benjamin A. and {Y{\`e}che}, Christophe and {Zou}, Hu and {Zhou}, Xu and {Abareshi}, Behzad and {Abbott}, T.~M.~C. and {Abolfathi}, Bela and {Aguilera}, C. and {Alam}, Shadab and {Allen}, Lori and {Alvarez}, A. and {Annis}, James and {Ansarinejad}, Behzad and {Aubert}, Marie and {Beechert}, Jacqueline and {Bell}, Eric F. and {BenZvi}, Segev Y. and {Beutler}, Florian and {Bielby}, Richard M. and {Bolton}, Adam S. and {Brice{\~n}o}, C{\'e}sar and {Buckley-Geer}, Elizabeth J. and {Butler}, Karen and {Calamida}, Annalisa and {Carlberg}, Raymond G. and {Carter}, Paul and {Casas}, Ricard and {Castander}, Francisco J. and {Choi}, Yumi and {Comparat}, Johan and {Cukanovaite}, Elena and {Delubac}, Timoth{\'e}e and {DeVries}, Kaitlin and {Dey}, Sharmila and {Dhungana}, Govinda and {Dickinson}, Mark and {Ding}, Zhejie and {Donaldson}, John B. and {Duan}, Yutong and {Duckworth}, Christopher J. and {Eftekharzadeh}, Sarah and {Eisenstein}, Daniel J. and {Etourneau}, Thomas and {Fagrelius}, Parker A. and {Farihi}, Jay and {Fitzpatrick}, Mike and {Font-Ribera}, Andreu and {Fulmer}, Leah and {G{\"a}nsicke}, Boris T. and {Gaztanaga}, Enrique and {George}, Koshy and {Gerdes}, David W. and {Gontcho}, Satya Gontcho A. and {Gorgoni}, Claudio and {Green}, Gregory and {Guy}, Julien and {Harmer}, Diane and {Hernandez}, M. and {Honscheid}, Klaus and {Huang}, Lijuan Wendy and {James}, David J. and {Jannuzi}, Buell T. and {Jiang}, Linhua and {Joyce}, Richard and {Karcher}, Armin and {Karkar}, Sonia and {Kehoe}, Robert and {Kneib}, Jean-Paul and {Kueter-Young}, Andrea and {Lan}, Ting-Wen and {Lauer}, Tod R. and {Le Guillou}, Laurent and {Le Van Suu}, Auguste and {Lee}, Jae Hyeon and {Lesser}, Michael and {Perreault Levasseur}, Laurence and {Li}, Ting S. and {Mann}, Justin L. and {Marshall}, Robert and {Mart{\'\i}nez-V{\'a}zquez}, C.~E. and {Martini}, Paul and {du Mas des Bourboux}, H{\'e}lion and {McManus}, Sean and {Meier}, Tobias Gabriel and {M{\'e}nard}, Brice and {Metcalfe}, Nigel and {Mu{\~n}oz-Guti{\'e}rrez}, Andrea and {Najita}, Joan and {Napier}, Kevin and {Narayan}, Gautham and {Newman}, Jeffrey A. and {Nie}, Jundan and {Nord}, Brian and {Norman}, Dara J. and {Olsen}, Knut A.~G. and {Paat}, Anthony and {Palanque-Delabrouille}, Nathalie and {Peng}, Xiyan and {Poppett}, Claire L. and {Poremba}, Megan R. and {Prakash}, Abhishek and {Rabinowitz}, David and {Raichoor}, Anand and {Rezaie}, Mehdi and {Robertson}, A.~N. and {Roe}, Natalie A. and {Ross}, Ashley J. and {Ross}, Nicholas P. and {Rudnick}, Gregory and {Safonova}, Sasha and {Saha}, Abhijit and {S{\'a}nchez}, F. Javier and {Savary}, Elodie and {Schweiker}, Heidi and {Scott}, Adam and {Seo}, Hee-Jong and {Shan}, Huanyuan and {Silva}, David R. and {Slepian}, Zachary and {Soto}, Christian and {Sprayberry}, David and {Staten}, Ryan and {Stillman}, Coley M. and {Stupak}, Robert J. and {Summers}, David L. and {Sien Tie}, Suk and {Tirado}, H. and {Vargas-Maga{\~n}a}, Mariana and {Vivas}, A. Katherina and {Wechsler}, Risa H. and {Williams}, Doug and {Yang}, Jinyi and {Yang}, Qian and {Yapici}, Tolga and {Zaritsky}, Dennis and {Zenteno}, A. and {Zhang}, Kai and {Zhang}, Tianmeng and {Zhou}, Rongpu and {Zhou}, Zhimin},
        title = "{Overview of the DESI Legacy Imaging Surveys}",
      journal = {\aj},
         year = 2019,
        month = may,
       volume = {157},
       number = {5},
          eid = {168},
        pages = {168},
          doi = {10.3847/1538-3881/ab089d},
archivePrefix = {arXiv},
       eprint = {1804.08657},
 primaryClass = {astro-ph.IM},
       adsurl = {https://ui.adsabs.harvard.edu/abs/2019AJ....157..168D}
}

@misc{PanSTARRS,
      title={The Pan-STARRS1 Surveys}, 
      author={K. C. Chambers and E. A. Magnier and N. Metcalfe and H. A. Flewelling and M. E. Huber and C. Z. Waters and L. Denneau and P. W. Draper and D. Farrow and D. P. Finkbeiner and C. Holmberg and J. Koppenhoefer and P. A. Price and A. Rest and R. P. Saglia and E. F. Schlafly and S. J. Smartt and W. Sweeney and R. J. Wainscoat and W. S. Burgett and S. Chastel and T. Grav and J. N. Heasley and K. W. Hodapp and R. Jedicke and N. Kaiser and R. -P. Kudritzki and G. A. Luppino and R. H. Lupton and D. G. Monet and J. S. Morgan and P. M. Onaka and B. Shiao and C. W. Stubbs and J. L. Tonry and R. White and E. Bañados and E. F. Bell and R. Bender and E. J. Bernard and M. Boegner and F. Boffi and M. T. Botticella and A. Calamida and S. Casertano and W. -P. Chen and X. Chen and S. Cole and N. Deacon and C. Frenk and A. Fitzsimmons and S. Gezari and V. Gibbs and C. Goessl and T. Goggia and R. Gourgue and B. Goldman and P. Grant and E. K. Grebel and N. C. Hambly and G. Hasinger and A. F. Heavens and T. M. Heckman and R. Henderson and T. Henning and M. Holman and U. Hopp and W. -H. Ip and S. Isani and M. Jackson and C. D. Keyes and A. M. Koekemoer and R. Kotak and D. Le and D. Liska and K. S. Long and J. R. Lucey and M. Liu and N. F. Martin and G. Masci and B. McLean and E. Mindel and P. Misra and E. Morganson and D. N. A. Murphy and A. Obaika and G. Narayan and M. A. Nieto-Santisteban and P. Norberg and J. A. Peacock and E. A. Pier and M. Postman and N. Primak and C. Rae and A. Rai and A. Riess and A. Riffeser and H. W. Rix and S. Röser and R. Russel and L. Rutz and E. Schilbach and A. S. B. Schultz and D. Scolnic and L. Strolger and A. Szalay and S. Seitz and E. Small and K. W. Smith and D. R. Soderblom and P. Taylor and R. Thomson and A. N. Taylor and A. R. Thakar and J. Thiel and D. Thilker and D. Unger and Y. Urata and J. Valenti and J. Wagner and T. Walder and F. Walter and S. P. Watters and S. Werner and W. M. Wood-Vasey and R. Wyse},
      year={2019},
      eprint={1612.05560},
      archivePrefix={arXiv},
      primaryClass={astro-ph.IM},
      url={https://arxiv.org/abs/1612.05560}, 
}

@ARTICLE{Golden-Marx+23,
       author = {{Golden-Marx}, Emmet and {Moravec}, E. and {Shen}, L. and {Cai}, Z. and {Blanton}, E.~L. and {Gendron-Marsolais}, M.~L. and {R{\"o}ttgering}, H.~J.~A. and {van Weeren}, R.~J. and {Buiten}, V. and {Grumitt}, R.~D.~P. and {Golden-Marx}, J. and {Pinjarkar}, S. and {Tang}, H.},
        title = "{The High-redshift Clusters Occupied by Bent Radio AGN (COBRA) Survey: Investigating the Role of Environment on Bent Radio AGNs Using LOFAR}",
      journal = {\apj},
         year = 2023,
        month = oct,
       volume = {956},
       number = {2},
          eid = {87},
        pages = {87},
          doi = {10.3847/1538-4357/acf46b},
archivePrefix = {arXiv},
       eprint = {2308.16238},
 primaryClass = {astro-ph.GA},
       adsurl = {https://ui.adsabs.harvard.edu/abs/2023ApJ...956...87G}
}

@ARTICLE{Ma+19,
       author = {{Ma}, Yik Ki and {Mao}, S.~A. and {Stil}, Jeroen and {Basu}, Aritra and {West}, Jennifer and {Heiles}, Carl and {Hill}, Alex S. and {Betti}, S.~K.},
        title = "{A broad-band spectro-polarimetric view of the NVSS rotation measure catalogue - I. Breaking the n{\ensuremath{\pi}}-ambiguity}",
      journal = {\mnras},
         year = 2019,
        month = aug,
       volume = {487},
       number = {3},
        pages = {3432-3453},
          doi = {10.1093/mnras/stz1325},
archivePrefix = {arXiv},
       eprint = {1905.04313},
 primaryClass = {astro-ph.GA},
       adsurl = {https://ui.adsabs.harvard.edu/abs/2019MNRAS.487.3432M}
}

@ARTICLE{Algeri+16,
       author = {{Algeri}, S. and {van Dyk}, D.~A. and {Conrad}, J. and {Anderson}, B.},
        title = "{On methods for correcting for the look-elsewhere effect in searches for new physics}",
      journal = {Journal of Instrumentation},
         year = 2016,
        month = dec,
       volume = {11},
       number = {12},
        pages = {P12010},
          doi = {10.1088/1748-0221/11/12/P12010},
archivePrefix = {arXiv},
       eprint = {1602.03765},
 primaryClass = {physics.data-an},
       adsurl = {https://ui.adsabs.harvard.edu/abs/2016JInst..11P2010A}
}

@ARTICLE{Guidetti+11,
       author = {{Guidetti}, D. and {Laing}, R.~A. and {Bridle}, A.~H. and {Parma}, P. and {Gregorini}, L.},
        title = "{Ordered magnetic fields around radio galaxies: evidence for interaction with the environment}",
      journal = {\mnras},
         year = 2011,
        month = jun,
       volume = {413},
       number = {4},
        pages = {2525-2544},
          doi = {10.1111/j.1365-2966.2011.18321.x},
archivePrefix = {arXiv},
       eprint = {1101.1807},
 primaryClass = {astro-ph.CO},
       adsurl = {https://ui.adsabs.harvard.edu/abs/2011MNRAS.413.2525G}
}

@ARTICLE{Osinga+25,
       author = {{Osinga}, E. and {van Weeren}, R.~J. and {Rudnick}, L. and {Andrade-Santos}, F. and {Bonafede}, A. and {Clarke}, T. and {Duncan}, K. and {Giacintucci}, S. and {R{\"o}ttgering}, H.~J.~A.},
        title = "{Probing cluster magnetism with embedded and background radio sources in Planck clusters}",
      journal = {\aap},
         year = 2025,
        month = feb,
       volume = {694},
          eid = {A44},
        pages = {A44},
          doi = {10.1051/0004-6361/202451885},
archivePrefix = {arXiv},
       eprint = {2408.07178},
 primaryClass = {astro-ph.CO},
       adsurl = {https://ui.adsabs.harvard.edu/abs/2025A&A...694A..44O}
}

@ARTICLE{Mingo+19,
       author = {{Mingo}, B. and {Croston}, J.~H. and {Hardcastle}, M.~J. and {Best}, P.~N. and {Duncan}, K.~J. and {Morganti}, R. and {Rottgering}, H.~J.~A. and {Sabater}, J. and {Shimwell}, T.~W. and {Williams}, W.~L. and {Brienza}, M. and {Gurkan}, G. and {Mahatma}, V.~H. and {Morabito}, L.~K. and {Prandoni}, I. and {Bondi}, M. and {Ineson}, J. and {Mooney}, S.},
        title = "{Revisiting the Fanaroff-Riley dichotomy and radio-galaxy morphology with the LOFAR Two-Metre Sky Survey (LoTSS)}",
      journal = {\mnras},
         year = 2019,
        month = sep,
       volume = {488},
       number = {2},
        pages = {2701-2721},
          doi = {10.1093/mnras/stz1901},
archivePrefix = {arXiv},
       eprint = {1907.03726},
 primaryClass = {astro-ph.GA},
       adsurl = {https://ui.adsabs.harvard.edu/abs/2019MNRAS.488.2701M}
}

@ARTICLE{Barkus+22,
       author = {{Barkus}, B. and {Croston}, J.~H. and {Piotrowska}, J. and {Mingo}, B. and {Best}, P.~N. and {Hardcastle}, M.~J. and {Mostert}, R.~I.~J. and {R{\"o}ttgering}, H.~J.~A. and {Sabater}, J. and {Webster}, B. and {Williams}, W.~L.},
        title = "{The application of ridgelines in extended radio source cross-identification}",
      journal = {\mnras},
         year = 2022,
        month = jan,
       volume = {509},
       number = {1},
        pages = {1-15},
          doi = {10.1093/mnras/stab2952},
archivePrefix = {arXiv},
       eprint = {2110.05254},
 primaryClass = {astro-ph.GA},
       adsurl = {https://ui.adsabs.harvard.edu/abs/2022MNRAS.509....1B}
}

@ARTICLE{Wezgowiec+24,
       author = {{We{\.z}gowiec}, M. and {Jamrozy}, M. and {Chy{\.z}y}, K.~T. and {Hardcastle}, M.~J. and {Ku{\'z}micz}, A. and {Heald}, G. and {Shimwell}, T.~W.},
        title = "{The twisted jets and magnetic fields of the extended radio galaxy 4C 70.19}",
      journal = {\aap},
         year = 2024,
        month = nov,
       volume = {691},
          eid = {A193},
        pages = {A193},
          doi = {10.1051/0004-6361/202451580},
archivePrefix = {arXiv},
       eprint = {2411.02121},
 primaryClass = {astro-ph.GA},
       adsurl = {https://ui.adsabs.harvard.edu/abs/2024A&A...691A.193W}
}

@ARTICLE{Vega-Garcia+19,
       author = {{Vega-Garc{\'\i}a}, L. and {Perucho}, M. and {Lobanov}, A.~P.},
        title = "{Derivation of the physical parameters of the jet in S5 0836+710 from stability analysis}",
      journal = {\aap},
         year = 2019,
        month = jul,
       volume = {627},
          eid = {A79},
        pages = {A79},
          doi = {10.1051/0004-6361/201935119},
archivePrefix = {arXiv},
       eprint = {1904.02030},
 primaryClass = {astro-ph.HE},
       adsurl = {https://ui.adsabs.harvard.edu/abs/2019A&A...627A..79V}
}

@ARTICLE{Perucho+12,
       author = {{Perucho}, M. and {Kovalev}, Y.~Y. and {Lobanov}, A.~P. and {Hardee}, P.~E. and {Agudo}, I.},
        title = "{Anatomy of Helical Extragalactic Jets: The Case of S5 0836+710}",
      journal = {\apj},
         year = 2012,
        month = apr,
       volume = {749},
       number = {1},
          eid = {55},
        pages = {55},
          doi = {10.1088/0004-637X/749/1/55},
archivePrefix = {arXiv},
       eprint = {1202.1182},
 primaryClass = {astro-ph.CO},
       adsurl = {https://ui.adsabs.harvard.edu/abs/2012ApJ...749...55P}
}

@ARTICLE{FilFinder,
   author = {{Koch}, E.~W. and {Rosolowsky}, E.~W.},
    title = "{Filament identification through mathematical morphology}",
  journal = {\mnras},
archivePrefix = "arXiv",
   eprint = {1507.02289},
     year = 2015,
    month = oct,
   volume = 452,
    pages = {3435-3450},
      doi = {10.1093/mnras/stv1521},
   adsurl = {http://adsabs.harvard.edu/abs/2015MNRAS.452.3435K}
}

@ARTICLE{Subramanian+06,
       author = {{Subramanian}, Kandaswamy and {Shukurov}, Anvar and {Haugen}, Nils Erland L.},
        title = "{Evolving turbulence and magnetic fields in galaxy clusters}",
      journal = {\mnras},
         year = 2006,
        month = mar,
       volume = {366},
       number = {4},
        pages = {1437-1454},
          doi = {10.1111/j.1365-2966.2006.09918.x},
archivePrefix = {arXiv},
       eprint = {astro-ph/0505144},
 primaryClass = {astro-ph},
       adsurl = {https://ui.adsabs.harvard.edu/abs/2006MNRAS.366.1437S}
}

@ARTICLE{Dolag+08,
       author = {{Dolag}, K. and {Bykov}, A.~M. and {Diaferio}, A.},
        title = "{Non-Thermal Processes in Cosmological Simulations}",
      journal = {\ssr},
         year = 2008,
        month = feb,
       volume = {134},
       number = {1-4},
        pages = {311-335},
          doi = {10.1007/s11214-008-9319-2},
archivePrefix = {arXiv},
       eprint = {0801.1048},
 primaryClass = {astro-ph},
       adsurl = {https://ui.adsabs.harvard.edu/abs/2008SSRv..134..311D}
}

@ARTICLE{Vazza+12,
       author = {{Vazza}, F. and {Roediger}, E. and {Br{\"u}ggen}, M.},
        title = "{Turbulence in the ICM from mergers, cool-core sloshing, and jets: results from a new multi-scale filtering approach}",
      journal = {\aap},
         year = 2012,
        month = aug,
       volume = {544},
          eid = {A103},
        pages = {A103},
          doi = {10.1051/0004-6361/201118688},
archivePrefix = {arXiv},
       eprint = {1202.5882},
 primaryClass = {astro-ph.CO},
       adsurl = {https://ui.adsabs.harvard.edu/abs/2012A&A...544A.103V}
}

@ARTICLE{Gabuzda+18,
       author = {{Gabuzda}, Denise},
        title = "{Evidence for Helical Magnetic Fields Associated with AGN Jets and the Action of a Cosmic Battery}",
      journal = {Galaxies},
         year = 2018,
        month = dec,
       volume = {7},
       number = {1},
          eid = {5},
        pages = {5},
          doi = {10.3390/galaxies7010005},
       adsurl = {https://ui.adsabs.harvard.edu/abs/2018Galax...7....5G}
}

@ARTICLE{ERASS1,
       author = {{Bulbul}, E. and {Liu}, A. and {Kluge}, M. and {Zhang}, X. and {Sanders}, J.~S. and {Bahar}, Y.~E. and {Ghirardini}, V. and {Artis}, E. and {Seppi}, R. and {Garrel}, C. and {Ramos-Ceja}, M.~E. and {Comparat}, J. and {Balzer}, F. and {B{\"o}ckmann}, K. and {Br{\"u}ggen}, M. and {Clerc}, N. and {Dennerl}, K. and {Dolag}, K. and {Freyberg}, M. and {Grandis}, S. and {Gruen}, D. and {Kleinebreil}, F. and {Krippendorf}, S. and {Lamer}, G. and {Merloni}, A. and {Migkas}, K. and {Nandra}, K. and {Pacaud}, F. and {Predehl}, P. and {Reiprich}, T.~H. and {Schrabback}, T. and {Veronica}, A. and {Weller}, J. and {Zelmer}, S.},
        title = "{The SRG/eROSITA All-Sky Survey. The first catalog of galaxy clusters and groups in the Western Galactic Hemisphere}",
      journal = {\aap},
         year = 2024,
        month = may,
       volume = {685},
          eid = {A106},
        pages = {A106},
          doi = {10.1051/0004-6361/202348264},
archivePrefix = {arXiv},
       eprint = {2402.08452},
 primaryClass = {astro-ph.CO},
       adsurl = {https://ui.adsabs.harvard.edu/abs/2024A&A...685A.106B}
}

@ARTICLE{Fanaroff&Riley,
       author = {{Fanaroff}, B.~L. and {Riley}, J.~M.},
        title = "{The morphology of extragalactic radio sources of high and low luminosity}",
      journal = {\mnras},
         year = 1974,
        month = may,
       volume = {167},
        pages = {31P-36P},
          doi = {10.1093/mnras/167.1.31P},
       adsurl = {https://ui.adsabs.harvard.edu/abs/1974MNRAS.167P..31F}
}

@ARTICLE{Bera+24,
       author = {{Bera}, Soumen Kumar and {Sasmal}, Tapan K. and {Mondal}, Soumen and {Fang}, Taotao and {Chen}, Xuelei},
        title = "{Strange and Odd Morphology Extragalactic Radio Sources (STROMERSs): A Faint Images of the Radio Sky at Twenty-Centimeters (FIRST) Look at the Strange and Odd Radio Sources}",
      journal = {Universe},
         year = 2024,
        month = aug,
       volume = {10},
       number = {9},
          eid = {347},
        pages = {347},
          doi = {10.3390/universe10090347},
       adsurl = {https://ui.adsabs.harvard.edu/abs/2024Univ...10..347B}
}

@ARTICLE{Rudnick&Owen1976,
       author = {{Rudnick}, L. and {Owen}, F.~N.},
        title = "{Head-tail radio sources in clusters of galaxies.}",
      journal = {\apjl},
         year = 1976,
        month = feb,
       volume = {203},
        pages = {L107-L111},
          doi = {10.1086/182030},
       adsurl = {https://ui.adsabs.harvard.edu/abs/1976ApJ...203L.107R}
}

@ARTICLE{Owen&Rudnick1979,
       author = {{Owen}, F.~N. and {Burns}, J.~O. and {Rudnick}, L. and {Greisen}, E.~W.},
        title = "{VLA observations of head-tail radio sources.}",
      journal = {\apjl},
         year = 1979,
        month = apr,
       volume = {229},
        pages = {L59-L63},
          doi = {10.1086/182930},
       adsurl = {https://ui.adsabs.harvard.edu/abs/1979ApJ...229L..59O}
}

@ARTICLE{Blanton+00,
       author = {{Blanton}, E.~L. and {Gregg}, M.~D. and {Helfand}, D.~J. and {Becker}, R.~H. and {White}, R.~L.},
        title = "{FIRST Bent-Double Radio Sources: Tracers of High-Redshift Clusters}",
      journal = {\apj},
         year = 2000,
        month = mar,
       volume = {531},
       number = {1},
        pages = {118-136},
          doi = {10.1086/308428},
archivePrefix = {arXiv},
       eprint = {astro-ph/9910099},
 primaryClass = {astro-ph},
       adsurl = {https://ui.adsabs.harvard.edu/abs/2000ApJ...531..118B}
}

@ARTICLE{Dehghan+14,
       author = {{Dehghan}, S. and {Johnston-Hollitt}, M. and {Franzen}, T.~M.~O. and {Norris}, R.~P. and {Miller}, N.~A.},
        title = "{Bent-tailed Radio Sources in the Australia Telescope Large Area Survey of the Chandra Deep Field South}",
      journal = {\aj},
         year = 2014,
        month = nov,
       volume = {148},
       number = {5},
          eid = {75},
        pages = {75},
          doi = {10.1088/0004-6256/148/5/75},
archivePrefix = {arXiv},
       eprint = {1405.6484},
 primaryClass = {astro-ph.GA},
       adsurl = {https://ui.adsabs.harvard.edu/abs/2014AJ....148...75D}
}

@ARTICLE{Sasmal+22,
       author = {{Sasmal}, Tapan K. and {Bera}, Soumen and {Mondal}, Soumen},
        title = "{Miscellaneous radio galaxies from LOFAR survey}",
      journal = {Astronomische Nachrichten},
         year = 2022,
        month = oct,
       volume = {343},
       number = {8},
          eid = {e20210083},
        pages = {e20210083},
          doi = {10.1002/asna.20210083},
       adsurl = {https://ui.adsabs.harvard.edu/abs/2022AN....34310083S}
}

@ARTICLE{Pal+23,
       author = {{Pal}, Sabyasachi and {Kumari}, Shobha},
        title = "{A new catalog of head{\textendash}tail radio galaxies from LoTSS DR1}",
      journal = {Journal of Astrophysics and Astronomy},
         year = 2023,
        month = jun,
       volume = {44},
       number = {1},
          eid = {17},
        pages = {17},
          doi = {10.1007/s12036-022-09892-x},
archivePrefix = {arXiv},
       eprint = {2103.15153},
 primaryClass = {astro-ph.GA},
       adsurl = {https://ui.adsabs.harvard.edu/abs/2023JApA...44...17P}
}

@ARTICLE{Missaglia+19,
       author = {{Missaglia}, V. and {Massaro}, F. and {Capetti}, A. and {Paolillo}, M. and {Kraft}, R.~P. and {Baldi}, R.~D. and {Paggi}, A.},
        title = "{WATCAT: a tale of wide-angle tailed radio galaxies}",
      journal = {\aap},
         year = 2019,
        month = jun,
       volume = {626},
          eid = {A8},
        pages = {A8},
          doi = {10.1051/0004-6361/201935058},
archivePrefix = {arXiv},
       eprint = {1904.02175},
 primaryClass = {astro-ph.HE},
       adsurl = {https://ui.adsabs.harvard.edu/abs/2019A&A...626A...8M}
}

@ARTICLE{Bera+20,
       author = {{Bera}, Soumen and {Pal}, Sabyasachi and {Sasmal}, Tapan K. and {Mondal}, Soumen},
        title = "{FIRST Winged Radio Galaxies with X and Z Symmetry}",
      journal = {\apjs},
         year = 2020,
        month = nov,
       volume = {251},
       number = {1},
          eid = {9},
        pages = {9},
          doi = {10.3847/1538-4365/abb367},
archivePrefix = {arXiv},
       eprint = {2011.03839},
 primaryClass = {astro-ph.GA},
       adsurl = {https://ui.adsabs.harvard.edu/abs/2020ApJS..251....9B}
}

@ARTICLE{Bera+22,
       author = {{Bera}, Soumen and {Sasmal}, Tapan K. and {Patra}, Dusmanta and {Mondal}, Soumen},
        title = "{``Winged'' Radio Sources from the LOFAR Two-meter Sky Survey First Data Release (LoTSS DR1)}",
      journal = {\apjs},
         year = 2022,
        month = may,
       volume = {260},
       number = {1},
          eid = {7},
        pages = {7},
          doi = {10.3847/1538-4365/ac5cc4},
       adsurl = {https://ui.adsabs.harvard.edu/abs/2022ApJS..260....7B}
}

@ARTICLE{Yang+19,
       author = {{Yang}, Xiaolong and {Joshi}, Ravi and {Gopal-Krishna} and {An}, Tao and {Ho}, Luis C. and {Wiita}, Paul J. and {Liu}, Xiang and {Yang}, Jun and {Wang}, Ran and {Wu}, Xue-Bing and {Yang}, Xiaofeng},
        title = "{Extended Catalog of Winged or X-shaped Radio Sources from the FIRST Survey}",
      journal = {\apjs},
         year = 2019,
        month = nov,
       volume = {245},
       number = {1},
          eid = {17},
        pages = {17},
          doi = {10.3847/1538-4365/ab4811},
archivePrefix = {arXiv},
       eprint = {1905.06356},
 primaryClass = {astro-ph.HE},
       adsurl = {https://ui.adsabs.harvard.edu/abs/2019ApJS..245...17Y}
}

@ARTICLE{Mahatma+19,
       author = {{Mahatma}, V.~H. and {Hardcastle}, M.~J. and {Williams}, W.~L. and {Best}, P.~N. and {Croston}, J.~H. and {Duncan}, K. and {Mingo}, B. and {Morganti}, R. and {Brienza}, M. and {Cochrane}, R.~K. and {G{\"u}rkan}, G. and {Harwood}, J.~J. and {Jarvis}, M.~J. and {Jamrozy}, M. and {Jurlin}, N. and {Morabito}, L.~K. and {R{\"o}ttgering}, H.~J.~A. and {Sabater}, J. and {Shimwell}, T.~W. and {Smith}, D.~J.~B. and {Shulevski}, A. and {Tasse}, C.},
        title = "{LoTSS DR1: Double-double radio galaxies in the HETDEX field}",
      journal = {\aap},
         year = 2019,
        month = feb,
       volume = {622},
          eid = {A13},
        pages = {A13},
          doi = {10.1051/0004-6361/201833973},
archivePrefix = {arXiv},
       eprint = {1811.08194},
 primaryClass = {astro-ph.GA},
       adsurl = {https://ui.adsabs.harvard.edu/abs/2019A&A...622A..13M}
}

@ARTICLE{Nandi+12,
       author = {{Nandi}, S. and {Saikia}, D.~J.},
        title = "{Double-double radio galaxies from the FIRST survey}",
      journal = {Bulletin of the Astronomical Society of India},
         year = 2012,
        month = jun,
       volume = {40},
       number = {2},
        pages = {121-137},
          doi = {10.48550/arXiv.1208.1941},
archivePrefix = {arXiv},
       eprint = {1208.1941},
 primaryClass = {astro-ph.CO},
       adsurl = {https://ui.adsabs.harvard.edu/abs/2012BASI...40..121N}
}

@ARTICLE{Begelman+79,
       author = {{Begelman}, M.~C. and {Rees}, M.~J. and {Blandford}, R.~D.},
        title = "{A twin-jet model for radio trails}",
      journal = {\nat},
         year = 1979,
        month = jun,
       volume = {279},
        pages = {770-773},
          doi = {10.1038/279770a0},
       adsurl = {https://ui.adsabs.harvard.edu/abs/1979Natur.279..770B}
}

@ARTICLE{Jones&Owen1979,
       author = {{Jones}, T.~W. and {Owen}, F.~N.},
        title = "{Hot gas in elliptical galaxies and the formation of head-tail radio sources}",
      journal = {\apj},
         year = 1979,
        month = dec,
       volume = {234},
        pages = {818-824},
          doi = {10.1086/157561},
       adsurl = {https://ui.adsabs.harvard.edu/abs/1979ApJ...234..818J}
}

@ARTICLE{Freeland+08,
       author = {{Freeland}, E. and {Cardoso}, R.~F. and {Wilcots}, E.},
        title = "{Bent-Double Radio Sources as Probes of Intergalactic Gas}",
      journal = {\apj},
         year = 2008,
        month = oct,
       volume = {685},
       number = {2},
        pages = {858-862},
          doi = {10.1086/591443},
archivePrefix = {arXiv},
       eprint = {0806.3971},
 primaryClass = {astro-ph},
       adsurl = {https://ui.adsabs.harvard.edu/abs/2008ApJ...685..858F}
}

@ARTICLE{Vallee+81,
       author = {{Vallee}, J.~P. and {Bridle}, A.~H. and {Wilson}, A.~S.},
        title = "{Orbital motion of the head-tail radio galaxy IC 708.}",
      journal = {\apj},
         year = 1981,
        month = nov,
       volume = {250},
        pages = {66-78},
          doi = {10.1086/159348},
       adsurl = {https://ui.adsabs.harvard.edu/abs/1981ApJ...250...66V}
}

@ARTICLE{Garon+19,
       author = {{Garon}, Avery F. and {Rudnick}, Lawrence and {Wong}, O. Ivy and {Jones}, Tom W. and {Kim}, Jin-Ah and {Andernach}, Heinz and {Shabala}, Stanislav S. and {Kapi{\'n}ska}, Anna D. and {Norris}, Ray P. and {de Gasperin}, Francesco and {Tate}, Jean and {Tang}, Hongming},
        title = "{Radio Galaxy Zoo: The Distortion of Radio Galaxies by Galaxy Clusters}",
      journal = {\aj},
         year = 2019,
        month = mar,
       volume = {157},
       number = {3},
          eid = {126},
        pages = {126},
          doi = {10.3847/1538-3881/aaff62},
archivePrefix = {arXiv},
       eprint = {1901.05480},
 primaryClass = {astro-ph.GA},
       adsurl = {https://ui.adsabs.harvard.edu/abs/2019AJ....157..126G}
}

@ARTICLE{Banfield+16,
       author = {{Banfield}, J.~K. and {Andernach}, H. and {Kapi{\'n}ska}, A.~D. and {Rudnick}, L. and {Hardcastle}, M.~J. and {Cotter}, G. and {Vaughan}, S. and {Jones}, T.~W. and {Heywood}, I. and {Wing}, J.~D. and {Wong}, O.~I. and {Matorny}, T. and {Terentev}, I.~A. and {L{\'o}pez-S{\'a}nchez}, {\'A}. R. and {Norris}, R.~P. and {Seymour}, N. and {Shabala}, S.~S. and {Willett}, K.~W.},
        title = "{Radio Galaxy Zoo: discovery of a poor cluster through a giant wide-angle tail radio galaxy}",
      journal = {\mnras},
         year = 2016,
        month = aug,
       volume = {460},
       number = {3},
        pages = {2376-2384},
          doi = {10.1093/mnras/stw1067},
archivePrefix = {arXiv},
       eprint = {1606.05016},
 primaryClass = {astro-ph.GA},
       adsurl = {https://ui.adsabs.harvard.edu/abs/2016MNRAS.460.2376B}
}

@ARTICLE{Golden-Marx+21,
       author = {{Golden-Marx}, Emmet and {Blanton}, E.~L. and {Paterno-Mahler}, R. and {Brodwin}, M. and {Ashby}, M.~L.~N. and {Moravec}, E. and {Shen}, L. and {Lemaux}, B.~C. and {Lubin}, L.~M. and {Gal}, R.~R. and {Tomczak}, A.~R.},
        title = "{The High-redshift Clusters Occupied by Bent Radio AGN (COBRA) Survey: Radio Source Properties}",
      journal = {\apj},
         year = 2021,
        month = feb,
       volume = {907},
       number = {2},
          eid = {65},
        pages = {65},
          doi = {10.3847/1538-4357/abcd96},
archivePrefix = {arXiv},
       eprint = {2011.12313},
 primaryClass = {astro-ph.GA},
       adsurl = {https://ui.adsabs.harvard.edu/abs/2021ApJ...907...65G}
}

@ARTICLE{Paterno-Mahler+17,
       author = {{Paterno-Mahler}, R. and {Blanton}, E.~L. and {Brodwin}, M. and {Ashby}, M.~L.~N. and {Golden-Marx}, E. and {Decker}, B. and {Wing}, J.~D. and {Anand}, G.},
        title = "{The High-redshift Clusters Occupied by Bent Radio AGN (COBRA) Survey: The Spitzer  Catalog}",
      journal = {\apj},
         year = 2017,
        month = jul,
       volume = {844},
       number = {1},
          eid = {78},
        pages = {78},
          doi = {10.3847/1538-4357/aa7b89},
archivePrefix = {arXiv},
       eprint = {1611.00746},
 primaryClass = {astro-ph.GA},
       adsurl = {https://ui.adsabs.harvard.edu/abs/2017ApJ...844...78P}
}

@ARTICLE{Sunyaev+21,
       author = {{Sunyaev}, R. and {Arefiev}, V. and {Babyshkin}, V. and {Bogomolov}, A. and {Borisov}, K. and {Buntov}, M. and {Brunner}, H. and {Burenin}, R. and {Churazov}, E. and {Coutinho}, D. and {Eder}, J. and {Eismont}, N. and {Freyberg}, M. and {Gilfanov}, M. and {Gureyev}, P. and {Hasinger}, G. and {Khabibullin}, I. and {Kolmykov}, V. and {Komovkin}, S. and {Krivonos}, R. and {Lapshov}, I. and {Levin}, V. and {Lomakin}, I. and {Lutovinov}, A. and {Medvedev}, P. and {Merloni}, A. and {Mernik}, T. and {Mikhailov}, E. and {Molodtsov}, V. and {Mzhelsky}, P. and {M{\"u}ller}, S. and {Nandra}, K. and {Nazarov}, V. and {Pavlinsky}, M. and {Poghodin}, A. and {Predehl}, P. and {Robrade}, J. and {Sazonov}, S. and {Scheuerle}, H. and {Shirshakov}, A. and {Tkachenko}, A. and {Voron}, V.},
        title = "{SRG X-ray orbital observatory. Its telescopes and first scientific results}",
      journal = {\aap},
         year = 2021,
        month = dec,
       volume = {656},
          eid = {A132},
        pages = {A132},
          doi = {10.1051/0004-6361/202141179},
archivePrefix = {arXiv},
       eprint = {2104.13267},
 primaryClass = {astro-ph.HE},
       adsurl = {https://ui.adsabs.harvard.edu/abs/2021A&A...656A.132S}
}

@ARTICLE{Predehl+21,
       author = {{Predehl}, P. and {Andritschke}, R. and {Arefiev}, V. and {Babyshkin}, V. and {Batanov}, O. and {Becker}, W. and {B{\"o}hringer}, H. and {Bogomolov}, A. and {Boller}, T. and {Borm}, K. and {Bornemann}, W. and {Br{\"a}uninger}, H. and {Br{\"u}ggen}, M. and {Brunner}, H. and {Brusa}, M. and {Bulbul}, E. and {Buntov}, M. and {Burwitz}, V. and {Burkert}, W. and {Clerc}, N. and {Churazov}, E. and {Coutinho}, D. and {Dauser}, T. and {Dennerl}, K. and {Doroshenko}, V. and {Eder}, J. and {Emberger}, V. and {Eraerds}, T. and {Finoguenov}, A. and {Freyberg}, M. and {Friedrich}, P. and {Friedrich}, S. and {F{\"u}rmetz}, M. and {Georgakakis}, A. and {Gilfanov}, M. and {Granato}, S. and {Grossberger}, C. and {Gueguen}, A. and {Gureev}, P. and {Haberl}, F. and {H{\"a}lker}, O. and {Hartner}, G. and {Hasinger}, G. and {Huber}, H. and {Ji}, L. and {Kienlin}, A. v. and {Kink}, W. and {Korotkov}, F. and {Kreykenbohm}, I. and {Lamer}, G. and {Lomakin}, I. and {Lapshov}, I. and {Liu}, T. and {Maitra}, C. and {Meidinger}, N. and {Menz}, B. and {Merloni}, A. and {Mernik}, T. and {Mican}, B. and {Mohr}, J. and {M{\"u}ller}, S. and {Nandra}, K. and {Nazarov}, V. and {Pacaud}, F. and {Pavlinsky}, M. and {Perinati}, E. and {Pfeffermann}, E. and {Pietschner}, D. and {Ramos-Ceja}, M.~E. and {Rau}, A. and {Reiffers}, J. and {Reiprich}, T.~H. and {Robrade}, J. and {Salvato}, M. and {Sanders}, J. and {Santangelo}, A. and {Sasaki}, M. and {Scheuerle}, H. and {Schmid}, C. and {Schmitt}, J. and {Schwope}, A. and {Shirshakov}, A. and {Steinmetz}, M. and {Stewart}, I. and {Str{\"u}der}, L. and {Sunyaev}, R. and {Tenzer}, C. and {Tiedemann}, L. and {Tr{\"u}mper}, J. and {Voron}, V. and {Weber}, P. and {Wilms}, J. and {Yaroshenko}, V.},
        title = "{The eROSITA X-ray telescope on SRG}",
      journal = {\aap},
         year = 2021,
        month = mar,
       volume = {647},
          eid = {A1},
        pages = {A1},
          doi = {10.1051/0004-6361/202039313},
archivePrefix = {arXiv},
       eprint = {2010.03477},
 primaryClass = {astro-ph.HE},
       adsurl = {https://ui.adsabs.harvard.edu/abs/2021A&A...647A...1P}
}

@ARTICLE{Loi+25,
       author = {{Loi}, Francesca and {Serra}, Paolo and {Murgia}, Matteo and {Govoni}, Federica and {Vacca}, Valentina and {Maccagni}, Filippo and {Kleiner}, Dane and {Kamphuis}, Peter},
        title = "{The MeerKAT Fornax Survey IV. A close look at the cluster physics through the densest rotation measure grid}",
      journal = {arXiv e-prints},
         year = 2025,
        month = jan,
          eid = {arXiv:2501.05519},
        pages = {arXiv:2501.05519},
          doi = {10.48550/arXiv.2501.05519},
archivePrefix = {arXiv},
       eprint = {2501.05519},
 primaryClass = {astro-ph.CO},
       adsurl = {https://ui.adsabs.harvard.edu/abs/2025arXiv250105519L}
}

@ARTICLE{Morris+22,
       author = {{Morris}, Melissa Elizabeth and {Wilcots}, Eric and {Hooper}, Eric and {Heinz}, Sebastian},
        title = "{How Does Environment Affect the Morphology of Radio AGN?}",
      journal = {\aj},
         year = 2022,
        month = jun,
       volume = {163},
       number = {6},
          eid = {280},
        pages = {280},
          doi = {10.3847/1538-3881/ac66db},
archivePrefix = {arXiv},
       eprint = {2204.08510},
 primaryClass = {astro-ph.GA},
       adsurl = {https://ui.adsabs.harvard.edu/abs/2022AJ....163..280M}
}

@ARTICLE{AOfluxcalref,
       author = {{Offringa}, A.~R. and {van de Gronde}, J.~J. and {Roerdink}, J.~B.~T.~M.},
        title = "{A morphological algorithm for improving radio-frequency interference detection}",
      journal = {\aap},
         year = 2012,
        month = mar,
       volume = {539},
          eid = {A95},
        pages = {A95},
          doi = {10.1051/0004-6361/201118497},
archivePrefix = {arXiv},
       eprint = {1201.3364},
 primaryClass = {astro-ph.IM},
       adsurl = {https://ui.adsabs.harvard.edu/abs/2012A&A...539A..95O}
}

@ARTICLE{Rudnick&Blundell03,
       author = {{Rudnick}, Lawrence and {Blundell}, Katherine M.},
        title = "{Lowering Inferred Cluster Magnetic Field Strengths: The Radio Galaxy Contributions}",
      journal = {\apj},
         year = 2003,
        month = may,
       volume = {588},
       number = {1},
        pages = {143-154},
          doi = {10.1086/373891},
archivePrefix = {arXiv},
       eprint = {astro-ph/0301260},
 primaryClass = {astro-ph},
       adsurl = {https://ui.adsabs.harvard.edu/abs/2003ApJ...588..143R}
}

@ARTICLE{Ensslin+03,
       author = {{Ensslin}, Torsten A. and {Vogt}, Corina and {Clarke}, T.~E. and {Taylor}, Greg B.},
        title = "{Are the Faraday Rotating Magnetic Fields Local to Intracluster Radio Galaxies?}",
      journal = {\apj},
         year = 2003,
        month = nov,
       volume = {597},
       number = {2},
        pages = {870-877},
          doi = {10.1086/378631},
archivePrefix = {arXiv},
       eprint = {astro-ph/0301552},
 primaryClass = {astro-ph},
       adsurl = {https://ui.adsabs.harvard.edu/abs/2003ApJ...597..870E}
}

@ARTICLE{PSZ1,
       author = {{Planck Collaboration} and {Ade}, P.~A.~R. and {Aghanim}, N. and {Armitage-Caplan}, C. and {Arnaud}, M. and {Ashdown}, M. and {Atrio-Barandela}, F. and {Aumont}, J. and {Aussel}, H. and {Baccigalupi}, C. and {Banday}, A.~J. and {Barreiro}, R.~B. and {Barrena}, R. and {Bartelmann}, M. and {Bartlett}, J.~G. and {Battaner}, E. and {Benabed}, K. and {Beno{\^\i}t}, A. and {Benoit-L{\'e}vy}, A. and {Bernard}, J. -P. and {Bersanelli}, M. and {Bielewicz}, P. and {Bikmaev}, I. and {Bobin}, J. and {Bock}, J.~J. and {B{\"o}hringer}, H. and {Bonaldi}, A. and {Bond}, J.~R. and {Borrill}, J. and {Bouchet}, F.~R. and {Bridges}, M. and {Bucher}, M. and {Burenin}, R. and {Burigana}, C. and {Butler}, R.~C. and {Cardoso}, J. -F. and {Carvalho}, P. and {Catalano}, A. and {Challinor}, A. and {Chamballu}, A. and {Chary}, R. -R. and {Chen}, X. and {Chiang}, H.~C. and {Chiang}, L. -Y. and {Chon}, G. and {Christensen}, P.~R. and {Churazov}, E. and {Church}, S. and {Clements}, D.~L. and {Colombi}, S. and {Colombo}, L.~P.~L. and {Comis}, B. and {Couchot}, F. and {Coulais}, A. and {Crill}, B.~P. and {Curto}, A. and {Cuttaia}, F. and {Da Silva}, A. and {Dahle}, H. and {Danese}, L. and {Davies}, R.~D. and {Davis}, R.~J. and {de Bernardis}, P. and {de Rosa}, A. and {de Zotti}, G. and {Delabrouille}, J. and {Delouis}, J. -M. and {D{\'e}mocl{\`e}s}, J. and {D{\'e}sert}, F. -X. and {Dickinson}, C. and {Diego}, J.~M. and {Dolag}, K. and {Dole}, H. and {Donzelli}, S. and {Dor{\'e}}, O. and {Douspis}, M. and {Dupac}, X. and {Efstathiou}, G. and {En{\ss}lin}, T.~A. and {Eriksen}, H.~K. and {Feroz}, F. and {Ferragamo}, A. and {Finelli}, F. and {Flores-Cacho}, I. and {Forni}, O. and {Frailis}, M. and {Franceschi}, E. and {Fromenteau}, S. and {Galeotta}, S. and {Ganga}, K. and {G{\'e}nova-Santos}, R.~T. and {Giard}, M. and {Giardino}, G. and {Gilfanov}, M. and {Giraud-H{\'e}raud}, Y. and {Gonz{\'a}lez-Nuevo}, J. and {G{\'o}rski}, K.~M. and {Grainge}, K.~J.~B. and {Gratton}, S. and {Gregorio}, A. and {Groeneboom}, E., N. and {Gruppuso}, A. and {Hansen}, F.~K. and {Hanson}, D. and {Harrison}, D. and {Hempel}, A. and {Henrot-Versill{\'e}}, S. and {Hern{\'a}ndez-Monteagudo}, C. and {Herranz}, D. and {Hildebrandt}, S.~R. and {Hivon}, E. and {Hobson}, M. and {Holmes}, W.~A. and {Hornstrup}, A. and {Hovest}, W. and {Huffenberger}, K.~M. and {Hurier}, G. and {Hurley-Walker}, N. and {Jaffe}, A.~H. and {Jaffe}, T.~R. and {Jones}, W.~C. and {Juvela}, M. and {Keih{\"a}nen}, E. and {Keskitalo}, R. and {Khamitov}, I. and {Kisner}, T.~S. and {Kneissl}, R. and {Knoche}, J. and {Knox}, L. and {Kunz}, M. and {Kurki-Suonio}, H. and {Lagache}, G. and {L{\"a}hteenm{\"a}ki}, A. and {Lamarre}, J. -M. and {Lasenby}, A. and {Laureijs}, R.~J. and {Lawrence}, C.~R. and {Leahy}, J.~P. and {Leonardi}, R. and {Le{\'o}n-Tavares}, J. and {Lesgourgues}, J. and {Li}, C. and {Liddle}, A. and {Liguori}, M. and {Lilje}, P.~B. and {Linden-V{\o}rnle}, M. and {L{\'o}pez-Caniego}, M. and {Lubin}, P.~M. and {Mac{\'\i}as-P{\'e}rez}, J.~F. and {MacTavish}, C.~J. and {Maffei}, B. and {Maino}, D. and {Mandolesi}, N. and {Maris}, M. and {Marshall}, D.~J. and {Martin}, P.~G. and {Mart{\'\i}nez-Gonz{\'a}lez}, E. and {Masi}, S. and {Massardi}, M. and {Matarrese}, S. and {Matthai}, F. and {Mazzotta}, P. and {Mei}, S. and {Meinhold}, P.~R. and {Melchiorri}, A. and {Melin}, J. -B. and {Mendes}, L. and {Mennella}, A. and {Migliaccio}, M. and {Mikkelsen}, K. and {Mitra}, S. and {Miville-Desch{\^e}nes}, M. -A. and {Moneti}, A. and {Montier}, L. and {Morgante}, G. and {Mortlock}, D. and {Munshi}, D. and {Murphy}, J.~A. and {Naselsky}, P. and {Nastasi}, A. and {Nati}, F. and {Natoli}, P. and {Nesvadba}, N.~P.~H. and {Netterfield}, C.~B. and {N{\o}rgaard-Nielsen}, H.~U. and {Noviello}, F. and {Novikov}, D. and {Novikov}, I. and {O'Dwyer}, I.~J. and {Olamaie}, M. and {Osborne}, S. and {Oxborrow}, C.~A. and {Paci}, F. and {Pagano}, L. and {Pajot}, F. and {Paoletti}, D. and {Pasian}, F. and {Patanchon}, G. and {Pearson}, T.~J. and {Perdereau}, O.},
        title = "{Planck 2013 results. XXXII. The updated Planck catalogue of Sunyaev-Zeldovich sources}",
      journal = {\aap},
         year = 2015,
        month = sep,
       volume = {581},
          eid = {A14},
        pages = {A14},
          doi = {10.1051/0004-6361/201525787},
archivePrefix = {arXiv},
       eprint = {1502.00543},
 primaryClass = {astro-ph.CO},
       adsurl = {https://ui.adsabs.harvard.edu/abs/2015A&A...581A..14P}
}

@ARTICLE{PSZ2,
       author = {{Planck Collaboration} and {Ade}, P.~A.~R. and {Aghanim}, N. and {Arnaud}, M. and {Ashdown}, M. and {Aumont}, J. and {Baccigalupi}, C. and {Banday}, A.~J. and {Barreiro}, R.~B. and {Barrena}, R. and {Bartlett}, J.~G. and {Bartolo}, N. and {Battaner}, E. and {Battye}, R. and {Benabed}, K. and {Beno{\^\i}t}, A. and {Benoit-L{\'e}vy}, A. and {Bernard}, J. -P. and {Bersanelli}, M. and {Bielewicz}, P. and {Bikmaev}, I. and {B{\"o}hringer}, H. and {Bonaldi}, A. and {Bonavera}, L. and {Bond}, J.~R. and {Borrill}, J. and {Bouchet}, F.~R. and {Bucher}, M. and {Burenin}, R. and {Burigana}, C. and {Butler}, R.~C. and {Calabrese}, E. and {Cardoso}, J. -F. and {Carvalho}, P. and {Catalano}, A. and {Challinor}, A. and {Chamballu}, A. and {Chary}, R. -R. and {Chiang}, H.~C. and {Chon}, G. and {Christensen}, P.~R. and {Clements}, D.~L. and {Colombi}, S. and {Colombo}, L.~P.~L. and {Combet}, C. and {Comis}, B. and {Couchot}, F. and {Coulais}, A. and {Crill}, B.~P. and {Curto}, A. and {Cuttaia}, F. and {Dahle}, H. and {Danese}, L. and {Davies}, R.~D. and {Davis}, R.~J. and {de Bernardis}, P. and {de Rosa}, A. and {de Zotti}, G. and {Delabrouille}, J. and {D{\'e}sert}, F. -X. and {Dickinson}, C. and {Diego}, J.~M. and {Dolag}, K. and {Dole}, H. and {Donzelli}, S. and {Dor{\'e}}, O. and {Douspis}, M. and {Ducout}, A. and {Dupac}, X. and {Efstathiou}, G. and {Eisenhardt}, P.~R.~M. and {Elsner}, F. and {En{\ss}lin}, T.~A. and {Eriksen}, H.~K. and {Falgarone}, E. and {Fergusson}, J. and {Feroz}, F. and {Ferragamo}, A. and {Finelli}, F. and {Forni}, O. and {Frailis}, M. and {Fraisse}, A.~A. and {Franceschi}, E. and {Frejsel}, A. and {Galeotta}, S. and {Galli}, S. and {Ganga}, K. and {G{\'e}nova-Santos}, R.~T. and {Giard}, M. and {Giraud-H{\'e}raud}, Y. and {Gjerl{\o}w}, E. and {Gonz{\'a}lez-Nuevo}, J. and {G{\'o}rski}, K.~M. and {Grainge}, K.~J.~B. and {Gratton}, S. and {Gregorio}, A. and {Gruppuso}, A. and {Gudmundsson}, J.~E. and {Hansen}, F.~K. and {Hanson}, D. and {Harrison}, D.~L. and {Hempel}, A. and {Henrot-Versill{\'e}}, S. and {Hern{\'a}ndez-Monteagudo}, C. and {Herranz}, D. and {Hildebrandt}, S.~R. and {Hivon}, E. and {Hobson}, M. and {Holmes}, W.~A. and {Hornstrup}, A. and {Hovest}, W. and {Huffenberger}, K.~M. and {Hurier}, G. and {Jaffe}, A.~H. and {Jaffe}, T.~R. and {Jin}, T. and {Jones}, W.~C. and {Juvela}, M. and {Keih{\"a}nen}, E. and {Keskitalo}, R. and {Khamitov}, I. and {Kisner}, T.~S. and {Kneissl}, R. and {Knoche}, J. and {Kunz}, M. and {Kurki-Suonio}, H. and {Lagache}, G. and {Lamarre}, J. -M. and {Lasenby}, A. and {Lattanzi}, M. and {Lawrence}, C.~R. and {Leonardi}, R. and {Lesgourgues}, J. and {Levrier}, F. and {Liguori}, M. and {Lilje}, P.~B. and {Linden-V{\o}rnle}, M. and {L{\'o}pez-Caniego}, M. and {Lubin}, P.~M. and {Mac{\'\i}as-P{\'e}rez}, J.~F. and {Maggio}, G. and {Maino}, D. and {Mak}, D.~S.~Y. and {Mandolesi}, N. and {Mangilli}, A. and {Martin}, P.~G. and {Mart{\'\i}nez-Gonz{\'a}lez}, E. and {Masi}, S. and {Matarrese}, S. and {Mazzotta}, P. and {McGehee}, P. and {Mei}, S. and {Melchiorri}, A. and {Melin}, J. -B. and {Mendes}, L. and {Mennella}, A. and {Migliaccio}, M. and {Mitra}, S. and {Miville-Desch{\^e}nes}, M. -A. and {Moneti}, A. and {Montier}, L. and {Morgante}, G. and {Mortlock}, D. and {Moss}, A. and {Munshi}, D. and {Murphy}, J.~A. and {Naselsky}, P. and {Nastasi}, A. and {Nati}, F. and {Natoli}, P. and {Netterfield}, C.~B. and {N{\o}rgaard-Nielsen}, H.~U. and {Noviello}, F. and {Novikov}, D. and {Novikov}, I. and {Olamaie}, M. and {Oxborrow}, C.~A. and {Paci}, F. and {Pagano}, L. and {Pajot}, F. and {Paoletti}, D. and {Pasian}, F. and {Patanchon}, G. and {Pearson}, T.~J. and {Perdereau}, O. and {Perotto}, L. and {Perrott}, Y.~C. and {Perrotta}, F. and {Pettorino}, V. and {Piacentini}, F. and {Piat}, M. and {Pierpaoli}, E. and {Pietrobon}, D. and {Plaszczynski}, S. and {Pointecouteau}, E. and {Polenta}, G. and {Pratt}, G.~W. and {Pr{\'e}zeau}, G. and {Prunet}, S. and {Puget}, J. -L.},
        title = "{Planck 2015 results. XXVII. The second Planck catalogue of Sunyaev-Zeldovich sources}",
      journal = {\aap},
         year = 2016,
        month = sep,
       volume = {594},
          eid = {A27},
        pages = {A27},
          doi = {10.1051/0004-6361/201525823},
archivePrefix = {arXiv},
       eprint = {1502.01598},
 primaryClass = {astro-ph.CO},
       adsurl = {https://ui.adsabs.harvard.edu/abs/2016A&A...594A..27P}
}

@ARTICLE{ESZ11,
       author = {{Planck Collaboration} and {Ade}, P.~A.~R. and {Aghanim}, N. and {Arnaud}, M. and {Ashdown}, M. and {Aumont}, J. and {Baccigalupi}, C. and {Balbi}, A. and {Banday}, A.~J. and {Barreiro}, R.~B. and {Bartelmann}, M. and {Bartlett}, J.~G. and {Battaner}, E. and {Battye}, R. and {Benabed}, K. and {Beno{\^\i}t}, A. and {Bernard}, J. -P. and {Bersanelli}, M. and {Bhatia}, R. and {Bock}, J.~J. and {Bonaldi}, A. and {Bond}, J.~R. and {Borrill}, J. and {Bouchet}, F.~R. and {Brown}, M.~L. and {Bucher}, M. and {Burigana}, C. and {Cabella}, P. and {Cantalupo}, C.~M. and {Cardoso}, J. -F. and {Carvalho}, P. and {Catalano}, A. and {Cay{\'o}n}, L. and {Challinor}, A. and {Chamballu}, A. and {Chary}, R. -R. and {Chiang}, L. -Y. and {Chiang}, C. and {Chon}, G. and {Christensen}, P.~R. and {Churazov}, E. and {Clements}, D.~L. and {Colafrancesco}, S. and {Colombi}, S. and {Couchot}, F. and {Coulais}, A. and {Crill}, B.~P. and {Cuttaia}, F. and {da Silva}, A. and {Dahle}, H. and {Danese}, L. and {Davis}, R.~J. and {de Bernardis}, P. and {de Gasperis}, G. and {de Rosa}, A. and {de Zotti}, G. and {Delabrouille}, J. and {Delouis}, J. -M. and {D{\'e}sert}, F. -X. and {Dickinson}, C. and {Diego}, J.~M. and {Dolag}, K. and {Dole}, H. and {Donzelli}, S. and {Dor{\'e}}, O. and {D{\"o}rl}, U. and {Douspis}, M. and {Dupac}, X. and {Efstathiou}, G. and {Eisenhardt}, P. and {En{\ss}lin}, T.~A. and {Feroz}, F. and {Finelli}, F. and {Flores-Cacho}, I. and {Forni}, O. and {Fosalba}, P. and {Frailis}, M. and {Franceschi}, E. and {Fromenteau}, S. and {Galeotta}, S. and {Ganga}, K. and {G{\'e}nova-Santos}, R.~T. and {Giard}, M. and {Giardino}, G. and {Giraud-H{\'e}raud}, Y. and {Gonz{\'a}lez-Nuevo}, J. and {Gonz{\'a}lez-Riestra}, R. and {G{\'o}rski}, K.~M. and {Grainge}, K.~J.~B. and {Gratton}, S. and {Gregorio}, A. and {Gruppuso}, A. and {Harrison}, D. and {Hein{\"a}m{\"a}ki}, P. and {Henrot-Versill{\'e}}, S. and {Hern{\'a}ndez-Monteagudo}, C. and {Herranz}, D. and {Hildebrandt}, S.~R. and {Hivon}, E. and {Hobson}, M. and {Holmes}, W.~A. and {Hovest}, W. and {Hoyland}, R.~J. and {Huffenberger}, K.~M. and {Hurier}, G. and {Hurley-Walker}, N. and {Jaffe}, A.~H. and {Jones}, W.~C. and {Juvela}, M. and {Keih{\"a}nen}, E. and {Keskitalo}, R. and {Kisner}, T.~S. and {Kneissl}, R. and {Knox}, L. and {Kurki-Suonio}, H. and {Lagache}, G. and {Lamarre}, J. -M. and {Lasenby}, A. and {Laureijs}, R.~J. and {Lawrence}, C.~R. and {Le Jeune}, M. and {Leach}, S. and {Leonardi}, R. and {Li}, C. and {Liddle}, A. and {Lilje}, P.~B. and {Linden-V{\o}rnle}, M. and {L{\'o}pez-Caniego}, M. and {Lubin}, P.~M. and {Mac{\'\i}as-P{\'e}rez}, J.~F. and {MacTavish}, C.~J. and {Maffei}, B. and {Maino}, D. and {Mandolesi}, N. and {Mann}, R. and {Maris}, M. and {Marleau}, F. and {Mart{\'\i}nez-Gonz{\'a}lez}, E. and {Masi}, S. and {Matarrese}, S. and {Matthai}, F. and {Mazzotta}, P. and {Mei}, S. and {Meinhold}, P.~R. and {Melchiorri}, A. and {Melin}, J. -B. and {Mendes}, L. and {Mennella}, A. and {Mitra}, S. and {Miville-Desch{\^e}nes}, M. -A. and {Moneti}, A. and {Montier}, L. and {Morgante}, G. and {Mortlock}, D. and {Munshi}, D. and {Murphy}, A. and {Naselsky}, P. and {Nati}, F. and {Natoli}, P. and {Netterfield}, C.~B. and {N{\o}rgaard-Nielsen}, H.~U. and {Noviello}, F. and {Novikov}, D. and {Novikov}, I. and {Olamaie}, M. and {Osborne}, S. and {Pajot}, F. and {Pasian}, F. and {Patanchon}, G. and {Pearson}, T.~J. and {Perdereau}, O. and {Perotto}, L. and {Perrotta}, F. and {Piacentini}, F. and {Piat}, M. and {Pierpaoli}, E. and {Piffaretti}, R. and {Plaszczynski}, S. and {Pointecouteau}, E. and {Polenta}, G. and {Ponthieu}, N. and {Poutanen}, T. and {Pratt}, G.~W. and {Pr{\'e}zeau}, G. and {Prunet}, S. and {Puget}, J. -L. and {Rachen}, J.~P. and {Reach}, W.~T. and {Rebolo}, R. and {Reinecke}, M. and {Renault}, C. and {Ricciardi}, S. and {Riller}, T. and {Ristorcelli}, I. and {Rocha}, G. and {Rosset}, C. and {Rubi{\~n}o-Mart{\'\i}n}, J.~A. and {Rusholme}, B. and {Saar}, E. and {Sandri}, M.},
        title = "{Planck early results. VIII. The all-sky early Sunyaev-Zeldovich cluster sample}",
      journal = {\aap},
         year = 2011,
        month = dec,
       volume = {536},
          eid = {A8},
        pages = {A8},
          doi = {10.1051/0004-6361/201116459},
archivePrefix = {arXiv},
       eprint = {1101.2024},
 primaryClass = {astro-ph.CO},
       adsurl = {https://ui.adsabs.harvard.edu/abs/2011A&A...536A...8P}
}

@ARTICLE{Planck18results,
       author = {{Planck Collaboration} and {Aghanim}, N. and {Akrami}, Y. and {Ashdown}, M. and {Aumont}, J. and {Baccigalupi}, C. and {Ballardini}, M. and {Banday}, A.~J. and {Barreiro}, R.~B. and {Bartolo}, N. and {Basak}, S. and {Battye}, R. and {Benabed}, K. and {Bernard}, J. -P. and {Bersanelli}, M. and {Bielewicz}, P. and {Bock}, J.~J. and {Bond}, J.~R. and {Borrill}, J. and {Bouchet}, F.~R. and {Boulanger}, F. and {Bucher}, M. and {Burigana}, C. and {Butler}, R.~C. and {Calabrese}, E. and {Cardoso}, J. -F. and {Carron}, J. and {Challinor}, A. and {Chiang}, H.~C. and {Chluba}, J. and {Colombo}, L.~P.~L. and {Combet}, C. and {Contreras}, D. and {Crill}, B.~P. and {Cuttaia}, F. and {de Bernardis}, P. and {de Zotti}, G. and {Delabrouille}, J. and {Delouis}, J. -M. and {Di Valentino}, E. and {Diego}, J.~M. and {Dor{\'e}}, O. and {Douspis}, M. and {Ducout}, A. and {Dupac}, X. and {Dusini}, S. and {Efstathiou}, G. and {Elsner}, F. and {En{\ss}lin}, T.~A. and {Eriksen}, H.~K. and {Fantaye}, Y. and {Farhang}, M. and {Fergusson}, J. and {Fernandez-Cobos}, R. and {Finelli}, F. and {Forastieri}, F. and {Frailis}, M. and {Fraisse}, A.~A. and {Franceschi}, E. and {Frolov}, A. and {Galeotta}, S. and {Galli}, S. and {Ganga}, K. and {G{\'e}nova-Santos}, R.~T. and {Gerbino}, M. and {Ghosh}, T. and {Gonz{\'a}lez-Nuevo}, J. and {G{\'o}rski}, K.~M. and {Gratton}, S. and {Gruppuso}, A. and {Gudmundsson}, J.~E. and {Hamann}, J. and {Handley}, W. and {Hansen}, F.~K. and {Herranz}, D. and {Hildebrandt}, S.~R. and {Hivon}, E. and {Huang}, Z. and {Jaffe}, A.~H. and {Jones}, W.~C. and {Karakci}, A. and {Keih{\"a}nen}, E. and {Keskitalo}, R. and {Kiiveri}, K. and {Kim}, J. and {Kisner}, T.~S. and {Knox}, L. and {Krachmalnicoff}, N. and {Kunz}, M. and {Kurki-Suonio}, H. and {Lagache}, G. and {Lamarre}, J. -M. and {Lasenby}, A. and {Lattanzi}, M. and {Lawrence}, C.~R. and {Le Jeune}, M. and {Lemos}, P. and {Lesgourgues}, J. and {Levrier}, F. and {Lewis}, A. and {Liguori}, M. and {Lilje}, P.~B. and {Lilley}, M. and {Lindholm}, V. and {L{\'o}pez-Caniego}, M. and {Lubin}, P.~M. and {Ma}, Y. -Z. and {Mac{\'\i}as-P{\'e}rez}, J.~F. and {Maggio}, G. and {Maino}, D. and {Mandolesi}, N. and {Mangilli}, A. and {Marcos-Caballero}, A. and {Maris}, M. and {Martin}, P.~G. and {Martinelli}, M. and {Mart{\'\i}nez-Gonz{\'a}lez}, E. and {Matarrese}, S. and {Mauri}, N. and {McEwen}, J.~D. and {Meinhold}, P.~R. and {Melchiorri}, A. and {Mennella}, A. and {Migliaccio}, M. and {Millea}, M. and {Mitra}, S. and {Miville-Desch{\^e}nes}, M. -A. and {Molinari}, D. and {Montier}, L. and {Morgante}, G. and {Moss}, A. and {Natoli}, P. and {N{\o}rgaard-Nielsen}, H.~U. and {Pagano}, L. and {Paoletti}, D. and {Partridge}, B. and {Patanchon}, G. and {Peiris}, H.~V. and {Perrotta}, F. and {Pettorino}, V. and {Piacentini}, F. and {Polastri}, L. and {Polenta}, G. and {Puget}, J. -L. and {Rachen}, J.~P. and {Reinecke}, M. and {Remazeilles}, M. and {Renzi}, A. and {Rocha}, G. and {Rosset}, C. and {Roudier}, G. and {Rubi{\~n}o-Mart{\'\i}n}, J.~A. and {Ruiz-Granados}, B. and {Salvati}, L. and {Sandri}, M. and {Savelainen}, M. and {Scott}, D. and {Shellard}, E.~P.~S. and {Sirignano}, C. and {Sirri}, G. and {Spencer}, L.~D. and {Sunyaev}, R. and {Suur-Uski}, A. -S. and {Tauber}, J.~A. and {Tavagnacco}, D. and {Tenti}, M. and {Toffolatti}, L. and {Tomasi}, M. and {Trombetti}, T. and {Valenziano}, L. and {Valiviita}, J. and {Van Tent}, B. and {Vibert}, L. and {Vielva}, P. and {Villa}, F. and {Vittorio}, N. and {Wandelt}, B.~D. and {Wehus}, I.~K. and {White}, M. and {White}, S.~D.~M. and {Zacchei}, A. and {Zonca}, A.},
        title = "{Planck 2018 results. VI. Cosmological parameters}",
      journal = {\aap},
         year = 2020,
        month = sep,
       volume = {641},
          eid = {A6},
        pages = {A6},
          doi = {10.1051/0004-6361/201833910},
archivePrefix = {arXiv},
       eprint = {1807.06209},
 primaryClass = {astro-ph.CO},
       adsurl = {https://ui.adsabs.harvard.edu/abs/2020A&A...641A...6P}
}

@ARTICLE{Pfrommer&Dursi2010,
       author = {{Pfrommer}, Christoph and {Dursi}, L.~J.},
        title = "{Detecting the orientation of magnetic fields in galaxy clusters}",
      journal = {Nature Physics},
         year = 2010,
        month = jul,
       volume = {6},
       number = {7},
        pages = {520-526},
          doi = {10.1038/nphys1657},
archivePrefix = {arXiv},
       eprint = {0911.2476},
 primaryClass = {astro-ph.CO},
       adsurl = {https://ui.adsabs.harvard.edu/abs/2010NatPh...6..520P}
}

@ARTICLE{Govoni&Feretti2004,
       author = {{Govoni}, Federica and {Feretti}, Luigina},
        title = "{Magnetic Fields in Clusters of Galaxies}",
      journal = {International Journal of Modern Physics D},
         year = 2004,
        month = jan,
       volume = {13},
       number = {8},
        pages = {1549-1594},
          doi = {10.1142/S0218271804005080},
archivePrefix = {arXiv},
       eprint = {astro-ph/0410182},
 primaryClass = {astro-ph},
       adsurl = {https://ui.adsabs.harvard.edu/abs/2004IJMPD..13.1549G}
}

@ARTICLE{vanderJagt+25,
       author = {{van der Jagt}, Stefan and {Osinga}, Erik and {van Weeren}, Reinout J. and {Miley}, George K. and {Roberts}, Ian D. and {Botteon}, Andrea and {Ignesti}, Alessandro},
        title = "{The phase-space of tailed radio galaxies in massive clusters}",
      journal = {arXiv e-prints},
         year = 2025,
        month = may,
          eid = {arXiv:2505.17334},
        pages = {arXiv:2505.17334},
          doi = {10.48550/arXiv.2505.17334},
archivePrefix = {arXiv},
       eprint = {2505.17334},
 primaryClass = {astro-ph.CO},
       adsurl = {https://ui.adsabs.harvard.edu/abs/2025arXiv250517334V}
}

@ARTICLE{Tonnesen&Bryan2008,
       author = {{Tonnesen}, Stephanie and {Bryan}, Greg L.},
        title = "{The Impact of ICM Substructure on Ram Pressure Stripping}",
      journal = {\apjl},
         year = 2008,
        month = sep,
       volume = {684},
       number = {1},
        pages = {L9},
          doi = {10.1086/592066},
archivePrefix = {arXiv},
       eprint = {0808.0007},
 primaryClass = {astro-ph},
       adsurl = {https://ui.adsabs.harvard.edu/abs/2008ApJ...684L...9T}
}

@ARTICLE{Marquart+12,
       author = {{Macquart}, J. -P. and {Ekers}, R.~D. and {Feain}, I. and {Johnston-Hollitt}, M.},
        title = "{On the Reliability of Polarization Estimation Using Rotation Measure Synthesis}",
      journal = {\apj},
         year = 2012,
        month = may,
       volume = {750},
       number = {2},
          eid = {139},
        pages = {139},
          doi = {10.1088/0004-637X/750/2/139},
archivePrefix = {arXiv},
       eprint = {1203.2706},
 primaryClass = {astro-ph.IM},
       adsurl = {https://ui.adsabs.harvard.edu/abs/2012ApJ...750..139M}
}

@INPROCEEDINGS{Johnston-Hollitt+15b,
       author = {{Johnston-Hollitt}, M. and {Dehghan}, S. and {Pratley}, L.},
        title = "{Using Tailed Radio Galaxies to Probe the Environment and Magnetic Field of Galaxy Clusters in the SKA Era}",
    booktitle = {Advancing Astrophysics with the Square Kilometre Array (AASKA14)},
         year = 2015,
        month = apr,
          eid = {101},
        pages = {101},
          doi = {10.22323/1.215.0101},
archivePrefix = {arXiv},
       eprint = {1501.00761},
 primaryClass = {astro-ph.GA},
       adsurl = {https://ui.adsabs.harvard.edu/abs/2015aska.confE.101J}
}

@ARTICLE{Morsony+13,
       author = {{Morsony}, Brian J. and {Miller}, Jacob J. and {Heinz}, Sebastian and {Freeland}, Emily and {Wilcots}, Eric and {Br{\"u}ggen}, Marcus and {Ruszkowski}, Mateusz},
        title = "{Simulations of bent-double radio sources in galaxy groups}",
      journal = {\mnras},
         year = 2013,
        month = may,
       volume = {431},
       number = {1},
        pages = {781-792},
          doi = {10.1093/mnras/stt210},
archivePrefix = {arXiv},
       eprint = {1210.1612},
 primaryClass = {astro-ph.CO},
       adsurl = {https://ui.adsabs.harvard.edu/abs/2013MNRAS.431..781M}
}

@ARTICLE{Gaensler+25,
       author = {{Gaensler}, B.~M. and {Heald}, G.~H. and {McClure-Griffiths}, N.~M. and {Anderson}, C.~S. and {Van Eck}, C.~L. and {West}, J.~L. and {Thomson}, A.~J.~M. and {Leahy}, J.~P. and {Rudnick}, L. and {Ma}, Y.~K. and {Akahori}, Takuya and {G{\"u}rkan}, G. and {Landecker}, T.~L. and {Mao}, S.~A. and {O'Sullivan}, S.~P. and {Raja}, W. and {Sun}, X. and {Vernstrom}, T. and {Baidoo}, Lerato and {Carretti}, Ettore and {Taylor}, A.~R. and {Willis}, A.~G. and {Osinga}, Erik and {Livingston}, J.~D. and {Alexander}, E.~L. and {Alonso-L{\'o}pez}, David and {Amaral}, A.~D. and {An}, T. and {Bracco}, Andrea and {Bradbury}, S. and {Br{\"u}ggen}, Marcus and {Eswaraiah}, Chakali and {En{\ss}lin}, Torsten and {Galvin}, T.~J. and {Haverkorn}, Marijke and {Hopkins}, A.~M. and {Hutschenreuter}, Sebastian and {Ideguchi}, Shinsuke and {Jaswanth}, S. and {Jung}, S. Lyla and {Kaczmarek}, J.~F. and {Kothes}, Roland and {Lazarevi{\'c}}, Sanja and {Leahy}, Denis and {Loi}, Francesca and {Marvil}, Joshua R. and {Norris}, Ray and {Pandhi}, Ayush and {Price}, Jason M. and {Riseley}, C.~J. and {Ryder}, P. and {Seta}, Amit and {Shaw}, Vasundhara and {Shen}, A.~X. and {Sobey}, C. and {Stil}, J. and {Stuardi}, Chiara and {Upasana}, Gupta and {Vanderwoude}, Shannon and {Velovi{\'c}}, Velibor},
        title = "{The Polarisation Sky Survey of the Universe's Magnetism (POSSUM): Science Goals and Survey Description}",
      journal = {Publications of the Astronomical Society of Australia},
         year = 2025,
        month = May,
          doi = {10.1017/pasa.2025.10031},
        pages = {1–32},
archivePrefix = {arXiv},
       eprint = {2505.08272},
 primaryClass = {astro-ph.GA},
       adsurl = {https://ui.adsabs.harvard.edu/abs/2025arXiv250508272G}
}

@ARTICLE{Leahy1987,
       author = {{Leahy}, J.~P.},
        title = "{Small-scale variations in the Galactic Faraday rotation.}",
      journal = {\mnras},
         year = 1987,
        month = may,
       volume = {226},
        pages = {433-446},
          doi = {10.1093/mnras/226.2.433},
       adsurl = {https://ui.adsabs.harvard.edu/abs/1987MNRAS.226..433L}
}

@ARTICLE{AndradeSantos+17,
       author = {{Andrade-Santos}, Felipe and {Jones}, Christine and {Forman}, William R. and {Lovisari}, Lorenzo and {Vikhlinin}, Alexey and {van Weeren}, Reinout J. and {Murray}, Stephen S. and {Arnaud}, Monique and {Pratt}, Gabriel W. and {D{\'e}mocl{\`e}s}, Jessica and {Kraft}, Ralph and {Mazzotta}, Pasquale and {B{\"o}hringer}, Hans and {Chon}, Gayoung and {Giacintucci}, Simona and {Clarke}, Tracy E. and {Borgani}, Stefano and {David}, Larry and {Douspis}, Marian and {Pointecouteau}, Etienne and {Dahle}, H{\r{a}}kon and {Brown}, Shea and {Aghanim}, Nabila and {Rasia}, Elena},
        title = "{The Fraction of Cool-core Clusters in X-Ray versus SZ Samples Using Chandra Observations}",
      journal = {\apj},
         year = 2017,
        month = jul,
       volume = {843},
       number = {1},
          eid = {76},
        pages = {76},
          doi = {10.3847/1538-4357/aa7461},
archivePrefix = {arXiv},
       eprint = {1703.08690},
 primaryClass = {astro-ph.CO},
       adsurl = {https://ui.adsabs.harvard.edu/abs/2017ApJ...843...76A}
}

@ARTICLE{Hardcastle+05,
       author = {{Hardcastle}, M.~J. and {Sakelliou}, I. and {Worrall}, D.~M.},
        title = "{A Chandra and XMM-Newton study of the wide-angle tail radio galaxy 3C465}",
      journal = {\mnras},
         year = 2005,
        month = may,
       volume = {359},
       number = {3},
        pages = {1007-1021},
          doi = {10.1111/j.1365-2966.2005.08966.x},
archivePrefix = {arXiv},
       eprint = {astro-ph/0502575},
 primaryClass = {astro-ph},
       adsurl = {https://ui.adsabs.harvard.edu/abs/2005MNRAS.359.1007H}
}

@ARTICLE{ODea&Baum2023,
       author = {{O'Dea}, Christopher P. and {Baum}, Stefi A.},
        title = "{Wide-Angle-Tail (WAT) Radio Sources}",
      journal = {Galaxies},
         year = 2023,
        month = may,
       volume = {11},
       number = {3},
          eid = {67},
        pages = {67},
          doi = {10.3390/galaxies11030067},
       adsurl = {https://ui.adsabs.harvard.edu/abs/2023Galax..11...67O}
}

@ARTICLE{Wing&Blanton2011,
       author = {{Wing}, Joshua D. and {Blanton}, Elizabeth L.},
        title = "{Galaxy Cluster Environments of Radio Sources}",
      journal = {\aj},
         year = 2011,
        month = mar,
       volume = {141},
       number = {3},
          eid = {88},
        pages = {88},
          doi = {10.1088/0004-6256/141/3/88},
archivePrefix = {arXiv},
       eprint = {1008.1099},
 primaryClass = {astro-ph.CO},
       adsurl = {https://ui.adsabs.harvard.edu/abs/2011AJ....141...88W}
}

@ARTICLE{Owen&Rudnick1978,
       author = {{Owen}, F.~N. and {Rudnick}, L.},
        title = "{Radio sources with wide-angle tails in Abell clusters of galaxies.}",
      journal = {\apjl},
         year = 1976,
        month = apr,
       volume = {205},
        pages = {L1-L4},
          doi = {10.1086/182077},
       adsurl = {https://ui.adsabs.harvard.edu/abs/1976ApJ...205L...1O}
}

@ARTICLE{Sakelliou&Merrifield2000,
       author = {{Sakelliou}, Irini and {Merrifield}, Michael R.},
        title = "{The origin of wide-angle tailed radio galaxies}",
      journal = {\mnras},
         year = 2000,
        month = jan,
       volume = {311},
       number = {3},
        pages = {649-656},
          doi = {10.1046/j.1365-8711.2000.03079.x},
archivePrefix = {arXiv},
       eprint = {astro-ph/9909511},
 primaryClass = {astro-ph},
       adsurl = {https://ui.adsabs.harvard.edu/abs/2000MNRAS.311..649S}
}

@ARTICLE{Blanton+01,
       author = {{Blanton}, E.~L. and {Gregg}, M.~D. and {Helfand}, D.~J. and {Becker}, R.~H. and {Leighly}, K.~M.},
        title = "{The Environments of a Complete Moderate-Redshift Sample of FIRST Bent-Double Radio Sources}",
      journal = {\aj},
         year = 2001,
        month = jun,
       volume = {121},
       number = {6},
        pages = {2915-2927},
          doi = {10.1086/321074},
archivePrefix = {arXiv},
       eprint = {astro-ph/0102499},
 primaryClass = {astro-ph},
       adsurl = {https://ui.adsabs.harvard.edu/abs/2001AJ....121.2915B}
}

@ARTICLE{Wing&Blanton2013,
       author = {{Wing}, Joshua D. and {Blanton}, Elizabeth L.},
        title = "{An Examination of the Optical Substructure of Galaxy Clusters Hosting Radio Sources}",
      journal = {\apj},
         year = 2013,
        month = apr,
       volume = {767},
       number = {2},
          eid = {102},
        pages = {102},
          doi = {10.1088/0004-637X/767/2/102},
archivePrefix = {arXiv},
       eprint = {1211.3399},
 primaryClass = {astro-ph.CO},
       adsurl = {https://ui.adsabs.harvard.edu/abs/2013ApJ...767..102W}
}

@ARTICLE{Golden-Marx+19,
       author = {{Golden-Marx}, Emmet and {Blanton}, E.~L. and {Paterno-Mahler}, R. and {Brodwin}, M. and {Ashby}, M.~L.~N. and {Lemaux}, B.~C. and {Lubin}, L.~M. and {Gal}, R.~R. and {Tomczak}, A.~R.},
        title = "{The High-redshift Clusters Occupied by Bent Radio AGN (COBRA) Survey: Follow-up Optical Imaging}",
      journal = {\apj},
         year = 2019,
        month = dec,
       volume = {887},
       number = {1},
          eid = {50},
        pages = {50},
          doi = {10.3847/1538-4357/ab5106},
archivePrefix = {arXiv},
       eprint = {1910.11884},
 primaryClass = {astro-ph.GA},
       adsurl = {https://ui.adsabs.harvard.edu/abs/2019ApJ...887...50G}
}

@ARTICLE{deVos+21,
       author = {{de Vos}, K. and {Hatch}, N.~A. and {Merrifield}, M.~R. and {Mingo}, B.},
        title = "{Clusters' far-reaching influence on narrow-angle tail radio galaxies}",
      journal = {\mnras},
         year = 2021,
        month = sep,
       volume = {506},
       number = {1},
        pages = {L55-L58},
          doi = {10.1093/mnrasl/slab075},
archivePrefix = {arXiv},
       eprint = {2107.00449},
 primaryClass = {astro-ph.GA},
       adsurl = {https://ui.adsabs.harvard.edu/abs/2021MNRAS.506L..55D}
}

@ARTICLE{Vardoulaki+21,
       author = {{Vardoulaki}, Eleni and {Vazza}, Franco and {Jim{\'e}nez-Andrade}, Eric F. and {Gozaliasl}, Ghassem and {Finoguenov}, Alexis and {Wittor}, Denis},
        title = "{Bent It Like FRs: Extended Radio AGN in the COSMOS Field and Their Large-Scale Environment}",
      journal = {Galaxies},
         year = 2021,
        month = nov,
       volume = {9},
       number = {4},
          eid = {93},
        pages = {93},
          doi = {10.3390/galaxies9040093},
archivePrefix = {arXiv},
       eprint = {2110.00034},
 primaryClass = {astro-ph.GA},
       adsurl = {https://ui.adsabs.harvard.edu/abs/2021Galax...9...93V}
}

@ARTICLE{Vardoulaki+25,
       author = {{Vardoulaki}, E. and {Back{\"o}fer}, V. and {Finoguenov}, A. and {Vazza}, F. and {Comparat}, J. and {Gozaliasl}, G. and {Whittam}, I.~H. and {Hale}, C.~L. and {Weaver}, J.~R. and {Koekemoer}, A.~M. and {Collier}, J.~D. and {Frank}, B. and {Heywood}, I. and {Sekhar}, S. and {Taylor}, A.~R. and {Pinjarkar}, S. and {Hardcastle}, M.~J. and {Shimwell}, T. and {Hoeft}, M. and {White}, S.~V. and {An}, F. and {Tabatabaei}, F. and {Randriamanakoto}, Z. and {Filipovic}, M.~D.},
        title = "{The jet paths of radio active galactic nuclei and their cluster weather}",
      journal = {\aap},
         year = 2025,
        month = mar,
       volume = {695},
          eid = {A178},
        pages = {A178},
          doi = {10.1051/0004-6361/202453180},
       adsurl = {https://ui.adsabs.harvard.edu/abs/2025A&A...695A.178V}
}

@ARTICLE{Becker+95,
       author = {{Becker}, Robert H. and {White}, Richard L. and {Helfand}, David J.},
        title = "{The FIRST Survey: Faint Images of the Radio Sky at Twenty Centimeters}",
      journal = {\apj},
         year = 1995,
        month = sep,
       volume = {450},
        pages = {559},
          doi = {10.1086/176166},
       adsurl = {https://ui.adsabs.harvard.edu/abs/1995ApJ...450..559B}
}
\bibliographystyle{aasjournal}

\end{document}